\newif\iffigs\figstrue
\documentclass[a4paper,11pt]{article}
\usepackage{latexsym,amssymb,lscape,graphics}
\pdfoutput=1
\usepackage{graphicx}        
\usepackage{longtable}
\usepackage{multirow}
\usepackage{color}
\usepackage{slashed,epsfig}
\usepackage{amsfonts}

\newtheorem{definizione}{Definition}[section]

\newcommand{\bd}{\begin{definizione}}
\newcommand{\ed}{\end{definizione}}

\def\IC{\relax\,\hbox{$\inbar\kern-.3em{\rm C}$}}
\def\IG{\relax\,\hbox{$\inbar\kern-.3em{\rm G}$}}
\def\IB{\relax{\rm I\kern-.18em B}}
\def\ID{\relax{\rm I\kern-.18em D}}
\def\IL{\relax{\rm I\kern-.18em L}}
\def\IF{\relax{\rm I\kern-.18em F}}
\def\IH{\relax{\rm I\kern-.18em H}}
\def\II{\relax{\rm I\kern-.17em I}}
\def\IN{\relax{\rm I\kern-.18em N}}
\def\IP{\relax{\rm I\kern-.18em P}}
\def\IQ{\relax\,\hbox{$\inbar\kern-.3em{\rm Q}$}}
\def\bfzero{\relax\,\hbox{$\inbar\kern-.3em{\rm 0}$}}
\def\IK{\relax{\rm I\kern-.18em K}}
\def\IG{\relax\,\hbox{$\inbar\kern-.3em{\rm G}$}}
 \font\cmss=cmss10 \font\cmsss=cmss10 at 7pt
\def\IR{\relax{\rm I\kern-.18em R}}
\def\ZZ{\relax\ifmmode\mathchoice
{\hbox{\cmss Z\kern-.4em Z}}{\hbox{\cmss Z\kern-.4em Z}}
{\lower.9pt\hbox{\cmsss Z\kern-.4em Z}} {\lower1.2pt\hbox{\cmsss
Z\kern-.4em Z}}\else{\cmss Z\kern-.4em Z}\fi}
\def\bfone{\relax{\rm 1\kern-.35em 1}}

\def\inbar{\vrule height1.5ex width.4pt depth0pt}
\def\bfzero{\relax{\rm I\kern-.18em 0}}
\def\bfone{\relax{\rm 1\kern-.35em 1}}

\DeclareFontFamily{U}{rsf}{} \DeclareFontShape{U}{rsf}{m}{n}{
  <5> <6> rsfs5 <7> <8> <9> rsfs7 <10-> rsfs10}{}
\DeclareMathAlphabet\Scr{U}{rsf}{m}{n}

\newcommand{\ft}[2]{{\textstyle\frac{#1}{#2}}}
\def\tilde{\widetilde}

\def\1bar{1\hskip -.275cm -}
\def\2bar{2\hskip -.275cm -}
\def\3bar{3\hskip -.275cm -}

\newsavebox{\uuunit}
\sbox{\uuunit}
                 {\setlength{\unitlength}{0.825em}
                      \begin{picture}(0.6,0.7)
                                      \thinlines
                                      \put(0,0){\line(1,0){0.5}}
                                      \put(0.15,0){\line(0,1){0.7}}
                                      \put(0.35,0){\line(0,1){0.8}}
                                     \multiput(0.3,0.8)(-0.04,-0.02){10}{\rule{0.5pt}{0.5pt}}
                      \end {picture}}

\makeatletter \@addtoreset{equation}{section} \makeatother

\def\bfone{\relax{\rm 1\kern-.35em 1}}

\def\bfone{\relax{\rm 1\kern-.35em 1}}
\font\cmss=cmss10 \font\cmsss=cmss10 at 7pt

\newcommand{\slal}{\mathfrak{sl}}

\def\IE{\relax{{\rm I\kern-.18em E}}}

\def\IGam{\relax{{\rm I}\kern-.18em \Gamma}}

\def\IA{\relax{\hbox{{\rm A}\kern-.82em {\rm A}}}}

\begin{document}
\begin{titlepage}
\vskip 0.2cm
\hskip 1.5cm
\vbox{\hbox{CERN-PH-TH/2014-001}}
\vfill
\vskip 0.5cm
\begin{center}
{\Large {\sc    On the  Gauged  K\"ahler Isometry \\
in Minimal Supergravity Models of Inflation}}\\[1cm]
{ \large Sergio Ferrara$^{a}$, Pietro Fr\'e$^{b}$\footnote{Prof. Fr\'e is presently fulfilling the duties of Scientific Counselor of the Italian Embassy in the Russian Federation, Denezhnij pereulok, 5, 121002 Moscow, Russia.}, Alexander S. Sorin$^{c}$ }
{}~\\
{}~\\
\quad
\quad \\{{\em $^a$ Physics Department, Theory Unit, CERN, CH 1211, Geneva 23, Switzerland, {\tt and}\\
INFN - Laboratori Nazionali di Frascati, Via Enrico Fermi 40, I-00044, Frascati, Italy}}, {\tt and}\quad \\
Department of Physics and Astronomy, University of California, Los Angeles, CA 90095-1547, USA~\quad \\
{\tt sergio.ferrara@cern.ch}
{}~\\
\quad \\
{{\em $^{b}$  Dipartimento di Fisica, Universit\'a di Torino,}}
\\
{{\em $\&$ INFN - Sezione di Torino}}\\
{\em via P. Giuria 1, I-10125 Torino, Italy}~\quad\\
{\tt pietro.fre@esteri.it}
{}~\\
\quad \\
{{\em $^{c}$ Bogoliubov Laboratory of Theoretical Physics, {\tt and} }}\\
{{\em  Veksler and Baldin Laboratory of High Energy Physics,}}\\
{{\em Joint Institute for Nuclear Research,}}\\
{\em 141980 Dubna, Moscow Region, Russia}~\quad\\
{\tt sorin@theor.jinr.ru}
\quad \\
\end{center}
~{}
~{}
\begin{abstract}
In this paper we address the question how to discriminate whether the  gauged isometry group $\mathcal{G}_\Sigma$ of the K\"ahler manifold $\Sigma$ that produces a D-type inflaton potential in a Minimal Supergravity Model is elliptic, hyperbolic or parabolic. We show that the classification of  isometries of symmetric cosets  can be extended to non  symmetric $\Sigma$.s if these manifolds satisfy additional mathematical restrictions. The classification criteria established in the mathematical literature are coherent with simple criteria formulated in terms of the asymptotic behavior of the K\"ahler potential $\mathcal{K}(C) \, = \, 2 \,J(C)$ where the real scalar field $C$ encodes the inflaton field. As a by product of our analysis we show that phenomenologically admissible potentials for the description of inflation and in particular $\alpha$-attractors are mostly obtained from the gauging of a parabolic isometry, this being, in particular the case of the Starobinsky model. Yet at least one exception exists of an elliptic $\alpha$-attractor, so that neither type of isometry can be a priori excluded. The requirement of regularity of the manifold $\Sigma$ poses instead strong constraints on the $\alpha$-attractors and reduces their space considerably. Curiously there is a unique integrable $\alpha$-attractor corresponding to a particular value of this parameter.
\end{abstract}
\end{titlepage}
\tableofcontents
\newpage
\section{Introduction}
Recent months have witnessed a lot of interest in the analysis of inflationary scenarios based on one scalar field $\phi$ (the inflaton). The vast theoretical literature on such a subject has been specially promoted by the publication and by the analysis of the results on the CMB power spectrum, firstly obtained  by the WMAP mission and more recently confirmed with a significantly enhanced accury by the PLANCK satellite \cite{Ade:2013uln,Ade:2013zuv,Hinshaw:2012aka}. Indeed according to observations the best theoretical modelling of the inflationary paradigm seems to be the simplest one in terms of a single inflaton  endowed with a suitable potential $V(\phi)$. In this way  the comparison with data is reduced to the framework provided by a spatially flat FLRW metric:
\begin{equation}\label{FLRW}
        ds^2 \ = \ - \   \mathrm{d}t^2 \ + \ a^2(t) \ \mathrm{d}\mathbf{x}\cdot \mathrm{d}\mathbf{x}
      \end{equation}
and by the Friedman equations that govern the evolution  of both the inflaton field $\phi(t)$ and the scale factor $a(t)$.
\par
A quite relevant question addressed by several research groups within this framework is whether some of the phenomenologically most favorite potentials can be embedded into supergravity theories. Such a question  has recently obtained a solution in terms of \textit{Minimal $\mathcal{N}=1$ supergravity models}. According to this scheme introduced in a series of recent papers \cite{minimalsergioKLP,Ferrara:2013wka,Ferrara:2013kca}, any inflation model based on a positive definite potential can be embedded into  $\mathcal{N}=1$  supergravity, coupled to a single Wess-Zumino multiplet and one massless vector multiplet, which may combine together in a massive vector multiplet with a lagrangian specified by a single real function $J(C)$\footnote{Please note that due to the Supergravity conventions utilized in this paper the sign of the funtion $J(C)$ is inverted with respect to the sign utilized in \cite{VanProeyen:1979ks}}  as shown in \cite{VanProeyen:1979ks} and, before, for a special case, in \cite{Freedman:1976uk}. The vector multiplet is utilized to gauge an isometry of the one-dimensional Hodge-Kahler manifold $\Sigma$ associated with the WZ multiplet. The catch of the method is the supergravity formulation of the Higgs mechanism, since the gauging introduces a D-term and the positive definite potential $V$ is the square of the momentum-map of of the Killing vector $k^{\mathfrak{z}}$
that generates the gauged isometry of $\Sigma$.
\par
As first conceptualized by two of us in \cite{primosashapietro,piesashatwo}, this scheme introduces a mathematical map, named
the \textit{$D$-map}, that,  to every positive definite potential $V(\phi)$ associates a two-dimensional Riemannian manifold $(\Sigma,g)$ endowed, by its very definition, with a continuous one-dimensional group $\mathcal{G}_\Sigma$ of isometries\footnote{Note that in a supersymmetric Lagrangian as given in eq.(\ref{minsugra}), the $D$-term is $D=g \, J^\prime(C)\, = \, \mathcal{P}(\phi)$, where the relations (\ref{gartoccio}) and (\ref{curlandia}) have been used.}. The local data encoded in the potential  provide the metric $g$ of the surface $\Sigma$:
\begin{equation}\label{corrusco}
    ds^2_{g} \, = \,   d\phi^2 \, + \, \underbrace{\left(\frac{\mathrm{d}\sqrt{V(\phi)}}{\mathrm{d}\phi}\right)^2}_{f^2(\phi)} \, dB^2
\end{equation}
but do not give immediate answer to following three mathematical questions that are crucial for the construction of the corresponding minimal supergravity model:
\begin{enumerate}
  \item What is the global topology of the manifold $\Sigma$ whose metric is (\ref{corrusco})?
  \item What is the appropriate complex structure of $\Sigma$ and the appropriate local complex coordinate $\mathfrak{z}$ by means of which (\ref{corrusco}) is turned into a standard K\"ahler metric:
      \begin{equation}\label{standardkalle}
        ds^2_{g} \, = \, \partial{\mathfrak{z}}\,\partial{\bar{\mathfrak{z}}} \, \mathcal{K}(\mathfrak{z}\, ,\, \bar{\mathfrak{z}}) \, d\mathfrak{z} \, d\bar{\mathfrak{z}} \quad ; \quad \mathcal{K}(\mathfrak{z}\, ,\, \bar{\mathfrak{z}}) \, = \, \mathcal{K}^\star(\mathfrak{z}\, ,\, \bar{\mathfrak{z}})
      \end{equation}
      having denoted by $\mathcal{K}(\mathfrak{z}\, ,\, \bar{\mathfrak{z}})$ the K\"ahler potential?
 \item What is the nature of the isometry group $\mathcal{G}_\Sigma$ whose elements are locally described by the shifts of the cyclic coordinate $B$, namely $B\to B+\alpha$? Is $\mathcal{G}_\Sigma$ a compact $\mathrm{U(1)}$ group or is it non-compact? Are the isometries encoded in $\mathcal{G}_\Sigma$, elliptic, hyperbolic or parabolic?
\end{enumerate}
These three questions are not independent since the answer to anyone of them determines the answer to the other two. The investigation of such an issue is the main goal and the main topic of the present paper.
In \cite{sergiosashapietroOne} the relevance of this point was emphasized by illustrating the quite different symmetry pattern of minimal supergravities respectively based on a compact and on a non-compact isometry group  $\mathcal{G}_\Sigma$. Furthermore in \cite{sergiosashapietroOne} we provided a complete analysis  of cosmological models based on constant curvature manifolds K\"ahler $\Sigma$. In the present paper we extend our analysis to cosmological potentials that emerge from K\"ahler manifolds whose metric is of type (\ref{corrusco}) but the curvature is not constant.
\par
As it was already done in \cite{sergiosashapietroOne} we will show that a decision about the nature of the isometry group $\mathcal{G}_\Sigma$ can be taken by inspecting the form of the asymptotic expansions of the function $J(C)$ for values of $C$ corresponding to the boundary and to the origin of $\Sigma$. The geometrical characterization of the coordinate $C(\phi)$, named by us  the Van Proeyen coordinate,   as the integral of the complex structure equation, was already pointed out both in \cite{piesashatwo} and in \cite{sergiosashapietroOne}, while the essential identification of $J(C)$ with the K\"ahler potential was introduced in \cite{minimalsergioKLP}. Hence the topology of all the K\"ahler manifolds $\Sigma$ that are located in the image of the $D$-map, is characterized by the analytic behavior of their K\"ahler potential on the boundary $\partial\Sigma$ and in the origin of $p_0 \, \in \,\Sigma$. The boundary is a clear geometrical concept for all manifolds, while what is the origin $p_0$ is clear for coset manifolds but much less clear for generic manifolds. A careful analysis shows that the correct extension of this concept to non-constant curvature manifolds is provided by the identification of $p_0$ with a fixed point of the isometry group $\mathcal{G}_\Sigma$:
\begin{equation}\label{fissopunto}
    \forall \, \Gamma \, \in \, \mathcal{G}_\Sigma \, \quad \Gamma.p_0 \, = \, p_0
\end{equation}
The existence of such a fixed point in the interior of the manifold is a necessary and sufficient condition to identify the group $\mathcal{G}_\Sigma$ with a compact $\mathrm{U(1)}$ and to determine the complex coordinate $\mathfrak{z}$ accordingly\footnote{As we will discuss more extensively in the sequel and illustrate with an explicit example in sect.\ref{dottorsottile}, what we stated here holds true if the manifold $\Sigma$ is simply connected, namely if $\pi_1(\Sigma)\, = \, 1$. If not, the existence of a fixed point $p_0$ is no longer implied by the elliptic character of the isometry group. }. When $p_0$ exists
it corresponds to either $C\to +\infty$ or $C\to -\infty$ and in the neighborhood of $p_0$ the $J(C)$ function behaves as follows:
\begin{equation}\label{cammello}
    J(C) \, \simeq \, \exp\left[ \mp \,|\delta| \,C \right] \, \stackrel{C\to \pm \infty}{\to} \, 0
\end{equation}
When $p_0$ does not exist the group $\mathcal{G}_\Sigma$ is necessarily non-compact and it describes isometries that are either parabolic or hyperbolic\footnote{Once again what we state here holds true for simply connected manifolds. When $\pi_1(\Sigma) \ne 1$ from the absence of a fixed point we cannot conclude that the isometry is non-compact. Indeed in section \ref{dottorsottile} we precisely discuss an example where the isometry is elliptic, yet there is no fixed point $p_0$ inside the manifold that is multiply connected.}.
\par
On the boundary $\partial\Sigma$ of the manifold, corresponding  either to $C\to C_\partial=0$ or to $C\to C_\partial=\infty$ or still to  $C\to C_\partial=-\,\infty$ the behavior of $J(C)$ is universal in the following sense.
\begin{enumerate}
  \item In the case of an elliptic isometry there are two possible topologies of $\Sigma$, either the topology of the disk which corresponds to a simply connected manifold $\pi_1(\Sigma)  \, = \, 1$, or the topology of the  circular corona which is not simply connected $\pi_1(\Sigma) \, = \, \mathbb{Z}$.
  \item In the case of the disk-topology, the boundary $\partial\Sigma$ is unique and diffeomorphic to the unit circle bounding the unit disk. On this boundary the limiting curvature can be either negative $R_\partial <0$ or vanishing.
  \item If $R_\partial <0$ the behavior of $J(C)$ for $C\to C_\partial$ is:
  \begin{equation}\label{cruccognocco}
    J(C) \, \stackrel{C\to C_{\partial}}{\approx} \,  \frac{1}{ R_{\partial}} \, \log\left[C\right]
\end{equation}
\item If $R_\partial =0$ the boundary is necessarily at $C =C_\partial \, = \, \infty$ or at $C =C_\partial \, = \,- \infty$ there are two possibilities, either:
 \begin{eqnarray}\label{cruccomalloreddu}
    J(C) & \stackrel{C\to C_{\partial}}{\approx} & \exp\left[\delta_\partial C\right]\nonumber\\
    & \mbox{or} &\nonumber\\
    J(C) & \stackrel{C\to C_{\partial}}{\approx} & \kappa_\partial \, C^2
\end{eqnarray}
where $\delta_\partial,\kappa_\partial$ are some appropriate real numbers. Indeed those above are the two possible solutions for a K\"ahler potential yielding a zero-curvature K\"ahler metric with a continuous isometry.
\item In the case of the corona topology there are two boundaries and they can be either $C_\partial \, = \, 0$ and $C_{\partial^\prime} \, = \, \infty$ or $C_\partial \, = \,  \infty$ and $C_{\partial^\prime} \, = - \, \infty$.
    \item A priori on each of the two boundaries of the corona topology, the curvature can be independently negative or zero. For negative curvatures we have:
        \begin{equation}\label{barilotto}
          J(C) \, \stackrel{C\to C_{\partial}}{\approx} \,  \frac{1}{ R_{\partial}} \, \log\left[C\right]  \quad ; \quad J(C) \, \stackrel{C\to C_{\partial^\prime}}{\approx} \,  \frac{1}{ R_{\partial^\prime}} \, \log\left[C\right]
        \end{equation}
        If on any or both of the two boundaries the limiting curvature is zero such boundary must be located at $C_\partial \, = \, \infty$ or $C_\partial \, = \, - \, \infty$  so that the behavior of $J(C)$ on such a boundary can take one of the two universal forms required by zero-curvature that were recalled in eq.(\ref{cruccomalloreddu}):
\end{enumerate}
\par
In the case of parabolic isometries the topology of the manifold $\Sigma$ must be that of the upper complex plane $\mathbb{H}_+$ to which $\Sigma$ should be diffeomorphic. In this case there are no interior fixed points but there are always two boundaries $\partial\Sigma$ and $\partial^{\,\prime}\Sigma$ that, in Poincar\'e plane geometry, respectively correspond to the real axis ($\mathrm{Im} t \, \to \, 0$) and to the point at infinity ($\mathrm{Im} t \, \to \, \infty$).
On both boundaries we have once again the options we had for the elliptic case. Either the limiting curvature is negative $R_{\partial} < 0$ or it vanishes $R_{\partial} = 0$. A priori all situations are possible:
\begin{equation}\label{frigolino}
  \left(R_{\partial} \, , \, R_{\partial^\prime}\right) \, = \, \left\{\begin{array}{c}
                                                                         (< 0, <0) \\
                                                                          (< 0, 0)  \\
                                                                          ( 0, 0)
                                                                       \end{array}
   \right.
\end{equation}
On any boundary where the limiting curvature is $R_{\partial} < 0$ the behavior of the $J(C)$ function must display a logarithmic divergence:
\begin{equation}\label{frigidario}
  J(C) \, \stackrel{C\to C_{\partial}}{\approx} \,  \frac{1}{ R_{\partial}} \, \log\left[C\right]
\end{equation}
This implies that necessarily we have either $C_{\partial}\, = \, 0$ or  $C_{\partial}\, = \, \infty$. On the other hand if on any boundary the limiting curvature is zero, there we must find one of the two options recalled in eq.(\ref{cruccomalloreddu}).
The first option makes sense for $C_\partial \, = \,  \pm \infty$ (depending on the sign of $\delta_\partial$) while the second makes sense for both  $C_\partial \, = \,  0$ and  $C_\partial \, = \,  \pm \infty$ (depending on the sign of $\delta_\partial$).
\par
Hyperbolic isometries are on the other hand characterized by the fact the VP coordinate $C$ exists only on a finite range $\left [C_{min}, C_{max}\right ]$.
\par
These criteria that can be extracted from the heuristic considerations put forward  in the sequel  and  that are confirmed by all the examples presented in this paper, can be given a solid mathematical ground if the considered manifolds $\Sigma$ are restricted to be  \textit{Hadamard manifolds}. As we recall in appendix \ref{mathtopo},  by definition Hadamard manifolds are \textit{smooth, simply connected manifolds whose scalar curvature is everywhere finite and non-positive definite}. In such manifolds the possible isometries can be unambiguously classified as elliptic, hyperbolic or parabolic and the existence of a fixed point $p_0$ in the interior of $\Sigma$ is, according to Gromov et al \cite{Gromov1985}, the defining property of an elliptic isometry group, namely $\mathrm{U(1)}$ \footnote{Once again let us remind the reader that the existence of  a fixed point $p_0$ is necessary for elliptic isometries of Hadamard manifolds which, by definition are simply connected. When $\pi_1(\Sigma) \ne 1$ (corona topology) we can have elliptic isometries without a fixed point. Yet simple connectedness seems a necessary property of manifolds apt to describe Wess-Zumino multiples in supergravity, so that the above statements are correct in the framework of supergravity.}.
\par
All our examples are consistent with Gromov's classification of isometries although the corresponding manifolds $\Sigma$ are not strictly Hadamard, since their curvature is mostly negative and tends to negative values at the boundaries, yet it attains some maximum value $R\, = \, k$ in the interior which can be positive. This suggests \textbf{the conjecture} that the characterization of isometries in terms of fixed points, at least in the case of two-dimensions, can be extended to the class of $\mathrm{CAT}(k)$ 2-manifolds according to the definition of Gromov \cite{Gromov1987}.  These are simply connected manifolds with an upper bound on the curvature,  and other distinguished convexity properties which allow  for the definition of a distance, Hadamard manifolds being just the $\mathrm{CAT}(0)$ manifolds where the maximum allowed value of $R$ is zero.  In view of such a conjecture  the acceptable cosmological potentials for minimal supergravity might be singled out as those that  $D$-map to   $\mathrm{CAT}(k)$ 2-manifolds.
\par
In this paper, after a general discussion of the main issue presented above (sect.s \ref{genOne},\ref{genTwo}), we consider (in
sect.\ref{trattoriagricoli}) the case of $\alpha$-attractors introduced by Kallosh et al in \cite{alfatrattori}, that incorporate in a systematic scheme also several results obtained by different groups of authors in various previous papers \cite{Kallosh:2013hoa,Kallosh:2013lkr,Farakos:2013cqa,Kallosh:2013maa,Kallosh:2013daa,Kallosh:2013tua,johndimitri,Ketov:2010qz,Ketov:2012jt}. Using these $\alpha$-attractors  as prototypes of cosmological inflaton potentials consistent with PLANCK data, we show that one of the integrable potentials classified in \cite{noicosmoitegr} and \cite{mariosashapietrocosmo}, whose list is recalled in tables \ref{tab:families} and \ref{Sporadic}, namely that identified as $I_6$, fits into the definition of $\alpha$-attractors with the value $\alpha \, = \, \ft 49$. This provides the challenging perspective of writing in an explicit analytic form the Mukhanov Sasaki equation \cite{sasakimukhanov} for a background compatible with experimental CMB data.
\par
Next in three separate sections
(sect.s \ref{ellittica},\ref{parabolica},\ref{iperbarico}), each containing several subsections, we present examples of K\"ahler manifolds $\Sigma$ in the image of the $D$-map where the isometry group $\mathcal{G}_\Sigma$ is respectively elliptic, parabolic and hyperbolic, as dictated  by  the criteria based on asymptotic expansions of the $J(C)$ function, which turn out to coincide with the criteria based on Gromov et al classification. In particular we show that many of the  $\alpha$-attractors and of the other potentials that might play a role in the description of inflation, or of its early phase, encoded in the phenomenon of climbing scalars \cite{Dudasprimo,Dudas:2012vv,Sagnotti:2013ica,SagnottiDubna}, belong to the group where the isometry is parabolic. Yet we also have some $\alpha$-attractors that are obtained from  the gauging of an elliptic isometry group, so that the issue whether parabolic or elliptic gaugings are favored by observations cannot be decided at the present time.
\par
Section \ref{concludo} contains our conclusions. Appendix \ref{integralnymodely}, contains, for the reader's convenience a summary of the bestiary of integrable inflaton potentials with the definition of what integrable potential actually means.
Appendix \ref{mathtopo} is mathematical. It summarizes the classification of isometries for Hadamard manifolds according to Gromov et al \cite{Gromov1985}.
\par
Finally before turning to the development of the program we have outlined let us make some comments on the used notations.
In those adopted in papers, \cite{mariosashapietrocosmo} and \cite{sergiosashapietroOne}, the Friedman equations that govern the time evolution  are written as follows:
\begin{equation}
 H^2 \ = \  \frac{1}{3} \ \dot{\phi}^2 \, + \, \frac{2}{3} \ V(\phi)  \quad ; \quad  \dot{H}  \ = \ - \, \dot{\phi}^2   \quad ; \quad \ddot{\phi} \,+ \,  3 \, H \, \dot{\phi} \, + \, V^{\,\prime}  \ = \ 0  \label{fridmano}
\end{equation}
where
$ H(t) \, \equiv \, \frac{\dot{a}(t)}{a(t)}$ is the Hubble function. Equations (2.1) and (2.2) of the recent review \cite{Encyclopaedia} of inflationary models coincide with eq.s (\ref{fridmano})  if one chooses the convention $2 \, M_{Pl}^2 \, = \, 1$. This observation, together with the statement that the kinetic term of the inflaton is canonical in our lagrangian :
\begin{equation}\label{gargamello}
  \mathcal{L}\, = \, \dots + \, \ft 12 \, \partial_\mu \phi \partial^\mu \phi \, + \, \dots
\end{equation}
 fixes completely all normalizations and allows the comparison of the results  presented there  with any other result  in  the vast literature on inflation. Indeed such a comparison is immediately done by using the following conversion rule to the notations where $ M_{Pl}^2 \, = \, 1$. Naming ${\hat \phi}$ the canonical scalar field in the physical units  $ M_{Pl}^2 \, = \, 1$, its relation with the field $\phi$ utilized in  \cite{primosashapietro,piesashatwo,mariosashapietrocosmo,sergiosashapietroOne} is the following one:
 \begin{equation}\label{ciurlacco}
  \phi \, = \, \frac{1}{\sqrt{2}} \, {\hat \phi}
 \end{equation}
This observation is quite relevant in relation with the specific form of the potentials that happen to be integrable. To make our results more easily readable we have summarized the final formulae for the various potentials discussed in this paper in terms of the physical field ${\hat \phi}$ (see in particular tables
\ref{tab:families},\ref{potenziallini}, \ref{Sporadic},\ref{integpotenziallini}). In the discussion of the geometrical structures related with the K\"ahler surface $\Sigma$ we mostly use the coordinate $\phi$, since this simplifies a lot of prefactors and makes formulae neater.
\section{Complex Structures and the Asymptotic Behavior of the K\"ahler potential $\propto J(C)$}
\label{genOne}
As we advocated in \cite{primosashapietro,piesashatwo,sergiosashapietroOne,mariosashapietrocosmo}, in the minimal $\mathcal{N}=1$ supergravity realizations of one--scalar cosmologies the central item of the construction is a K\"ahler surface with a one-dimensional isometry group whose metric can be written as follows:
\begin{equation}\label{metraxia}
  ds^2_\Sigma \, = \, p(U) \, dU^2 \, + \, q(U) \, dB^2
\end{equation}
$p(U),q(U)$ being two positive definite functions of their argument. The isometry group  of the manifold $\Sigma$ is generated by  the Killing vector $\vec{k}_{[B]} \, = \, \partial_B$. This isometry is fundamental since it is by means of its gauging that one produces a $D$-type positive definite scalar potential that can encode the inflaton dynamics.  At the level of the supergravity model that is built by using the K\"ahler space $\Sigma$ as the target manifold where  the two scalar fields of the inflatonic Wess Zumino multiplet take values, a fundamental question is whether $\vec{k}_{[B]}$ generates a \textbf{compact rotation symmetry}, or \textbf{non compact symmetry either parabolic or hyperbolic}. As we showed in \cite{sergiosashapietroOne}  the supergravity lagrangian in general and its fermionic sector in particular,  display quite different features in the two cases, leading to a different pattern of physical charges and symmetries. Actually when $\Sigma \, = \, \Sigma_{max}$ is a constant curvature surface namely the coset manifold $\frac{\mathrm{SU(1,1)}}{\mathrm{U(1)}}\sim \frac{\mathrm{SL(2,\mathbb{R})}}{\mathrm{O(2)}}$, there is also a third possibility. In such a situation the killing vector  $\vec{k}_{[B]}$ can be the generator of a \textbf{dilatation}, namely it can correspond to a non-compact but semi-simple element $\mathbf{d}\, = \, \left(
                                                                                 \begin{array}{cc}
                                                                                   1 & 0 \\
                                                                                   0 & -1 \\
                                                                                 \end{array}
                                                                               \right)
$ of the Lie algebra $\mathrm{SL(2,\mathbb{R})}$ rather then to a nilpotent one $\mathbf{t} \, = \,\left(
                                     \begin{array}{cc}
                                       0 & 1 \\
                                       0 & 0 \\
                                     \end{array}
                                   \right)
$.
As it was explained in \cite{piesashatwo}, the standard presentation of the geometry of $\Sigma$ in terms of a complex coordinate and of a K\"ahler potential is obtained by means of the following steps. First one singles out the unique complex structure with vanishing Nienhuis tensor with respect to which the metric is hermitian:
$
   \mathfrak{J}_\alpha^\beta \,\mathfrak{ J}_\beta^\gamma \, = \, - \, \delta^\gamma_\alpha \quad ;\quad \partial_{[\alpha } \, \mathfrak{J}^\gamma_{\beta]} \, - \, \mathfrak{J}^\mu_\alpha \,
  \mathfrak{J}^\nu_\beta \,  \partial_{[\mu } \, \mathfrak{J}^\gamma_{\nu]} \, = \, 0 \quad ; \quad  g_{\alpha\beta} \, = \, \mathfrak{ J}_\alpha^\gamma \, \mathfrak{ J}_\beta^\delta \, g_{\gamma\delta}
$.
In terms of the metric coefficients, such a complex structure is  given by the following tensor $\mathfrak{J}$ and leads to the following closed K\"ahler 2-form $\mathrm{K}$:
\begin{equation}\label{forbito}
 \mathfrak{ J} \, = \, \left( \begin{array}{cc} 0 & \sqrt{\frac{p(U)}{q(U)}} \\
                              - \, \sqrt{\frac{q(U)}{p(U)}} & 0 \\
                     \end{array} \right) \quad \Rightarrow \quad \mathrm{K} \, = \, g_{\alpha\mu}\,\mathfrak{J}^\mu_\beta \, \, dx^\alpha \, \wedge \, dx^\beta \, =\, - \, \sqrt{p(U) \, q(U) } \, dU \, \wedge \, dB
\end{equation}
Next one aims at reproducing the K\"ahlerian metric (\ref{metraxia}) in terms of a complex coordinate $\mathfrak{z}\, = \,\mathfrak{z}(U,B) $  and a K\"ahler potential $\mathcal{K}(\mathfrak{z} \, , \, \bar{\mathfrak{z}})\, = \,\mathcal{K}^\star(\mathfrak{z} \, , \, \bar{\mathfrak{z}}) $ such that:
\begin{equation}\label{foxterry}
  \mathrm{K} \, = \, {\rm i} \, \partial \, \overline{\partial} \, \mathcal{K} \, = \, {\rm i} \partial_{\mathfrak{z}} \, \partial_{\bar{\mathfrak{z}}} \, \mathcal{K} \, d\mathfrak{z} \, \wedge \, d\bar{\mathfrak{z}} \quad ; \quad ds^2_{\Sigma} \, = \, \partial_{\mathfrak{z}} \, \partial_{\bar{\mathfrak{z}}} \, \mathcal{K} \, d\mathfrak{z} \, \otimes \, d\bar{\mathfrak{z}}
\end{equation}
As explained in \cite{piesashatwo} the complex coordinate $\mathfrak{z}$ is necessarily a solution of the complex structure equation:
\begin{equation}\label{golosina}
  \mathfrak{ J}_\alpha^\beta \, \partial_\beta \, \mathfrak{z} \, = \, {\rm i} \partial_\alpha \, \mathfrak{z} \quad \Rightarrow \quad \sqrt{\frac{p(U)}{q(U)}} \, \partial_B \, \mathfrak{z}(U,B) \, = \, {\rm i} \, \partial_U \, \mathfrak{z}(U,B)
\end{equation}
The general solution of such an equation is easily found. Define the linear combination \footnote{As it follows from the present discussion the Van Proeyen coordinate $C(U)$ has an intrinsic geometric characterization as that one which solves the differential equation of the complex structure.} :
\begin{equation}\label{gomorra}
    w \, \equiv \, {\rm i} \, C(U) \, - \, B \quad ; \quad C(U) \, = \, \int \, \sqrt{\frac{p(U)}{q(U)}} \, dU
\end{equation}
and consider any holomorphic function $f(w)$. As one can immediately verify, the position $ \mathfrak{z}(U,B) \, = \, f(w)$
solves eq.(\ref{golosina}). What is the appropriate choice of the holomorphic function $f(w)$? Locally (in an open neighborhood) this is an empty question, since  the holomorphic function $f(w)$ simply corresponds to a change of coordinates and gives rise to the same K\"ahler metric in a different basis.
Globally, however, there are significant restrictions that concern the range of the variables $B$ and $C(U)$, namely the global topology of the manifold $\Sigma$. By definition $B$ is the coordinate that, within $\Sigma$, parameterizes  points along the $\mathcal{G}_\Sigma$-orbits, having denoted by $\mathcal{G}_\Sigma$ the isometry group. If $\mathcal{G}_\Sigma$ is compact, then $B$ is a  coordinate on the circle and it must be defined up to identifications $B\simeq B + 2\, n \, \pi $, where $n$ is an integer. On the other hand if $B$ is non compact its range extends on the full real line $\mathbb{R}$. Furthermore, in order to obtain a presentation of the K\"ahler geometry of $\Sigma$ that allows to single out a \textit{canonical inflaton} field $\phi$ with a potential $V(\phi)$  we aim at a K\"ahler potential $\mathcal{K}(\mathfrak{z},\bar{\mathfrak{z}})$ that in terms of the variables $C(U)$ and $B$ should actually depend only on $C$, being constant on the $\mathcal{G}$-orbits. Starting from the metric (\ref{metraxia}) we can always choose a canonical variable $\phi$ defined by the position:
 \begin{equation}\label{gonzallo}
  \phi \, = \, \phi(U) \, = \, \int \,  \sqrt{p(U)} \, dU \quad ; \quad d\phi \, = \, \sqrt{p(U)} \, dU
\end{equation}
and assuming that $\phi(U)$ can be inverted $U \, = \, U(\phi)$ we can rewrite (\ref{metraxia}) in the following canonical form:
\begin{equation}\label{solarium}
  ds^2_{can} \, = \,   d\phi^2 \, + \, \left(\mathcal{ P}^\prime(\phi)\right)^2 \, dB^2  \quad ; \quad
  \mathcal{ P}^\prime(\phi) \, = \, \sqrt{ q\left(U(\phi)\right)} \quad ; \quad \underbrace{\sqrt{p(U(\phi))} \, \frac{dU}{d\phi} \, = \, 1}_{\mbox{by construction}}
  \end{equation}
The reason to call the square root of $q\left(U(\phi)\right)$ with the name $\mathcal{ P}^\prime(\phi)$  is the  interpretation  of such a function as the derivative with respect to the canonical variable $\phi$ of the momentum map of the Killing vector  $\vec{k}_{[B]}$. As it was pointed out in \cite{piesashatwo} and \cite{sergiosashapietroOne}, such interpretation is crucial  for the construction of the corresponding supergravity model but it is intrinsic to the geometry of the surface $\Sigma$.
\par
According to an analysis first introduced in section 4 of \cite{minimalsergioKLP}, by using the canonical variable $\phi$, the VP coordinate $C$ defined in equation (\ref{gomorra}) becomes:
\begin{equation}\label{sodoma}
    C(\phi) \, = \, C\left(U(\phi)\right) \, = \, \int \, \frac{d\phi}{\mathcal{ P}^\prime(\phi)}
\end{equation}
and the metric $ds^2_{\Sigma} \, = \, ds^2_{can}$ of the K\"ahler surface $\Sigma$  can be rewritten as:
\begin{equation}\label{Jmet}
   ds^2_{\Sigma} \, = \, \ft 12 \, \frac{\mathrm{d}^2J}{\mathrm{d}C^2} \left( \mathrm{d}C^2 \, + \mathrm{d}B^2\right)
\end{equation}
where the function $J(C)$ is defined as follows\footnote{See \cite{piesashatwo} for more details.}:
\begin{equation}
    \mathcal{J}(\phi) \, \equiv \, 2\, \int \, \frac{\mathcal{P}(\phi)}{\mathcal{P}^\prime(\phi)} \, d\phi \quad ; \quad J(C) \, \equiv \,  \mathcal{J}\left(\phi(C)\right) \label{gartoccio}
\end{equation}
It appears from the above formula that the crucial step in working out the analytic form of the function $J(C)$ is the ability of inverting the relation between the VP coordinate $C$, defined by the integral (\ref{sodoma}), and the canonical one $\phi$, a task which, in the general case, is quite hard in both directions. The indefinite integral (\ref{sodoma}) can be expressed in terms of special functions only in certain cases and even less frequently one has at his own disposal  inverse functions. In any case the problem is reduced to quadratures and one can proceed further. As we showed in \cite{sergiosashapietroOne}, having already established in eq.(\ref{gomorra}) the general solution of the complex structure equations, there are three possibilities that correspond, in the case of constant curvature manifolds $\Sigma_{max}$,  to the three conjugacy classes of $\mathrm{SL(2,\mathbb{R})}$ elements (elliptic, hyperbolic and parabolic). In the three cases $J(C)$ is identified with the K\"ahler potential $\mathcal{K}(\mathfrak{z},\bar{\mathfrak{z}})$, but it remains to be decided whether the VP coordinate $C$ is to be identified with the imaginary part of the complex coordinate $C \, = \, \mbox{Im} \, \mathfrak{z}$,  with the logarithm of its modulus $C \, = \, \ft 12 \, \log \, |\mathfrak{z}|^2$, or with a third combination of $\mathfrak{z}$ and $\bar{\mathfrak{z}}$, namely whether we choose the first  the second or the third  of the options listed below:
\begin{equation}\label{curlandia}
 \mathfrak{z} \, = \, \left \{ \begin{array}{rccclc}
                                 \zeta & \equiv & \exp\left[\,-\,{\rm i}\, w\right] &=&\underbrace{\exp\left[C(\phi)\right]}_{\rho(\phi)} \, \exp\left[ {\rm i} B\right] \\
                                 t & \equiv &  w &=& {\rm i} \, C(\phi) - B \\
                                 \hat{\zeta} & \equiv & {\rm i} \, \tanh \left(- \,\ft 12 \, w \right)&=& {\rm i} \, \tanh \left(-\,\ft 12  \, ({\rm i} \, C(\phi) - B )\right)\\
                               \end{array} \right| \quad \quad C(\phi) \, \equiv \, \int \,  \frac{1}{\mathcal{P}^\prime(\phi)}  \, d\phi
\end{equation}
If we choose the first solution $\mathfrak{z} \, = \, \zeta$, that in \cite{piesashatwo} was named  of \textbf{Disk-type}, we obtain that the basic isometry generated by the Killing vector $\vec{k}_{[B]}$ is a compact rotation symmetry and this implies a series of consequences on the supergravity lagrangian and its symmetries that we discussed in \cite{sergiosashapietroOne}. Choosing  the second solution $\mathfrak{z} \, = \, t$, that was named  of \textbf{Plane-type} in \cite{piesashatwo}, is appropriate instead to the case of a non compact shift symmetry and leads to different symmetries of the supergravity lagrangian. The third possibility mentioned above certainly occurs in the case of constant curvature surfaces $\Sigma_{max}$ and leads to the interpretation of the $B$-shift as an $\mathrm{SO(1,1)}$-hyperbolic transformation.
\par
In appendix \ref{mathtopo} we recall that the classification of a one dimensional isometry group as ellitptic, parabolic or hyperbolic exists also for non maximally symmetric manifolds and it can be unambiguosly formulated for \textit{Hadamard manifolds} that are, by definition, simply conected, smooth Riemannian manifolds with a non positive definite curvature, \textit{i.e.} $R(x) \le 0$, $\forall \, x \, \in \, \Sigma$, having denoted by $R(x)$ the scalar curvature at the point $x$.
\par
In the three cases mentioned in eq.(\ref{curlandia}) the analytic form of the holomorphic Killing vector  $\vec{k}_{[B]}$ is quite different:
\begin{equation}\label{familione}
    \vec{k}_{[B]} \, = \, \left \{ \begin{array}{lclccclcl}
                                     {\rm i} \zeta \, \partial_\zeta & \equiv &k^{\mathfrak{z}}\partial_{\mathfrak{z}} & \Rightarrow & k^{\mathfrak{z}} & = & {\rm i} \, \mathfrak{z} & ; & \mbox{Disk-type, compact rotation} \\
                                     \partial_t& \equiv &k^{\mathfrak{z}}\partial_{\mathfrak{z}}& \Rightarrow & k^{\mathfrak{z}} & = & 1 & ; & \mbox{Plane-type, non-compact shift}\\
                                     {\rm i} \left(1+\hat{\zeta}^2\right) \partial_{\hat{\zeta}}& \equiv &k^{\mathfrak{z}}\partial_{\mathfrak{z}} & \Rightarrow & k^{\mathfrak{z}} & = & {\rm i} \, \left(1 \, + \, \mathfrak{z}^2\right)& ; & \mbox{Disk-type, hyperbolic boost}
                                   \end{array} \right.
\end{equation}
As shown in \cite{sergiosashapietroOne} this has important consequences on the structure of the momentum map leading to the $D$-type scalar potential and on the transformation properties of the fermions.
\par
Choosing the complex structure amounts to the same as introducing one half of the missing information on the global structure of $\Sigma$, namely the range of the coordinate $B$. The other half is the range of the coordinate $U$ or $C$. Actually, as we showed in \cite{sergiosashapietroOne}  by means of the constant curvature examples,  a criterion able to discriminate the relevant topologies is encoded in the asymptotic behavior of the function $\partial_C^2 J(C)$ for large and small values of its argument, namely in the center of the bulk and on the boundary of the surface $\Sigma$.
The main conclusions that we reached in \cite{sergiosashapietroOne} are those summarized below and are also encoded in table \ref{potenziallini}:
\begin{description}
  \item[I)] The physical properties of the minimal supergravity models that encode one-field cosmologies with a positive definite potential depend in a crucial way on the global topology of the group $\mathcal{G}_\Sigma$ that is gauged in order to produce them. When it is compact we have a certain pattern of symmetries and charge assignments, when it is non-compact we have a different pattern.
  \item[II)] The global topology of the group $\mathcal{G}_\Sigma$ reflects into a different asymptotic behavior of the function $\partial_C^2 J(C)$ in the region that we can call the origin of the manifold. In the compact case the complex field $\mathfrak{z}$ is charged with respect to $\mathrm{U(1)}$ and, for consistency, this symmetry should exist at all orders in an expansion of the scalar field $\sigma$-model for small fields. Hence for $\mathfrak{z} \to 0$ the kinetic term of the scalars should go to the standard canonical one:
      \begin{equation}\label{gomoroid}
      \mathcal{L}^{(can)}_{kin} \, \propto \,  \partial_\mu \mathfrak{z} \, \partial^\mu \bar{ \mathfrak{z}}
      \end{equation}
      Assuming, as it is necessary for the $\mathrm{U(1)}$ interpretation of the $B$-shift symmetry, that $\mathfrak{z} \, = \, \zeta \, \, = \, \exp\left[\delta( C \, + \,{\rm i}\, B)\right]$, where $\delta$ is some real coefficient, eq.(\ref{gomoroid}) can be satisfied if and only if we have:
      \begin{equation}\label{canolicchio}
        \lim_{C \, \to  \, - \, \infty} \, \exp\left[ - \, 2 \, \delta \, C\right] \, \partial_C^2 J(C) \, = \, \mbox{const.}
      \end{equation}
      or more precisely:
      \begin{eqnarray}
        \partial_C^2 J(C) & \stackrel{C \, \to  \, -\, \infty}{\approx} & \mbox{const} \, \times \, \exp\left[  \, 2 \, \delta \, C\right] \, + \, \mbox{subleading}\nonumber\\
        J(C) & \stackrel{C \, \to  \, - \,\infty}{\approx} & \mbox{const} \, \times \, \exp\left[  \, 2 \, \delta \, C\right] \, + \, \mbox{subleading} \label{pagnocorto}
      \end{eqnarray}
      The above stated  is an intrinsic clue to establish the global topology of the inflaton K\"ahler surface $\Sigma$. In appendix \ref{mathtopo} we present some rigorous mathematical results that justify the above criterion to establish the compact nature of the gauged isometry. Indeed what, in physical jargon we call  the origin of the manifold is, in mathematical language, the fixed point for all $\Gamma \, \in \, \mathcal{G}_\Sigma$, located in the interior of the manifold, whose existence is a necessary defining feature of an elliptic\footnote{Let us stress that this is true for Hadamard manifolds and possibly for $\mathrm{CAT}(k)$ manifolds, in any case for simple connected manifolds. In the presence of a non trivial fundamental group the presence of a fixed point is not necessary in order to establish the compact nature of the isometry group.}isometry group $\mathcal{G}$.
      Furthermore with reference to the bosonic lagrangian of a Minimal Supergravity Model written in eq.(2.8) of \cite{Ferrara:2013kca}, namely:
      \begin{equation}\label{minsugra}
        L \, = \, -\ft12 R \, - \, \ft 14 \, F_{\mu\nu}(B) \,F^{\mu\nu}(B) \, - \, \ft{g^2}{2} \, J^{\prime\prime}(C) \, B_\mu \, B^\mu \, - \, \ft 12 \, J^{\prime\prime}(C) \,\partial_\mu C\, \partial^\mu C \, - \, \ft{g^2}{2} \, J^{\prime2}(C)
      \end{equation}
      let us note that  $J^{\prime\prime}(C)$ is the mass of the vector boson $B_\mu$. Hence where $J^{\prime\prime}(C)$ vanishes there the vector boson mass vanishes as well and the $\mathrm{U(1)}$ symmetry is restored. For a linearly realized symmetry, like a phase transformation of the complex scalar field, the existence of such a restoration point is necessary. Hence the existence of a fixed point is indeed necessary  in case the isometry is elliptic. On the contrary if the symmetry is non compact it has to be broken everywhere in order for the theory to be unitary. Hence no restoration point should exist.
      \item[III)] The Fayet Iliopoulos terms \cite{Fayet:1974jb},\cite{Komargodski:2009pc} always identified as linear terms in the VP coordinate $C$ in the function $J(C)$ are rather different in the complex variable $\mathfrak{z}$, depending on which is the appropriate topology.
       \item[IV)]  The above properties are general and apply to all inflaton models embedded into   a minimal $\mathcal{N}=1$ supergravity description. In the particular case of constant curvature K\"ahler surfaces there are five models, two associated with a flat K\"ahler manifold and three with the unique negative curvature two-dimensional symmetric space $\mathrm{SL(2,R)/O(2)}$. Of these five models three correspond to known inflationary potentials: the Higgs potential and  the chaotic inflation  quadratic potential, coming from a zero curvature K\"ahler manifold  and the Starobinsky-like potentials \cite{Starobinsky:1980te}, coming  from the gauging of parabolic subgroups of $\mathrm{SL(2,R)}$.
             Strictly speaking the Starobinsky model, which is dual to an $R+R^2$ model correspond to a fixed value of the curvarture of $\mathrm{SL(2,R)/O(2)}$, normalized as $R \, = \, \hat{\nu}^2 \,=\, \ft 23$. Other values of $\hat{\nu}$ give rise to the Starobinsky-like models.
             The remaining two potentials, respectively associated with the gauging of elliptic and hyperbolic subgroups so far have not yet been utilized as candidate inflationary potentials and, possibly, they are incompatible with PLANCK data.
        \item[V)]   Global topology amounts, at the end of the day, to giving the precise range of the coordinates $C$ and $B$ labeling the points of $\Sigma$. In the five constant curvature cases  these ranges are as follows. In the elliptic and parabolic case  $C$  is in the range $[-\infty,0]$, while it is in the range  $[-\infty,+\infty]$  for the flat case and it is periodic in the hyperbolic case. The cooordinate $B$ instead is periodic in the elliptic case, while it is unrestricted in the hyperbolic and parabolic cases. The manifold $\Sigma$ in the flat case with $B$ periodic is just a strip. It is instead the full plane in the flat  parabolic case.
\end{description}
\begin{table}[h!]
{\scriptsize
\centering
\begin{tabular}{|c|c|c|c|c|c|}
\hline
Curv. & Gauge Group & $V(\phi)$ & $V(C)$ & $V(\mathfrak{z} )$ & Comp. Struct. \\
\hline
\hline
\null & \null & \null & \null & \null & \null \\
$-{\hat \nu}^2$ & $\mathrm{U(1)}$ & $\left( \cosh\left(\hat{\nu} \, \hat{\phi} \right) \, + \, \mu\right)^2 $ & $ \left(\mu +\frac{2 \, e^{4\, C
   \hat{\nu} ^2}}{1-e^{4 \, C \,\hat{\nu}
   ^2}}+1\right)^2$ & $\frac{1}{\nu^4} \, \left( \frac{\mu+1 \, - \, \mu \, \zeta \, \bar{\zeta}}{1 \, - \, \zeta \, \bar{\zeta}}\right)^2$ & $\zeta \, = \, e^{C \, - \, {\rm i}B}$  \\
\null & \null & \null & \null & \null & \null \\
\hline
\null & \null & \null & \null & \null & \null \\
$-{\hat \nu}^2$ & $\mathrm{SO(1,1)}$ & $\left( \sinh\left(\hat{\nu} \, \hat{\phi} \right) \, + \, \mu\right)^2 $ & $ \left(\mu +\tan \left(2 \,C \,\hat{\nu}^2\right) \right)^2 $&   $ \left(\frac{{\bar \zeta} (\zeta
   + {\bar \zeta }) \zeta
   +\zeta +{\bar \zeta}+2
   \mu  (\zeta  \,{\bar \zeta}-1)}{4 \zeta \,{\bar \zeta}\,-\,4}\right)^2 $ &
   $\zeta \, = \, {\rm i}\, \tanh \left(\ft 12 (B-{\rm i}C)\, \nu^2 \right)$ \\
\null & \null & \null & \null & \null & \null \\
\hline
\null & \null & \null & \null & \null & \null \\
$-{\hat \nu}^2$ & $\mathrm{parabolic}$ & $\left( \exp\left(\hat{\nu} \, \hat{\phi} \right) \, + \, \mu\right)^2 $ & $ \left(\mu \, + \, \frac{1}{2 \,\hat{\nu}^2 \, C}\right)^2 $ & $ \left( \ft 12 \, \mu \, + \, \frac{{\rm i}}{2\,\hat{\nu}^2} \, \left(t-\bar{t}\right)^{-1}\right)^2 $ &
   $t \, = \, {\rm i}\,C \, - \, B$ \\
\null & \null & \null & \null & \null & \null \\
\hline
\null & \null & \null & \null & \null & \null \\
$0$ & $\mathrm{U(1)}$ & $M^4 \left[ \left(\frac{\phi}{\phi_0} \right)^2 \, \pm \, 1 \right]^2$ &  $M^4 \left[
\frac{e^{2 a_2 C}}{\phi_0^2}  \, \pm \, 1 \right]^2$  & $ \frac{1}{4} \, \left(  \mathfrak{z} \, \bar{\mathfrak{z}} \, - \, \frac{2\, a_0}{a_2} \, \right)^2 $ &
   $\mathfrak{z} \, = \, \exp \left [ a_2 \left(C \, - \, {\rm i} B\right)\right]$  \\
\null & \null & \null & \null & \null & \null \\
\hline
\null & \null & \null & \null & \null & \null \\
$0$ & $\mathrm{parabolic}$ & $\left(a_0 \, + \, a_1\, \phi\right)^2$ &  $\left(a_1 \, C \, +\, \beta\right)^2$  & $ \frac{1}{2} \, \left(  a_1 \mbox{Im} \mathfrak{z}  \, + \beta \right)^2 $ &
   $\mathfrak{z} \, = \, {\rm i} C \, - \, B$  \\
\null & \null & \null & \null & \null & \null \\
\hline
\end{tabular}
}
\caption{Summary of the potentials of $D$-type  obtained from constant curvature K\"ahler manifolds by gauging either a compact or a non compact isometry. Note that for reader's convenience we have  rewritten the results of \cite{sergiosashapietroOne} in terms of the standard physical field $\hat \phi$ and of the renormalized coefficient $\hat \nu$ introduced in eq.(\ref{formidabileBis}).}
\label{potenziallini}
\end{table}
\par
The goal of this paper is to extend the above results to examples, relevant to inflationary cosmology, where the curvature of the K\"ahler surface $\Sigma$ is not constant. In particular we will consider integrable examples from the list of integrable potentials compiled in \cite{noicosmoitegr} and further discussed in \cite{primosashapietro,piesashatwo,mariosashapietrocosmo} and in addition examples from the generically non integrable class of models introduced under the name of  $\alpha$-attractors  by the authors of \cite{alfatrattori}. One intersection between the list of integrable models and that of $\alpha$-attractors exists and is provided by the remarkable case of the $\arctan$-potential on which we shall dwell a little bit.
Utilizing such examples we will establish the criterion that singles out the interpretation of the $B$-shift isometry as a parabolic shift-symmetry. In all such cases the range of the VP coordinate is $\left[ -\infty ,0\right]$\footnote{Note that $\left[ -\infty ,0\right]$ as range of the $C$-coordinate is conventional. Were it to be $\left[ \infty ,0\right]$, we could just replace $C \,\to\, -\,C$ which is always possible since the K\"ahler metric is given by eq.(\ref{Jmet})} or $\left[ -\infty ,\infty\right]$. The limit $C\, \to \, 0$ always correspond to a boundary of  the K\"ahler manifold $\Sigma$  irrespectively whether the isometry group $\mathcal{G}_\Sigma$ is elliptic or parabolic. If the curvature is negative  we  always have:
\begin{eqnarray}
        \partial_C^2 J(C) & \stackrel{C \, \to  \, 0}{\approx} & \mbox{const} \, \times \, \frac{1}{C^2}\, + \, \mbox{subleading}\nonumber\\
        J(C) & \stackrel{C \, \to  \, 0}{\approx} & \mbox{const} \, \times \, \log \left[C\right] \, + \, \mbox{subleading} \label{pagnolungo}
      \end{eqnarray}
In case the curvature at $C=0$ is zero, we necessarily have the second option of eq.(\ref{cruccomalloreddu}), namely:
\begin{eqnarray}
        \partial_C^2 J(C) & \stackrel{C \, \to  \, 0}{\approx} & \mbox{const} \, + \, \mbox{subleading}\nonumber\\
        J(C) & \stackrel{C \, \to  \, 0}{\approx} & \mbox{const} \, \times \, C^2 \, + \, \mbox{subleading} \label{oselettistufati}
      \end{eqnarray}
Indeed we cannot organize an exponential behavior of $J(C)$  for $C\to 0$.
\par
In the case of a parabolic structure of the isometry group $\mathcal{G}_\Sigma$, the locus $C\, = \, -\,\infty$ is always a boundary and not an interior fixed point which does not exist. Differently from eq.(\ref{pagnocorto}) the asymptotic behavior of the metric and of the $J$-function is either:
 \begin{eqnarray}
        \partial_C^2 J(C) & \stackrel{C \, \to  \, - \, \infty}{\approx} & \mbox{const} \, \times \, \frac{1}{C^2}\, + \, \mbox{subleading}\nonumber\\
        J(C) & \stackrel{C \, \to  \, - \, \infty}{\approx} & \frac{1}{R_\infty} \, \times \, \log \left[C\right] \, + \, \mbox{subleading} \label{polentaconcia}
      \end{eqnarray}
      or
\begin{eqnarray}
        \partial_C^2 J(C) & \stackrel{C \, \to  \, - \, \infty}{\approx} & \mbox{const} \, + \, \mbox{subleading}\nonumber\\
        J(C) & \stackrel{C \, \to  \, - \, \infty}{\approx} & \mbox{const} \, \times \, C^2 \, + \, \mbox{subleading} \label{oseletti}
      \end{eqnarray}
The  asymptotic behavior (\ref{polentaconcia}) obtains when the limit of the curvature for $C\to \, - \,\infty$ is $R_\infty < 0$.
On the other hand, the exceptional asymptotic behavior (\ref{oseletti}) occurs when the limit of the curvature for $C\to \, - \, \infty$ is $R_\infty = 0$.
As we did for the compact case, also for the parabolic case, in appendix \ref{mathtopo} we present rigorous mathematical arguments that sustain the heuristic criteria (\ref{polentaconcia}) and (\ref{oseletti}). Hence in the case where we gauge a parabolic isometry group, the K\"ahler potential has typically two logarithmic divergences one at $C=0$,  and one at $C=-\infty$, the two boundaries of the manifold $\Sigma$. One logarithm can be replaced by $C^2$ (or by a suitable exponential $\exp[\delta \, C]$) in case the limiting curvature on the corresponding boundary is zero. In other regions the behavior of $J$ is different from logarithmic because of the non constant curvature.
\par
It will appear from the analysis of our examples that  most  of the potentials that are useful to describe \textbf{inflation} belong to the class obtained as $D$-terms from the gauging of a \textbf{parabolic group}. One exception to this rule is provided by a specific case of $\alpha$-attractor and it is presented in section \ref{bububu}. On the contrary most of the potentials obtained from the gauging of a compact isometry display an asymptotic stable de Sitter phase corresponding to an absolute minimum. Whether they are useful for inflation is an open question. Certainly, apart from the exception already mentioned,  they do not belong to the family of $\alpha$-attractors.
\par
Finally we can wonder what is the criterion to single out a hyperbolic characterization of the isometry group $\mathcal{G}_\Sigma$. A very simple answer arises from the example in the second line of table \ref{potenziallini} that was analyzed in depth in \cite{sergiosashapietroOne}. The hallmark of such models is a periodic VP coordinate $C$ or anyhow a $C$ that takes values in a finite range $\left[C_{min} \, , \, C_{max}\right ]$. The inspection of integrable potentials provides a few more examples where one of the two above conditions is verified, yet, with the exception of the constant curvature cases analyzed in \cite{sergiosashapietroOne} all such models present a serious pathology. The curvature of the corresponding K\"ahler manifold $\Sigma$ has a singularity in one of the two extrema  $C_{min} \, , \, C_{max}$. Turning to models that are not integrable we are able to find examples of K\"ahler surfaces with a hyperbolic group of isometries that are diffeomorphic to the constant curvature homogeneous space $\mathrm{SL(2,\mathbb{R})/O(2)}$. An example of a non constant curvature K\"ahler surface with a hyperbolic isometry is presented in section \ref{iperbarico}.
\par
There is still one subtle case of which we briefly discuss an example in sect.\ref{dottorsottile}. As we know from the results of our previous paper \cite{sergiosashapietroOne}, summarized in table \ref{potenziallini}, there are two versions of flat models, one where the gauged isometry is a compact $\mathrm{U(1)}$ and one where it is a parabolic translation. In both cases the curvature is zero but in the former case the $J(C)$ function is:
\begin{equation}\label{ellopiatto}
    J(C) \, \propto \, \exp \left[ \delta \, C\right] \quad ; \quad \mbox{elliptic case}
\end{equation}
while in the latter case we have
\begin{equation}\label{ellopiattoBis}
    J(C) \, \propto \, C^2\quad ; \quad \mbox{parabolic case}
\end{equation}
Hence the following question arises. For $\Sigma$ surfaces with a parabolic isometry group we foresaw the possibility, realized for instance in the example discussed in sect. \ref{piattocsquare}, that the limiting curvature might be zero on one of the boundaries so that the asymptotic behavior (\ref{polentaconcia}) is replaced by (\ref{oseletti}). In a similar way we might expect that there are elliptic models where the asymptotic behavior at $C\to \pm \infty$ is:
\begin{equation}\label{sinuhe}
    J(C) \, \stackrel{C\to \pm \infty}{\approx} \, \exp\left[\delta_\pm \, C\right]
\end{equation}
 one of the limits being interpreted as the symmetric fixed point in the interior of the manifold,  the other being interpreted as the boundary on which the curvature should be zero. In sect.\ref{dottorsottile} we will briefly sketch a model that realizes the above forseen situation. The corresponding manifold $\Sigma$ has the topology of the disk. In the same section, as a counterexample, we consider a case where the same asymptotic (\ref{sinuhe}) is realized in presence of an elliptic symmetry, yet $C\to -\infty$ no longer corresponds to an interior point, rather to a boundary. This is due to the non trivial homotopy group $\pi_1(\Sigma)$ of the surface which realizes such an asymptotic behavior. Being non-simply connected such K\"ahler surface is not a Hadamard manifold and presents a pathologies both from the mathematical and the physical stand-point.
\par
An even more intriguing situation can be realized. We can have  K\"ahler manifolds $\Sigma$ endowed with an elliptic isometry group and a vanishing limiting curvature at the boundary $C\to \pm \infty$, yet the asymptotic behavior of the $J(C)$ function approaching such a boundary is not (\ref{sinuhe}), as one might naively expect given the elliptic character of the isometry, rather it is the second option in eq.(\ref{cruccomalloreddu}), namely:
\begin{equation}\label{egiziano}
    J(C) \, \stackrel{C\to \pm \infty}{\approx} \, \kappa_\partial \, C^2
\end{equation}
An explicit and very simple example of this situation is described in section \ref{barabba}

It is interesting to anticipate that the requirement of non singularity of the curvature of the associated $\Sigma$ surface poses constraints also on the $\alpha$-attractor potentials. For instance, as we are going to show, the simplest attractors $V(\hat{\phi})\, \propto \, \left(\tanh\left[\frac{\phi}{\sqrt{6 \alpha}}\right]\right)^{2n}$ introduced in \cite{alfatrattori} are non singular only for $n\, = \, 1,2$.
\par
Hence let us turn to the analysis of the curvature.
\section{The Curvature and the K\"ahler Potential of the surface $\Sigma$}
\label{genTwo}
The curvature of  a two-dimensional  K\"ahler manifold with a one-dimensional isometry group can be written in two different ways in terms of the canonical coordinate $\phi$ or the VP coordinate $C$. In terms of the VP coordinate $C$ we have the following formula:
\begin{eqnarray}\label{curvatta}
  R & = &  R(C) \, = \, - \, \ft 12 \, \frac{J^{''''}(C) \, - \, J^{'''}(C)^2}{ J''(C)^3}
    \, = \, -\, \ft {1}{2} \, \partial_C^2 \, \log \left[ \partial_C^2 J(C) \right ] \, \frac{1}{\partial_C^2 J(C)}
\end{eqnarray}
which can be derived from the standard structural equations of the manifold \footnote{The factor $2$ introduced in this equation is chosen in order to have a normalization of what we name curvature that agrees with the normalization used in other papers of the inflationary literature.}:
\begin{eqnarray}
0 &=&  \mathrm{d}E^1 \, + \, \omega \, \wedge \, E^2 \nonumber\\
0 &=&  \mathrm{d}E^2 \, - \, \omega \, \wedge \, E^1  \nonumber\\
\mathfrak{R} & \equiv & \mathrm{d}\omega \, \equiv \, 2 \, R \, E^1 \, \wedge \, E^2 \label{garducci}
\end{eqnarray}
by inserting into them the appropriate form of the zweibein:
\begin{equation}\label{zweibeinOne}
    E^1 \, = \,  \sqrt{\frac{J''(C)}{2}} \, dC \quad ; \quad E^2 \, = \,  \sqrt{\frac{J''(C)}{2}} \, dB \quad \Rightarrow \quad ds^2 \, = \, \ft 12 \,  J''(C) \left( dC^2 \, + \, dB^2\right)
\end{equation}
Alternatively we can write the curvature in terms of the momentum map $\mathcal{P}(\phi)$ or of the D-type potential
 $V(\phi) \, \propto \, \mathcal{P}^2(\phi)$ if we use the canonical coordinate $\phi$ and the corresponding appropriate zweibein:
\begin{equation}\label{zweibeinTwo}
    E^1 \, = \,  d\phi \quad ; \quad E^2 \, = \, \mathcal{P}^\prime(\phi) \, dB \quad \Rightarrow \quad ds^2 \, = \,\left( d\phi^2 \, + \, \left(\mathcal{P}^\prime(\phi) \right)^2 \, dB^2\right)
\end{equation}
Upon insertion of eq.s (\ref{zweibeinTwo}) into (\ref{garducci}) we get:
\begin{eqnarray}\label{giunone}
     R(\phi) & = & -\, \ft 12 \, \frac{\mathcal{P}^{\prime\prime\prime}(\phi)}{\mathcal{P}^{\prime}(\phi)} \,= \, -\, \ft 12 \, \left ( \frac{V^{\prime\prime\prime}}{V^{\prime}} \, - \, \ft 32 \, \frac{V^{\prime\prime}}{V} \, - \, \ft  34 \, \left( \frac{V^\prime}{V}\right)^2 \right)
\end{eqnarray}
The zero curvature and constant curvature models were analyzed in \cite{sergiosashapietroOne}. We just recall the three constant curvature cases that provide the models for our generalization to non constant curvature.
In eq. (3.16) of \cite{piesashatwo}  the general solution of the  equation\footnote{In the equation below we introduce the coefficient $\nu$ that pairs with the geometrical field $\phi$ and the coefficient ${\hat \nu}$ which pairs with the physical field ${\hat \phi}$.}:
\begin{equation}\label{formidabileBis}
    R(\phi) \, = \, - \, \ft 12  \, \nu^2 \, \equiv \, - \, {\hat \nu}^2
\end{equation}
was presented in terms of the momentum map $\mathcal{P}(\phi)$ and of the canonical variable $\phi$. We have \footnote{Note that for the sake of our following arguments the solution of \cite{piesashatwo} is rewritten here in terms of exponentials rather than in terms of hyperbolic functions  $\cosh$ and $\sinh$.}:
\begin{equation}\label{gordino}
    \mathcal{P}(\phi) \, = \, a \, \exp(\nu \,\phi) \, + \,  b \, \exp (-\, \nu \,\phi) \, + \, c \quad ; \quad a,b,c \, \in \, \mathbb{R}
\end{equation}
In order to convert this solution in terms of the Jordan function $J(C)$ of the VP coordinate $C$, it is convenient to remark that, up to constant shift redefinitions and sign flips of the canonical variable $\phi \to \pm \phi + \kappa$, which leave its kinetic term invariant,  there are only three relevant cases:
\begin{description}
  \item[A)] $a \ne 0, \,  b \ne 0$ and $a/b >0$. In this case, up to an overall constant, we can just set:
  \begin{equation}\label{corrupziaA}
    \mathcal{P}(\phi) \, = \, \cosh (\nu \,\phi) \, + \, \gamma \quad \Rightarrow \quad V(\phi) \, \propto \,  \left(\cosh (\nu \,\phi) \, + \, \gamma\right)^2
  \end{equation}
  \item[B)] $a \ne 0, \, b \ne 0$ and $a/b <0$. In this case we can just set:
  \begin{equation}\label{corrupziaB}
    \mathcal{P}(\phi) \, = \, \sinh (\nu \,\phi) \, + \, \gamma\quad \Rightarrow \quad V(\phi) \, \propto \,  \left(\sinh (\nu \,\phi) \, + \, \gamma\right)^2
  \end{equation}
  \item[C)] $a \ne 0, \, b = 0$. In this case we can just set:
  \begin{equation}\label{corrupziaC}
    \mathcal{P}(\phi) \, = \, \exp (\nu \,\phi) \, + \, \gamma \quad \Rightarrow \quad V(\phi) \, \propto \,  \left(\exp (\nu \,\phi) \, + \, \gamma\right)^2
  \end{equation}
\end{description}
Since our main goal is to understand the topology of the inflaton K\"ahler surface $\Sigma$ and possibly to generalize the above three-fold classification of gauged isometries to the non constant curvature case, it is very useful to recall how, in the above three cases, the corresponding (euclidian) metric $ds^2_{\phi}$ is realized  as the pull-back on  the hyperboloid surface
\begin{equation}\label{sistuloA}
    X_1^2 \, + \, X_2^2 \, - \, X_3^2 \, = \, - \, 1
\end{equation}
of the flat Lorentz metric in the three-dimensional Minkowski space of coordinates $\{X_1,X_2,X_3\}$. The manifold is always the same but the three different  parameterization single out different gaussian curves on the same surface.
\subsection{Embedding of case A)}
\begin{figure}[!hbt]
\begin{center}
\iffigs
\includegraphics[height=95mm]{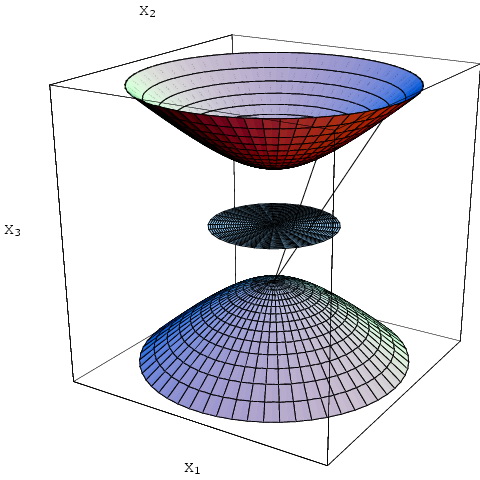}
\else
\end{center}
 \fi
\caption{\it  In this figure we show the hyperboloid ruled by lines of constant $\phi$ that are circles and of constant $B$ that are hyperbolae. In this figure we also show the stereographic projection of points of the hyperboloid onto points of the unit disk}
\label{hyperboloide}
 \iffigs
 \hskip 1cm \unitlength=1.1mm
 \end{center}
  \fi
\end{figure}

Let us  consider the case of the momentum map of eq.(\ref{corrupziaA}). The corresponding two-dimensional metric is:
\begin{equation}\label{giroscopioABis}
    ds^2_\phi \, = \, d\phi^2 \, + \, \sinh^2\left(\nu \,\phi\right) \, dB^2
\end{equation}
It  is the pull-back of the $(2,1)$-Lorentz metric onto the  hyperboloid surface (\ref{sistuloA}). Indeed setting:
\begin{eqnarray}
  X_1 &=& \sinh (\nu  \phi ) \cos (B \nu )\nonumber\\
  X_2 &=& \sinh (\nu  \phi ) \sin (B \nu ) \nonumber\\
  X_3 &=& \pm   \cosh (\nu  \phi ) \label{figliuttoA}
\end{eqnarray}
we obtain a parametric covering of the algebraic locus (\ref{sistuloA}) and we can verify that:
\begin{equation}\label{galloA}
    \frac{1}{\nu^2} \, \left(dX_1^2 \, + \, dX_2^2 \, - \, dX_3^2 \right) \, = \, d\phi^2 \, + \, \sinh^2\left(\nu \,\phi\right) \, dB^2 \, = \, ds^2_\phi
\end{equation}
A picture of the hyperboloid ruled by lines of constant $\phi$ and constant $B$ according to the parametrization (\ref{figliuttoA}) is depicted in fig.\ref{hyperboloide}.
In case of non constant curvature  with a momentum map which gives rise to a consistent $\mathrm{U(1)}$ interpretation of the isometry, the surface $\Sigma$ is also a revolution surface but of a different curve than the hyperbola.
\par
Setting:
\begin{equation}\label{ellipticuscasus}
    f(\phi) \, = \, \mathcal{P}^\prime(\phi)
\end{equation}
we consider the parametric surface:
\begin{eqnarray}
  X_1  &=& f(\phi) \, \cos B\nonumber\\
  X_2  &=& f(\phi) \, \sin B \nonumber\\
  X_3  &=& \pm \, g(\phi )\label{pianinidivano2}
\end{eqnarray}
where  $g(\phi)$ is a function that satisfies the differential equation:
\begin{equation}\label{granolatoBis}
     g'(\phi ) \, = \, \sqrt{\left(f'(\phi ) \right)^2 - 1} \quad \Rightarrow \quad g(\phi) \, = \,
    \int \mathrm{d}\phi \,\sqrt{\left(f'(\phi ) \right)^2 - 1}
\end{equation}
The pull back on the parametric surface (\ref{pianinidivano2}) of the flat Minkowski metric:
\begin{equation}\label{minkiometra}
    ds^2_M \, = \, dX_1^2 + dX_2^2 - dX_3^2
\end{equation}
reproduces the metric of the surface $\Sigma$ under analysis:
\begin{equation}\label{barlattus}
 ds^2_\Sigma \, = \, \mathrm{d}\phi^2 \, + \, f^2(\phi) \, \mathrm{d}B^2
\end{equation}
Hence the revolution surface (\ref{pianinidivano2}) is generically an explicit geometrical model of the K\"ahler manifolds $\Sigma$ where the   gauged isometry utilized as  mechanism to produce a $D$-term scalar potential is elliptic, namely a compact $\mathrm{U(1)}$.
Note that the last integral in eq.(\ref{granolatoBis}) can  be performed and yields a real function  only for those functions $f(\phi)$ that satisfy the  condition $\left(f'(\phi ) \right)^2 >1$. Hence the condition:
\begin{equation}\label{ellipticumomentum}
    \left(\mathcal{P}^{\prime\prime}(\phi ) \right)^2 >1
\end{equation}
is a necessary requirement\footnote{Note that if $\left(\mathcal{P}^{\prime\prime}(\phi ) \right)^2 > \Lambda^2$ is true rather than eq.(\ref{ellipticumomentum}), we can always reduce the case to (\ref{ellipticumomentum}) by appropriately choosing the gauge coupling constant $g$ in front of the momentum map. Hence condition (\ref{ellipticumomentum}) is universal.} for the $\mathrm{U(1)}$ interpretation of the gauged isometry which has to be true together with  the  asymptotic expansion criterion (\ref{pagnocorto}).
\par
Applying to the present constant curvature case the general rule given in eq.(\ref{sodoma})  that defines the VP coordinate $C$ we get:
\begin{equation}\label{CifunzionoA}
    C(\phi) \, = \, \int \frac{d\phi}{\mathcal{P}^\prime(\phi)} \, = \,  \frac{\log \left(\tanh
   \left(\frac{\nu  \phi
   }{2}\right)\right)}{\nu ^2} \quad \Leftrightarrow \quad \phi = \frac{2 \mbox{Arctanh}
   \,\left(e^{C \nu
   ^2}\right)}{\nu }
\end{equation}
from which we deduce that the allowed range of the flat variable $C$, in which the canonical variable $\phi$ is real and goes from $0$ to $\infty$, is the following one:
\begin{equation}\label{dorolatteA}
    C \, \in \, \left[ -\, \infty \, , \, 0 \right ]
\end{equation}
The K\"ahler potential function was calculated in \cite{sergiosashapietroOne} and we got\footnote{Note that in the present paper we have changed the normalization of the function $J(C)$ with respect to the normalization utilized in previous papers \cite{primosashapietro},\cite{piesashatwo},\cite{sergiosashapietroOne}. This has been done in order to facilitate the comparison of our results with those obtained in several other papers of the recent literature. In agreement with the latter the present function $J(C)$ is one-half of the K\"ahler potential (see eq.(\ref{Jmet})).}:
\begin{equation}\label{goliardinaBis}
    J(C) \, = \, 2 \,  (\gamma +1) \, C\,-\, 2 \, \frac{\log
   \left(1-e^{2 C \nu
   ^2}\right)}{\nu ^2}+ \, 2 \, \frac{\log
   (2)}{\nu ^2}
\end{equation}
In this case the appropriate relation between $\zeta$ in the unit circle and the real variables $C,B$ is the following:
\begin{equation}\label{zetosaA}
   \zeta \, = \,  e^{\nu^2  ({\rm i} B  +C )}
\end{equation}
\subsection{Embedding of case B)}
\begin{figure}[!hbt]
\begin{center}
\iffigs
\includegraphics[height=95mm]{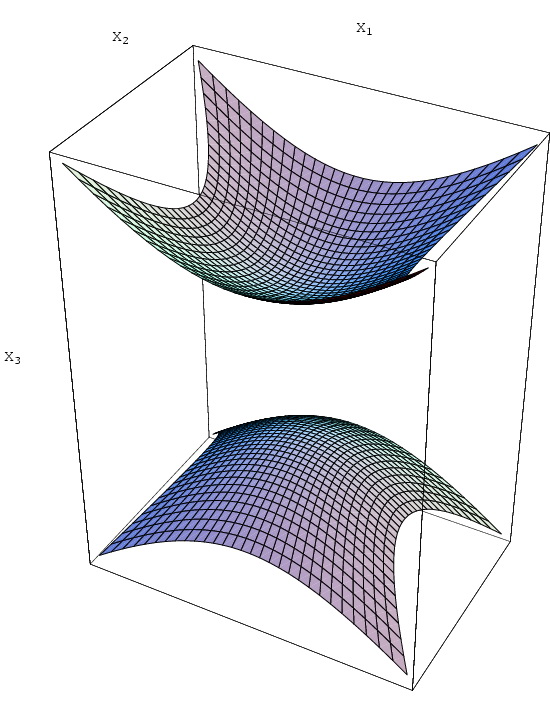}
\caption{\it  The hyperboloid surface displayed in the parametrization (\ref{figliutto}). The lines drawn on the hyperboloid surface are those of constant $B$ and constant $\phi$ respectively. Both of them are  hyperbolae, in this case. }
\label{iperbolonetwo}
 \iffigs
 \hskip 1cm \unitlength=1.1mm
 \end{center}
  \fi
\end{figure}

\begin{figure}[!hbt]
\begin{center}
\iffigs
\includegraphics[height=95mm]{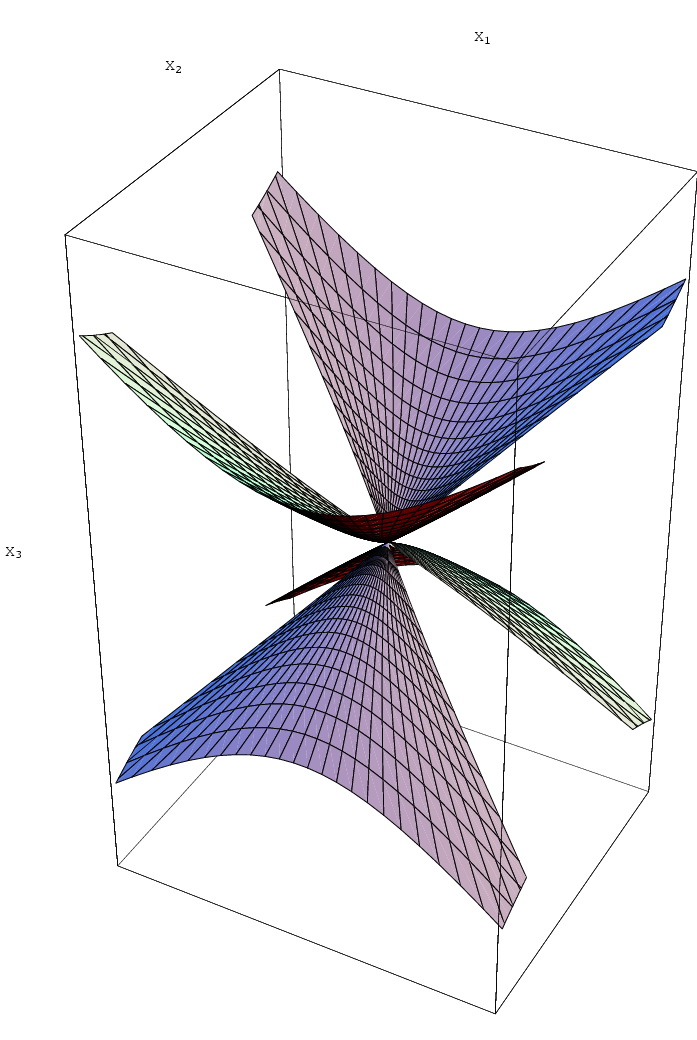}
\caption{\it In this figure, as a counterexample, we display the surface of eq.(\ref{pianinidivano3}) with $f(\phi) \, = \, \sinh(\phi)$ and $g(\phi)\, = \, -i \sqrt{2}\, E\left(i \phi \left|\frac{1}{2}\right.\right)$. The choice of $f(\phi)$ corresponds to the case of a compact $\mathrm{U(1)}$-symmetry and its improper use to realize a surface with a non compact $\mathrm{SO(1,1)}$ hyperbolic isometry leads to a singularity at the line $\phi \, = \,0$ whose points are all mapped  into the
point $X_1\, = \, X_2 \,= \, X_3 \, = \, 0$. In this conical point the tangent plane to the surface is undefined.}
\label{counterexample}
 \iffigs
 \hskip 1cm \unitlength=1.1mm
 \end{center}
  \fi
\end{figure}

Consider the case  of eq.(\ref{corrupziaB}). The corresponding two-dimensional metric is:
\begin{equation}\label{giroscopio}
    ds^2_\phi \, = \, \left(d\phi^2 \, + \, \cosh^2\left(\nu \,\phi\right) \, dB^2\right)
\end{equation}
which can be shown to be another form of the pull-back of the Lorentz metric onto a hyperboloid surface. Indeed setting:
\begin{eqnarray}
  X_1 &=& \cosh (\nu  \phi ) \sinh (B \nu )\nonumber\\
  X_2 &=& \sinh (\nu  \phi ) \nonumber\\
  X_3 &=& \pm \cosh (B \nu ) \cosh (\nu  \phi ) \label{figliutto}
\end{eqnarray}
we obtain a parametric covering of the algebraic locus (\ref{sistuloA}) and we can verify that:
\begin{equation}\label{gallo}
    \frac{1}{\nu^2} \, \left(dX_1^2 \, + \, dX_2^2 \, - \, dX_3^2 \right) \, = \,  \left(d\phi^2 \, + \, \cosh^2\left(\nu \,\phi\right) \, dB^2\right) \, = \, ds^2_\phi
\end{equation}
A three-dimensional picture of the hyperboloid ruled by lines of constant $\phi$ and constant $B$ is displayed in fig.\ref{iperbolonetwo}.
For other surfaces $\Sigma$ (if they exist and are regular) possessing a hyperbolic isometry we can realize their geometrical model considering the following parametric surface:
\begin{eqnarray}
  X_1  &=& f(\phi) \, \sinh B\nonumber\\
  X_2  &=&  g(\phi)  \nonumber\\
  X_3  &=& \pm \, f(\phi ) \, \cosh B  \label{pianinidivano3}
\end{eqnarray}
where:
\begin{equation}\label{funzionnoconP}
    f(\phi) \, = \, \mathcal{P}^\prime(\phi)
\end{equation}
and where  $g(\phi)$ is a function that satisfies the following differential equation:
\begin{equation}\label{granolatoHyp}
     g'(\phi ) \, = \, \sqrt{1 \, + \, \left(f'(\phi ) \right)^2 } \quad \Rightarrow \quad g(\phi) \, = \,
    \int \, \mathrm{d}\phi \,\sqrt{ 1\, + \, \left(f'(\phi ) \right)^2 }
\end{equation}
Once again the pull-back of the flat Minkowski metric (\ref{minkiometra}) on the parametric surface (\ref{pianinidivano3}) reproduces the looked for metric of the $\Sigma$-surface:
\begin{equation}\label{barlattusHyp}
 ds^2_\Sigma \, = \, \mathrm{d}\phi^2 \, + \, f^2(\phi) \, \mathrm{d}B^2
\end{equation}
The last integral in eq.(\ref{granolatoHyp}) can  be performed for any function so that from this point of view there emerges no condition for the $\mathrm{SO(1,1)}$ interpretation of the gauged isometry. This might seem a paradox. For whatever function $f(\phi)$ the hyperbolic $\mathrm{SO(1,1)}$ interpretation of the isometry group seems available. The paradox is resolved by looking at the details where, as usual, the devil is hidden. Let us consider the plane tangent to the surface (\ref{pianinidivano3}) in any of its points. In the three-dimensional embedding that we consider, if we name $\vec{X}(\phi,B)\, = \,\left\{X_1(\phi,B), \, X_2(\phi,B), \, X_3(\phi,B) \right\}$, the tangent plane is the span of the following two vectors:
\begin{eqnarray}\label{gudrunBis}
    \vec{v}_\phi & \equiv & \partial_\phi \vec{X}(\phi,B) \, = \, \left\{ f^\prime(\phi) \, \sinh ( B ), \,  g^\prime(\phi),  \, \pm \,f^\prime(\phi) \, \cosh ( B)\right\}\nonumber\\
    \vec{v}_B & \equiv & \partial_B \vec{X}(\phi,B) \, = \, \left\{ f(\phi) \, \cosh (B) , \,  0,  \, \pm \,f(\phi) \,
    \sinh ( B)\right\}
\end{eqnarray}
The surface (\ref{pianinidivano3}) is non singular and corresponds to a true smooth manifold if in every of its points the tangent plane is well defined, namely if for all $\{\phi, \, B\}$ the two vectors $\vec{v}_\phi, \, \vec{v}_B$ are regular and linear independent. This is what happens for the case of the surface defined in eq.(\ref{figliutto}). We obtain:
\begin{eqnarray}\label{gudrun}
    \vec{v}_\phi & \equiv & \partial_\phi \vec{X}(\phi,B) \, = \, \left\{ \sinh(\phi) \, \sinh(B) , \,  \cosh(\phi),  \, \pm \,\sinh(\phi) \, \cosh(B)\right\}\nonumber\\
    \vec{v}_B & \equiv & \partial_B \vec{X}(\phi,B) \, = \, \left\{ \cosh(\phi) \, \cosh (B) , \,  0,  \, \pm \,\cosh(\phi) \, \sinh (B)\right\}
\end{eqnarray}
and we can easily verify that two vectors are linear independent at all points $\{\phi, \, B\}$ of the surface. The same regularity of the embedding is verified by the parametric surface (\ref{figliuttoA}). Now, as a counterexample, suppose that, using the function $f(\phi) \, = \, \sinh(\phi)$ we defined the parametric surface (\ref{pianinidivano3}), namely we tried to interpret the shift of $B$, which in this case is a compact $\mathrm{U(1)}$ as a non compact $\mathrm{SO(1,1)}$. What is the price that we are going to pay for this stubbornness? The answer is simple: a singular point were the tangent plane becomes undefined. In this case
the integral defining the $g(\phi)$-function can be easily performed and we obtain:
\begin{equation}\label{francescopirla}
   g(\phi)\, = \, -i \sqrt{2}\, E\left(i \phi \left|\frac{1}{2}\right.\right)
\end{equation}
where $E\left(x \left|m\right.\right)$ denotes the elliptic integral of the second kind. The corresponding basis of vectors for the tangent plane takes the simple form:
\begin{eqnarray}\label{gudrunfalso}
    \vec{v}_\phi & \equiv & \partial_\phi \vec{X}(\phi,B) \, = \, \left\{\cosh (\phi ) \sinh (B),\sqrt{\sinh ^2(\phi )+2},\cosh (B) \cosh
   (\phi )\right\}\nonumber\\
    \vec{v}_B & \equiv & \partial_B \vec{X}(\phi,B) \, = \,\{\cosh (B) \sinh (\phi ),0,\sinh (B) \sinh (\phi )\}
\end{eqnarray}
The presence  of a singular point,  is evident from formulae (\ref{gudrunfalso}). At $\phi \, = \, 0$ the second vector vanishes (independently from the value of $B$).
This singularity is clearly visible in the picture of the surface displayed in fig.\ref{counterexample}.
We conclude that the appropriate interpretation of the $B$-symmetry is that one which leads to a non singular surface in the $3D$-geometric realization. Which is the appropriate interpretation is dictated by the asymptotic behavior of the $J(C)$ function and of its second derivative, or alternatively by the equivalent mathematical criteria discussed in section \ref{mathtopo}.
\par
Applying to the present constant curvature case  the general rule given in eq. (\ref{sodoma})  that defines the VP coordinate $C$ we get:
\begin{equation}\label{Cifunziono}
    C(\phi) \, = \, \int \frac{d\phi}{\mathcal{P}^\prime(\phi)} \, = \, \frac{2 \mbox{Arctan} \left(\tanh
   \left(\frac{\nu  \phi
   }{2}\right)\right)}{\nu ^2} \quad \Leftrightarrow \quad \phi = \frac{2 \mbox{Arctanh}
   \left(\tan \left(\frac{C
   \nu ^2}{2}\right)\right)}{\nu}
\end{equation}
from which we deduce that the allowed range of the flat variable $C$, in which the canonical variable $\phi$ is real and goes from $-\infty$ to $\infty$, is the following one:
\begin{equation}\label{dorolatte}
    C \, \in \, \left[ -\, \frac{\pi}{2 \, \nu^2} \, , \, \frac{\pi}{2 \, \nu^2} \right ]
\end{equation}
The  K\"ahler function $J(\phi)$ was calculated in \cite{sergiosashapietroOne} and we obtained:
\begin{equation}\label{goliardina}
    J(C) \, = \, 2\, \gamma \, C \,- \,\frac{2}{\nu ^2} \,\log
   \left(\cos \left(C \nu
   ^2\right)\right)
\end{equation}
In this case the appropriate relation between $\zeta$ in the unit circle and the real variables $C,B$ is different; as it was shown in \cite{sergiosashapietroOne} it is:
\begin{equation}\label{zetosa}
   \zeta \, = \, {\rm i} \tanh \left(\frac{1}{2} (B-{\rm i}\, C) \nu ^2\right)
\end{equation}
\subsection{Embedding of case C)} 
\begin{figure}[!hbt]
\begin{center}
\iffigs
\includegraphics[height=90mm]{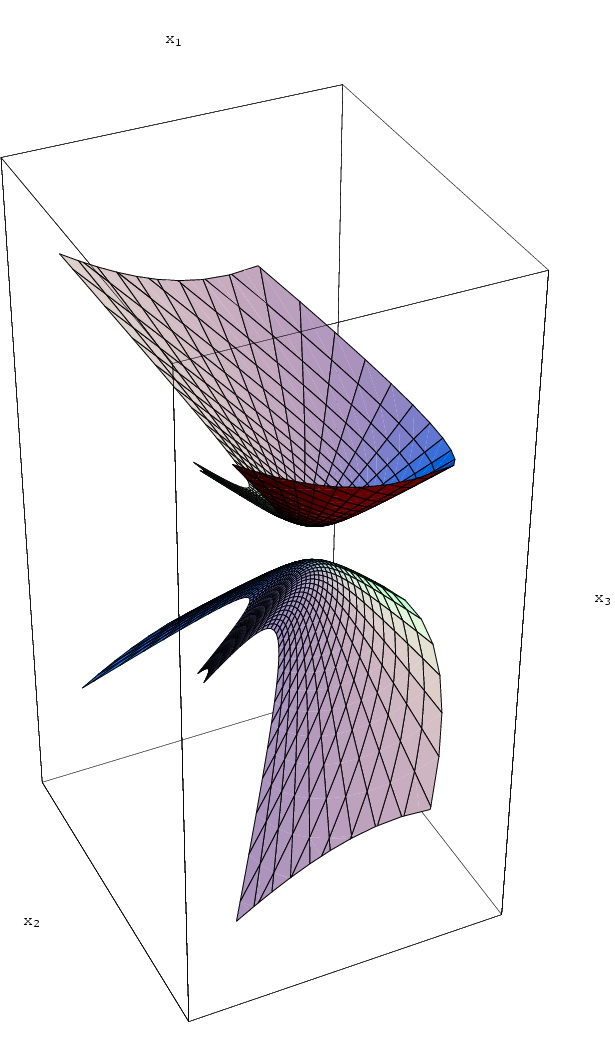}
\caption{\it  The hyperboloid surface displayed in the parametrization (\ref{pianinialetto}). The lines drawn on the hyperboloid surface are those of constant $B$ and constant $\phi$ respectively. The constant $\phi$ curves are parabolae and they are the orbits of the translation group.}
\label{bartacullo}
 \iffigs
 \hskip 1cm \unitlength=1.1mm
 \end{center}
  \fi
\end{figure}

In the case the momentum map is given by eq.(\ref{corrupziaC}) the parameterization of the hyperboloid is the following one:
\begin{eqnarray}
  X_1  &=& \frac{1}{2} \left(-e^{\nu  \phi } B^2+e^{\nu  \phi
   }-\frac{e^{-\nu  \phi }}{\nu ^2}\right) \nu \nonumber\\
  X_2  &=& B e^{\nu  \phi } \nu \nonumber\\
  X_3  &=& \frac{1}{2} \left(e^{\nu  \phi } B^2+e^{\nu  \phi
   }+\frac{e^{-\nu  \phi }}{\nu ^2}\right) \nu \label{pianinialetto}
\end{eqnarray}
Indeed upon insertion of eq.(\ref{pianinialetto}) into (\ref{sistuloA}) we see that for all values of $B$ and $\phi$ the constraint defining the algebraic locus is satisfied. At the same by an immediate calculation one finds:
\begin{equation}\label{tacchino}
    \frac{1}{\nu^2} \, \left(dX_1^2 \, + \, dX_2^2 \, - \, dX_3^2 \right) \, = \,   d\phi^2 \, + \, e^{2 \nu  \phi }\, dB^2  \, = \, ds^2_\phi
\end{equation}
so that the considered metric is the pull-back of the three-dimensional Lorentz metric on the surface $\Sigma$ parameterized as in eq.(\ref{pianinialetto}). The integration of  eq.(\ref{sodoma}) in this case is immediate and  the VP coordinate $C(\phi)$ takes the following very simple invertible form:
\begin{equation}\label{siccatus}
    C(\phi) \, = \, -\frac{e^{-\nu  \phi }}{\nu ^2} \, \Leftrightarrow \, \phi(C) \, = \,-\frac{\log
   \left(-C \nu ^2\right)}{\nu
   }
\end{equation}
The range of definition of $C$ is:
\begin{equation}\label{dorolatteC}
    C \, \in \, \left[ -\, \infty \, , \, 0 \right ]
\end{equation}
A three-dimensional picture of the hyperboloid ruled by lines of constant $\phi$ and constant $B$, according to eq.(\ref{pianinialetto}) is displayed in fig.\ref{bartacullo}.
\par
As it was shown in \cite{sergiosashapietroOne}, the  integration of eq.(\ref{gartoccio}) for the K\"ahler potential is equally immediate and using the inverse function $\phi(C)$ we obtain:
\begin{equation}
 J(C) \, = \, 2 \, \gamma \, C \, -\, \frac{2}{\nu ^2} \,\log \left(-C\right)\, + \, \mbox{const}
 \label{filantropone}
\end{equation}
From the form of equation (\ref{filantropone}) we conclude that in this case the appropriate solution of the complex structure equation is:
\begin{equation}\label{curiaceo}
    \mathfrak{z} \, = \, t \, = \, - \, {\rm i} C \, + \, B
\end{equation}
so that the K\"ahler metric becomes proportional to the Poincar\'e metric in the upper complex plane (note that $C$ is negative definite for the whole range of the canonical variable $\phi$):
\begin{equation}\label{cicculini}
    ds^2 \, = \, \ft 12 \, \frac{\mathrm{d}^2J}{\mathrm{d}C^2} \, \left(dC^2 + dB^2\right)\, = \, \frac{1}{4\, \nu^2} \, \frac{d\bar{t} \, dt}{\left(\mbox{Im}t\right)^2}
\end{equation}
As a consequence of equation (\ref{curiaceo}), we see that the $B$-translation happens to be, in this case, a non-compact parabolic symmetry.
\par
More generally for any surface $\Sigma$ where the isometry of the metric:
\begin{equation}\label{fishycase}
 \mathrm{d}s^2_\Sigma \, = \,  \mathrm{d}\phi^2 \, + \, f^2(\phi )\, \mathrm{d}B^2
\end{equation}
is interpreted as a parabolic shift-symmetry we can construct a geometric model of $\Sigma$ in three-dimensional Minkowski  space by considering the following parametric surface:
\begin{eqnarray}
  X_1  &=& \frac{1}{2} \left(-f(\phi ) B^2+f(\phi)+g(\phi )\right)\nonumber\\
  X_2  &=& B f(\phi ) \nonumber\\
  X_3  &=& \frac{1}{2} \left(f(\phi ) B^2+f(\phi)-g(\phi )\right)\label{pianinidivanoparab}
\end{eqnarray}
where  $g(\phi)$ is a function that satisfies the differential equation:
\begin{equation}\label{granolatoparab}
    f'(\phi ) \, g'(\phi ) \, = \, 1 \quad \Rightarrow \quad g(\phi) \, = \,
    \int \frac{1}{{f}^{\prime}(\phi)} \, d\phi
\end{equation}
The pull-back of the flat metric (\ref{minkiometra}) onto the surface (\ref{pianinidivanoparab}) is indeed the desired metric (\ref{fishycase}). Obviously we have by now learnt the lesson and this modeling is correct only if there are no singularities where the tangent plane becomes undefined. For any parametric surface of type (\ref{pianinidivanoparab})
we have:
\begin{eqnarray}\label{gudrun2}
    \vec{v}_\phi & \equiv & \partial_\phi \vec{X}(\phi,B) \, = \, \left\{\frac{1}{2} \left(-f'(\phi ) B^2+f'(\phi )+g'(\phi
   )\right),B f'(\phi ),\frac{1}{2} \left(f'(\phi ) B^2+f'(\phi
   )-g'(\phi )\right)\right\}\nonumber\\
    \vec{v}_B & \equiv & \partial_B \vec{X}(\phi,B) \, = \, \{-B f(\phi ),f(\phi ),B f(\phi )\}
\end{eqnarray}
In the particular case of eq.(\ref{pianinialetto}) we get:
\begin{eqnarray}\label{gudrun3}
    \vec{v}_\phi & \equiv & \partial_\phi \vec{X}(\phi,B) \, = \, \left\{\frac{1}{2} \left(e^{-\phi }-\left(B^2-1\right) e^{\phi
   }\right),B e^{\phi },\frac{e^{\phi } B^2}{2}+\sinh (\phi
   )\right\}\nonumber\\
    \vec{v}_B & \equiv & \partial_B \vec{X}(\phi,B) \, = \, \left\{-B e^{\phi },e^{\phi },B e^{\phi }\right\}
\end{eqnarray}
which provide a basis of two independent tangent vectors at any point of the surface excluding the existence of singular points. As another instance of counterexample suppose that we tried the parabolic interpretation of the surface in the case where $f(\phi) \, = \, \cosh(\phi)$ that, as we know corresponds instead to a compact $\mathrm{U(1)}$ isometry. In that case we would get:
\begin{equation}\label{cromolotto}
    g(\phi) \, = \, \log \left(\tanh \left(\frac{\phi }{2}\right)\right)
\end{equation}
and the two vectors generating the tangent plane would be:
\begin{eqnarray}\label{gudrun4}
    \vec{v}_\phi & \equiv & \partial_\phi \vec{X}(\phi,B) \, = \,\left\{\frac{1}{4} \left(\coth \left(\frac{\phi
   }{2}\right)-2 \left(B^2-1\right) \sinh (\phi )-\tanh
   \left(\frac{\phi }{2}\right)\right),B \sinh (\phi
   ), \right. \nonumber\\
   &&\left.\frac{1}{4} \left(-\coth \left(\frac{\phi
   }{2}\right)+2 \left(B^2+1\right) \sinh (\phi )+\tanh
   \left(\frac{\phi }{2}\right)\right)\right\}  \nonumber\\
    \vec{v}_B & \equiv & \partial_B \vec{X}(\phi,B) \, = \, \{-B \cosh (\phi ),\cosh (\phi ),B \cosh (\phi )\}
\end{eqnarray}
This basis is singular because the components of $\vec{v}_\phi$ diverge at $\phi\, = \, 0$ and so does $\vec{X}(\phi,B)$. The conceived parametric surface reproduces the considered metric but only in an open subspace of the coordinates $\left\{\phi , \, B\right\}$. The appropriate compact embedding (the hyperboloid) reproduces instead without singularity the considered metric in all points.
\section{$\alpha$-attractors}
\label{trattoriagricoli}
In a recent paper \cite{alfatrattori}, Kallosh, Linde and Roest have shown the good properties of a class of inflationary potentials with the following structure that they have named $\alpha$-attractors:
\begin{enumerate}
  \item The potential $V({\hat \phi})$ is the square of another function $\mathcal{P}(\hat{\phi})$, as it is appropriate for a $D$-term realization in a minimal supergravity model:
   \begin{equation}\label{trattorialfaOneBis}
    V({\hat \phi}) \, = \, \left [\mathcal{P}\left({\hat \phi}\right) \right]^2
\end{equation}
  \item For very large negative values of $\hat{\phi}$ the potential admits the following asymptotic expansion
  \begin{equation}\label{trattorialfaOne}
    V({\hat \phi}) \, \stackrel{\hat{\phi}\, \to\,   \,\infty}{\approx}\, V_0 \left(1\, - \,  \exp\left [-\, \sqrt{\frac{2}{3\alpha}}\, \hat{\phi}\right] \, + \,  \mathcal{O}\left( \exp\left [-\, \sqrt{\frac{2}{3\alpha}}\, \hat{\phi}\right]\right)^2\right)
\end{equation}
\end{enumerate}
Naming $N$ the number of $e$-foldings, the authors of \cite{alfatrattori} have shown that for large $N$ the predictions of such potentials on the two main observables of the CMB spectrum, namely the primordial scalar tilt $n_s$ and the ratio $r$ of tensor to scalar perturbations, are, at leading order in $1/N$, the following universal ones:
\begin{equation}\label{imodium}
  n_s \, = \, 1 \, - \, \frac{2}{N} \quad ; \quad r \, = \, \alpha \, \frac{12}{N^2}
\end{equation}
The value $\alpha \, = \,1$ corresponds to the pure Starobinsky model and the  predictions converge to the above limit (\ref{imodium})  for $\alpha$  small. On the other hand for $\alpha$ large the predictions tend to those of the chaotic inflationary model. This is not surprising, since a pure Starobinsky-like momentm map, as that in the third line of table \ref{potenziallini} with $\mu \, = \, -1$, realizes both points 1) and 2) of the above definition of $\alpha$-attractors with:
\begin{equation}\label{goliardico}
  \hat{\nu} \, = \, \sqrt{\frac{2}{3 \, \alpha}}
\end{equation}
so that $\alpha \, \to \, \infty$ is the flat space limit for the K\"ahler manifold which is associated with the momentum-maps and the potentials in the last two lines of table \ref{potenziallini}. The original Starobinsky model which is dual to a $R+R^2$ supergravity theory corresponds to $\alpha = 1$
\par
In view of these good universal properties the authors of \cite{alfatrattori} have proposed a definition of $\alpha$-attractors  by means of the following change of variable. Let:
\begin{equation}\label{ceccusmiobeppus}
U \, = \,  U_\alpha\left(\hat{\phi}\right) \,\equiv \, \tanh \left( \frac{\hat \phi}{\sqrt{6\alpha}}\right) \, = \, \tanh \left( \frac{ \phi}{\sqrt{3\alpha}}\right) \quad \Leftrightarrow \quad \phi \, = \, \sqrt{3 \alpha} \, \mbox{ArcTanh} \left(U\right)
\end{equation}
which maps the infinite ${\phi}$-range $\left[-\,\infty\, , \, + \, \infty\right]$ into the finite $U$-range $\left[-1\, , \, + \,1\right]$. Expressing the momentum map function in terms of the new variable $U$ we have:
\begin{equation}\label{franconeri}
  \mathcal{P}(\phi) \, = \,  \mathcal{P}\left(\sqrt{3 \alpha} \, \mbox{ArcTanh} \left(U\right)\right) \, \equiv \, \mathfrak{P}(U)
\end{equation}
and the potential is
\begin{equation}\label{trattorialfa}
    V({\hat \phi}) \, = \, \left [ \mathfrak{P}(U) \right]^2
\end{equation}
$\mathfrak{P}(U)$ being a regular function that yields realistic cosmological predictions when it has a zero some where and then  monotonically increases to a finite value $p_0 \, = \, \mathfrak{P}(1) $ as  $U \, \to \, 1$. Indeed the point $U=1$ correspond to the limit
 $\hat{\phi} \, \to \, \infty$ and for $\hat{\phi}$ very large we have:
 \begin{equation}\label{frullatodibanana}
   U \, \sim \, 1 \, - \, 2 \, \exp\left[ \, - \, \sqrt{\frac{2}{3\, \alpha}}\,\hat{ \phi} \right]
 \end{equation}
 Hence if $\mathfrak{P}(U)$ is a regular function with a finite value in $U=1$ it admits a series development:
 \begin{equation}\label{Umenouno}
   \mathfrak{P}(U) \, = \, p_0 \, + \, p_1\,  (1-U) \, + \, p_2\,  (1-U)^2 \, + \, \mathcal{O}\left[(1-U)^3\right]
 \end{equation}
 and combining (\ref{Umenouno}) with (\ref{frullatodibanana}) in (\ref{trattorialfa}) we obtain the  expansion (\ref{trattorialfaOne}) that defines the $\alpha$-attractors. The only condition is that the sign of  $ p_1$ in (\ref{Umenouno}) should be negative. Its absolute value is irrilevant because it can always be reabsorbed by a constant shift of $\hat \phi$. If $p_1$ is positive rather than negative it means that the asymptotic value $p_0$ is reached by the potential function from above rather then from below.
\par
In view of this  the authors of \cite{alfatrattori} have considered in some more detail the \textit{simplest models} provided by the case where $\mathfrak{P}(U)\, = \, \lambda U^n$ is just a simple power.
\par
It is clear that all potentials of the form (\ref{trattorialfa}) can be included in minimal supergravity models as D-term potentials by identifying $\mathfrak{P}\left(U({\hat \phi})\right)\, = \, \mathcal{P}(\phi)$  with the momentum map of a Killing vector ${\vec{k}} \, = \, \partial_B$, as we already did in eq.(\ref{franconeri}). Hence in section \ref{simplicio} we plan to study the geometry of the K\"ahler surfaces $\Sigma$ associated with such simplest attractors.
\par
Another interesting question is whether the list of integrable potentials elaborated in \cite{noicosmoitegr}, which for the reader's convenience we report in tables \ref{tab:families} and \ref{Sporadic} already rewritten in terms of the canonical field $\hat \phi$, has any intersection with the set of $\alpha$-attractors. Since all the potentials listed in these tables are linear combinations of exponentials, the momentum map is the square root of such combinations and the asymptotic expansion (\ref{trattorialfaOne}) is immediate with an $\alpha$ that is also immediately read  off from the very definition of the potential.  This simple analysis singles out just one integrable potential that is at the same time an $\alpha$-attractor. It is the potential  $I_6$ of table \ref{tab:families}.
\subsection{The $I_6$  integrable potential as an $\alpha$-attractor}. In this case, up to an overall constant, the potential can be written as folllows:
\begin{equation}\label{frigna}
  V(\hat{\phi}) \, = \, \pi  \beta +\arctan\left(e^{-\sqrt{6}\, \hat{\phi} }\right)
\end{equation}
where $\beta$ is some real constant. For $\phi\, \to \, \infty$ this potential tends to the finite  limit $\pi \, \beta$
and if we  choose:
\begin{equation}\label{gromallotto}
{\hat \phi} \quad ; \quad {\hat \phi} \, = \, \sqrt{\frac{1}{6}}\, \mathrm{Arctanh}[U] \quad
   \Leftrightarrow \, U\quad = \quad \tanh \left[\sqrt{6} \, {\hat \phi}\right]
\end{equation}
we obtain:
\begin{equation}\label{folocoquilici}
 \mathfrak{P}(U) \, = \, \sqrt{\pi  \beta +\mbox{Arccot }\left(e^{\mbox{Arctanh}(U)}\right)}
\end{equation}
Correspondingly in the neighborhood of $U=1$ the potential admits the following series development:
\begin{eqnarray}\label{colonnadigas}
  V& = & \pi \,\beta \,
  -\frac{\sqrt{1-U}}{\sqrt{2}}-\frac{(
   1-U)^{3/2}}{12 \sqrt{2}}+\frac{3
   (1-U)^{5/2}}{160
   \sqrt{2}}-\frac{5
   (1-U)^{7/2}}{896 \sqrt{2}} \, + \, \mathcal{O}\left[(1-U)^{\ft 92}\right] \nonumber\\
   & = & \, \pi \,\beta \,
  -\frac{\exp\left[-\sqrt{\frac{3}{2}}\, \hat{\phi} \right] }{\sqrt{2}}-\frac{\exp\left[- \, 3 \,\sqrt{\frac{3}{2}}\, \hat{\phi}\right]}{12 \sqrt{2}}+\frac{3
   \exp\left[- \, 5 \, \sqrt{\frac{3}{2}}\, \hat{\phi}\right]}{160
   \sqrt{2}}-\frac{5
   \exp\left[- \, 7 \, \sqrt{\frac{3}{2}}\, \hat{\phi}\right]}{896 \sqrt{2}}\nonumber\\
    &&\, + \, \mathcal{O}\left[ \exp\left[- \, 9 \, \sqrt{\frac{3}{2}}\, \hat{\phi}\right]\right]
\end{eqnarray}
which characterizes it as an $\alpha$-attractor with the following value of $\alpha$ :
\begin{equation}\label{quattrononi}
  \alpha \, = \, \frac{4}{9}
\end{equation}
which is inside the range $\left[\ft 13 \, , \, 3\right]$ considered by the authors of \cite{alfatrattori} as consistent with Planck Data.
In section \ref{arcotanno} we plan to analyze the properties of this model in some more detail in particular in connection with the associated K\"ahler surface.
\par
Irrespectively of whether they are $\alpha$-attractors or not, in the next three sections we consider examples of comological potentials that can be produced as $D$-terms in minimal supergravity models and that, depending on the asymptotic behavior of their $J(C)$ function correspond to non constant curvature $\Sigma$ surfaces with elliptic, parabolic or hyperbolic isometry.
\section{Examples of non maximally symmetric K\"ahler surfaces with an elliptic $\mathrm{U(1)}$-isometry group}
\label{ellittica}
As examples of non maximally symmetric K\"ahler surfaces with an elliptic $\mathrm{U(1)}$ isometry group we consider in this section the $D$-map image of some integrable potentials. In particular we analyse the images of a couple of sporadic integrable potentials and some potentials from the $I_7$ series of table \ref{tab:families}.
\subsection{Two Sporadic Integrable Models}
Let us consider table \ref{Sporadic} and the sixs sporadic integrable potentials $V_{II}({\hat \phi})$. Performing the transformation to the coordinate $U$ defined in eq.(\ref{ceccusmiobeppus}) with $\alpha = 1$, we obtain:
\begin{equation}\label{condorite}
 V_{II}(U) \, = \, \frac{\left(b U^4+c U^2+a\right) \lambda
   }{\left(U^2-1\right)^2}
\end{equation}
Substituting the six sets of coefficients ${a,b,c}$ mentioned in table \ref{Sporadic} that guarantee integrability we find six functions $V(U)$ of which only the last two are positive definite:
\begin{equation}\label{potenti56}
  V_5(U) \, = \, \frac{U^4-8 U^2+8}{8 \left(U^2-1\right)^2} \quad ; \quad V_6(U) \, = \, \frac{\frac{U^4}{16}-U^2+1}{\left(U^2-1\right)^2}
\end{equation}
Correspondingly we have two integrable models where the function $f(U)$ is:
\begin{equation}\label{cinziano}
  f(U) \, = \, \left \{ \begin{array}{ccc}
                                              f_{5}(U)  & = & \pm \,\frac{\sqrt{U^4-8 U^2+8}}{2 \sqrt{2} \left(U^2-1\right)} \\
                                              f_{6}(U)   & = & \pm \,\frac{\sqrt{U^4-16 U^2+16}}{4 \left(U^2-1\right)}
                                           \end{array}
  \right.
\end{equation}
\par
The  metrics on the corresponding  K\"ahler surfaces are respectively given by:
\begin{eqnarray}
  ds^2_5 &=&\mbox{d$\phi $}^2+\frac{\mbox{dB}^2 \left(\cosh
   \left(\frac{2 \phi }{\sqrt{3}}\right)+7\right)^2
   \sinh ^2\left(\frac{2 \phi }{\sqrt{3}}\right)}{12
   \left(28 \cosh \left(\frac{2 \phi
   }{\sqrt{3}}\right)+\cosh \left(\frac{4 \phi
   }{\sqrt{3}}\right)+35\right)}\nonumber \\
   &=&\frac{3}{\left(1-U^2\right)^2} \, \mbox{dU}^2 \, + \, \frac{U^2 \left(4-3 U^2\right)^2}{6
   \left(U^2-1\right)^2 \left(U^4-8 U^2+8\right)}\, \mbox{dB}^2 \label{cricetus5}\\
  ds^2_6 &=& \mbox{d$\phi $}^2+\frac{\mbox{dB}^2 \left(\cosh
   \left(\frac{2 \phi }{\sqrt{3}}\right)+15\right)^2
   \sinh ^2\left(\frac{2 \phi }{\sqrt{3}}\right)}{24
   \left(60 \cosh \left(\frac{2 \phi
   }{\sqrt{3}}\right)+\cosh \left(\frac{4 \phi
   }{\sqrt{3}}\right)+67\right)} \nonumber \\
   & = & \frac{3}{\left(1-U^2\right)^2} \, \mbox{dU}^2 \, + \, \frac{U^2 \left(8-7 U^2\right)^2}{12
   \left(U^2-1\right)^2 \left(U^4-16 U^2+16\right)}\, \mbox{dB}^2 \label{cricetus6}
\end{eqnarray}
It is quite remarkable that in these two cases the integration defining the VP coordinate $C(\phi)$ can be explicitly performed and we get:
\begin{eqnarray}
  C_5(\phi) &=& \frac{1}{2} \left(-\mbox{Arctan}\left(\frac{2 \sqrt{2}\,
   \mbox{sech}^2\left(\frac{\phi
   }{\sqrt{3}}\right)}{\sqrt{\mbox{sech}^4\left(\frac
   {\phi }{\sqrt{3}}\right)+6\,
   \mbox{sech}^2\left(\frac{\phi }{\sqrt{3}}\right)+1}}\right)+6 \, \log \left(\coth \left(\frac{\phi }{\sqrt{3}}\right)\right) \right.
   \nonumber\\
   &&\left.
   +3 \log \left(2\, \mbox{sech}^2\left(\frac{\phi }{\sqrt{3}}\right)+\sqrt{2} \sqrt{\mbox{sech}^4\left(\frac{\phi
   }{\sqrt{3}}\right)+6\, \mbox{sech}^2\left(\frac{\phi
   }{\sqrt{3}}\right)+1}+2\right)\right.\nonumber\\
   &&\left.+\sqrt{2}\, \log
   \left(\tanh ^2\left(\frac{\phi
   }{\sqrt{3}}\right)+\sqrt{\mbox{sech}^4\left(\frac{
   \phi }{\sqrt{3}}\right)+6\,
   \mbox{sech}^2\left(\frac{\phi
   }{\sqrt{3}}\right)+1}-4\right)\right) \label{ciuffoC5}\\
  C_6(\phi) &=& -\frac{3}{14} \left(\sqrt{3} \mbox{Arctan}\left(\frac{4
   \sqrt{3} \mbox{sech}^2\left(\frac{\phi
   }{\sqrt{3}}\right)}{\sqrt{\mbox{sech}^4\left(\frac
   {\phi }{\sqrt{3}}\right)+14
   \mbox{sech}^2\left(\frac{\phi
   }{\sqrt{3}}\right)+1}}\right) \right. \nonumber\\
   &&\left. -7 \log \left(2
   \mbox{sech}^2\left(\frac{\phi
   }{\sqrt{3}}\right)+\sqrt{\mbox{sech}^4\left(\frac{
   \phi }{\sqrt{3}}\right)+14
   \mbox{sech}^2\left(\frac{\phi
   }{\sqrt{3}}\right)+1}+2\right)+14 \log \left(\tanh
   \left(\frac{\phi }{\sqrt{3}}\right)\right)\right. \nonumber\\
   &&\left.-2 \log
   \left(\tanh ^2\left(\frac{\phi
   }{\sqrt{3}}\right)+\sqrt{\mbox{sech}^4\left(\frac{
   \phi }{\sqrt{3}}\right)+14
   \mbox{sech}^2\left(\frac{\phi
   }{\sqrt{3}}\right)+1}-8\right)\right) \label{ciuffoC6}
\end{eqnarray}
\begin{figure}[!hbt]
\begin{center}
\iffigs
\includegraphics[height=50mm]{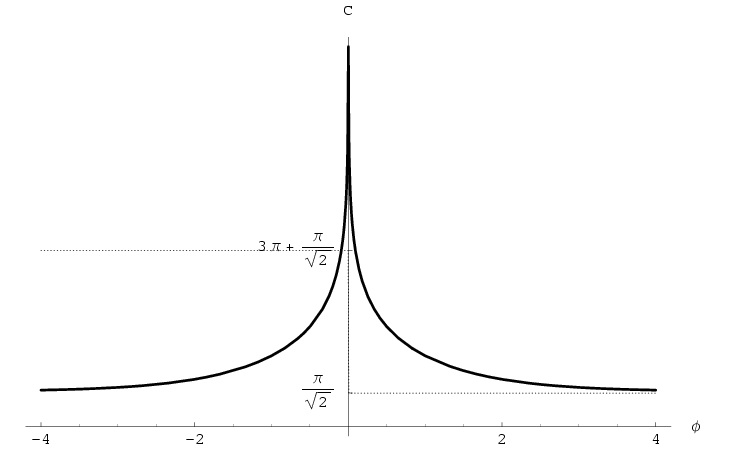}
\includegraphics[height=50mm]{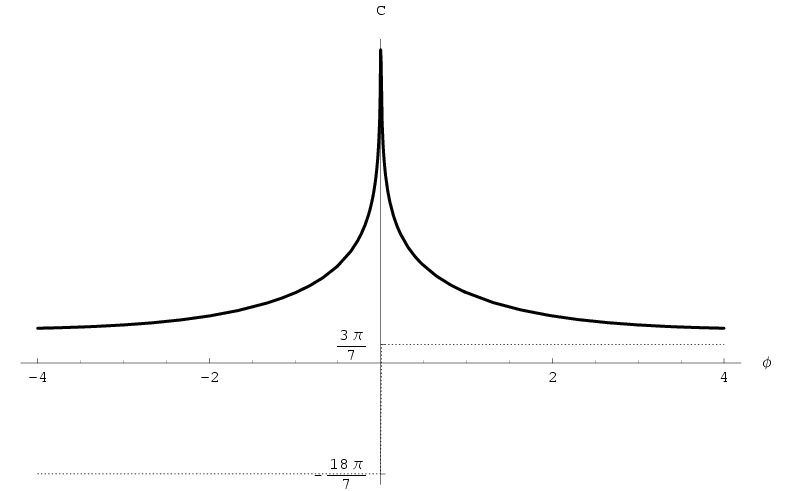}
\caption{\it  Plots of the real and imaginary parts of the VP coordinate $C_{5,6}(\phi)$ for the two sporadic integrable models discussed in the present section. The figure on the left refers to the model $5$, while that on the right refers to the model $6$. In both cases the solid line represents the real part while the dotted line is the imaginary part. As it is evident from the graphs the imaginary part (due to logarithms) is constant and can be eliminated by adding an integration constant. Yet it has a discontinuity at $\phi =0$, where the real part has a pole. It means that $C(\phi)$ can be defined either in the range $\phi \in [-\infty,0]$ or in the range $\phi \in [0,\infty]$.}
\label{C56funzie}
 \iffigs
 \hskip 1cm \unitlength=1.1mm
 \end{center}
  \fi
\end{figure}
The plots of the functions $C_{5,6}(\phi)$ are presented in fig.\ref{C56funzie}. In both cases there is a constant imaginary part due to the logarithm that can be disposed off since $C(\phi)$ is defined up to an additive constant. It should be noted that such integration constant is different for $\phi >0$ and for $\phi <0$. This simply emphasizes that there are two equivalent branches of the function $C(\phi)$ for $\phi$ positive and negative and in both branches $C$ goes from $0$ to $\infty$. Hence we conclude that
\begin{equation}\label{fuschiato}
  - \, C \, \in \, \left [0\, ,\, -\infty \right ]
\end{equation}
which is the prerequisite to interpret $\zeta \, = \, \exp \left[ \, - \, \delta \, C \, + \, {\rm i} \, B \right]$ as the correct complex structure for the K\"ahler manifold. The deciding criterion however is that the K\"ahler metric for $|\zeta| \to 0$ should behave as:
\begin{equation}\label{sordidus}
  g_{\zeta\bar{\zeta}} \, \stackrel{|\zeta| \to 0}{\simeq} \,  \mbox{const}
\end{equation}
Assuming the proposed complex strucuture, in terms of the $J(C)$ function the above condition is the following one:
\begin{equation}\label{corbezzolo}
  \lim_{C \, \to \, \infty} \, \exp\left[\, - \, 2 \, \delta \,  C \right] \, \times \, \frac{d^2J}{dC^2} (C) \, = \, \mbox{const}
\end{equation}
that was already anticipated in eq.(\ref{canolicchio}) with the opposite sign of $C$. Recalling that in terms of the field $\phi$ we have
$\frac{d^2J}{dC^2} (C) \, = \, \left(\mathcal{P}^\prime(\phi)\right)^2$ we conclude that (\ref{corbezzolo}) can be reformulated as:
\begin{equation}\label{ginepro}
  \lim_{ \phi \, \to \, \phi_0} \, \exp\left[\, - \, 2 \, \delta \,  C_{5,6}(\phi) \right] \, \times \, \left(\mathcal{P}_{5,6}^\prime(\phi)\right)^2 \, = \, \mbox{const}
\end{equation}
where $\phi_0$ is the value of $\phi$ for which $C(\phi)$ tends to $\infty$. In our case such a value  is $\phi_0 \, = \, 0$ and for $\delta \, = \, \ft 13$ we have:
\begin{eqnarray}
 e^{-\frac{2}{3} \left(C_5({\phi })\, -\,\frac{{\rm i} \pi }{\sqrt{2}}\right)}&=&\frac{1}{24} \left(4-2
   \sqrt{2}\right)^{-\frac{\sqrt{2}}{3}} e^{\pi /12}
   \phi ^2-\frac{5}{864} \left(\left(4-2
   \sqrt{2}\right)^{-\frac{\sqrt{2}}{3}} e^{\pi
   /12}\right) \phi ^4+\mathcal{O}\left(\phi ^5\right) \label{girco5} \\
  e^{-\frac{2}{3} \left(C_6(\phi )-\frac{3 i \pi
   }{7}\right)} &=& \frac{e^{\frac{\pi }{7 \sqrt{3}}} \phi ^2}{24
   2^{4/7}}-\frac{7 e^{\frac{\pi }{7 \sqrt{3}}} \phi
   ^4}{1728 2^{4/7}}+\mathcal{O}\left(\phi ^5\right) \label{girco6}
\end{eqnarray}
while, if we expand in series the metric we have:
\begin{eqnarray}
  \left(\mathcal{P}_5^\prime(\phi)\right)^2&=& \frac{\phi ^2}{9}+\frac{5 \phi ^4}{162}+O\left(\phi
   ^5\right) \label{cirimel5} \\
   \left(\mathcal{P}_6^\prime(\phi)\right)^2 &=&\frac{\phi ^2}{9}+\frac{7 \phi ^4}{324}+O\left(\phi
   ^5\right) \label{cirmel6}
\end{eqnarray}
This suffices to conclude that:
\begin{equation}
\begin{array}{lclcr}
 {\lim_{C \, \to \, \infty}} \,\exp\left[- 2 \, \ft 13 \, \left( C -\,\frac{{\rm i} \pi }{\sqrt{2}}\right)\right] \,\times \,  \frac{d^2J}{dC^2} (C)  &=&
 \frac{8}{3} \left(4-2
   \sqrt{2}\right)^{\frac{\sqrt{2}}{3}} e^{-\pi /12} & ; & \mbox{(case 5)}  \\
   \lim_{C \, \to \, \infty} \, \exp\left[- 2 \, \ft 13 \, \left( C -\,\frac{{\rm i}\,3\, \pi }{\sqrt{7}}\right)\right] \,  \times \, \frac{d^2J}{dC^2} (C) &=&\frac{8}{3} 2^{4/7} e^{-\frac{\pi }{7 \sqrt{3}}} & ; & \mbox{(case 6)}\\
 \end{array}
\end{equation}
So that in the two considered integrable models the interpretation of the isometry group is that of a compact $\mathrm{U(1)}$ and the appropriate complex coordinate spanning a unit circle is:
\begin{eqnarray}\label{ficicatto}
    \zeta & = & \exp\left[-  \ft 13 \, \left( C -\,\frac{{\rm i} \pi }{\sqrt{2}}\, + \, {\rm i} \, B\right)\right] \quad ; \quad \mbox{case 5} \nonumber\\
    \zeta & = & \exp\left[-  \ft 13 \, \left( C -\,3 \frac{{\rm i} \pi }{7}\, + \, {\rm i} \, B\right)\right] \quad ; \quad \mbox{case 6}
\end{eqnarray}
Next in order to get a complete understanding of the K\"ahler manifolds  $\Sigma$ of which (\ref{cricetus5}) and (\ref{cricetus6}) are the corresponding metrics we inspect the behavior of their curvatures and we try to realize $\Sigma$ as a parameterized surfaces in $\mathbb{R}^3$.
\begin{figure}[!hbt]
\begin{center}
\iffigs
\includegraphics[height=70mm]{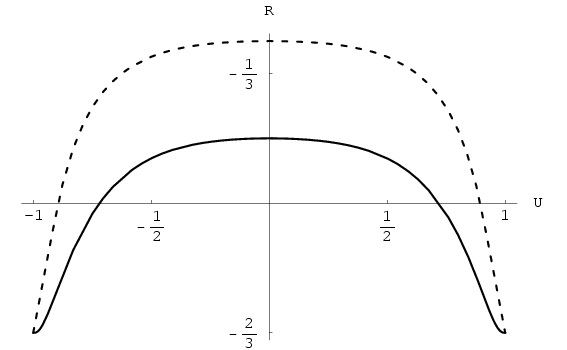}
\caption{\it  The curvature of the $\Sigma_{5,6}$ K\"ahler surfaces associated with the two sporadic integrable "quasi attractors". For both models $R$ is everywhere finite and negative. It increases from the boundary of the manifold $U\, = \, \pm 1$ to the center $U\, = \, 0$. The solid line corresponds to the model $5$, while the dashed line corresponds to the model $6$.}
\label{curvesporadiche}
 \iffigs
 \hskip 1cm \unitlength=1.1mm
 \end{center}
  \fi
\end{figure}
Utilizing the formulae (\ref{giunone}) for the curvature as a function of $\phi$ and reverting then to the variable $U$ we obtain the following result:
\begin{eqnarray}
  R_5(U) &=& -\frac{2 \left(15 U^{10}-88 U^8+244 U^6-428 U^4+416
   U^2-160\right)}{3 \left(3 U^2-4\right) \left(U^4-8
   U^2+8\right)^2} \label{curvatta5} \\
  R_6(U) &=& -\frac{2 \left(U^2-2\right) \left(79 U^8-290 U^6+756 U^4-992
   U^2+448\right)}{3 \left(7 U^2-8\right) \left(U^4-16
   U^2+16\right)^2} \label{curvatta6}
\end{eqnarray}
The plots of the curvature in the two cases is displayed in fig.\ref{curvesporadiche}. As we see the curvature is always negative but it increases from the border $U\, = \, \pm 1$ to the center of the manifold $U=0$. This shows that the two considered surfaces $\Sigma_{5,6}$ are Hadamard manifolds and the criteria discussed in Appendix \ref{mathtopo} apply. They yield the answer that the isometry is elliptic as the already considered asymptotic expansion tests have established. Indeed at $\phi \, = \, 0 \, \Leftrightarrow \, U=0$ we have the rquired fixed poin t in the interior of the manifold.
\par
\begin{figure}[!hbt]
\begin{center}
\iffigs
\includegraphics[height=50mm]{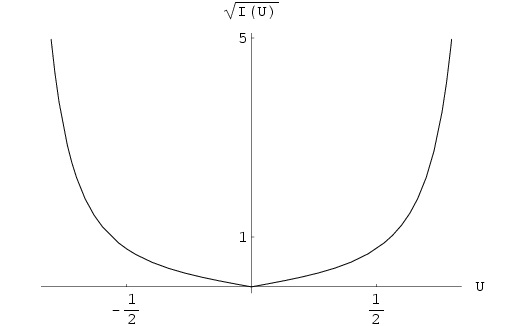}
\includegraphics[height=50mm]{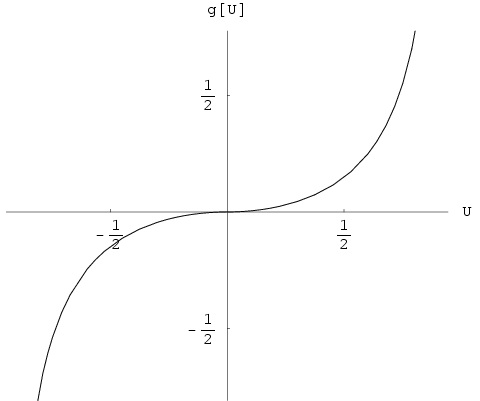}
\caption{\it  In relation with the sporadic integrable "quasi attractor" $5$ we display in this figure the plots of the integrand $\sqrt{I(U)}$ and of the integral $g(U)\, = \, \int \sqrt{I(U)}\, dU$ which allows to view the corresponding K\"ahler manifold $\Sigma_5$ as a parameterized surface in $\mathbb{R}^3$ encoded in eq.s (\ref{x1}-\ref{x3}).}
\label{Iuplotta}
 \iffigs
 \hskip 1cm \unitlength=1.1mm
 \end{center}
  \fi
\end{figure}
\begin{figure}[!hbt]
\begin{center}
\iffigs
\includegraphics[height=100mm]{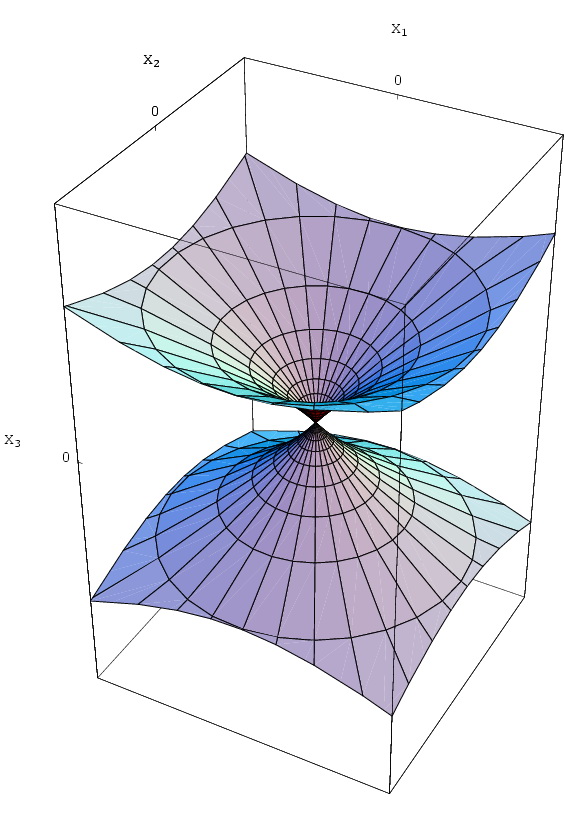}
\caption{\it   In this figure we display the $3D$-plot of the surface $\Sigma_5$ associated with the sporadic integrable potential $5$. This non-constant negative curvature surface replaces the hyperboloid of fig.\ref{hyperboloide} that instead corresponds to the constant curvature manifold $\mathrm{SL(2,\mathbb{R})/O(2)}$. In both cases the interpretation of the $B$-shift isometry as a $U(1)$-symmetry is reflected in the structure of the gaussian curves on the surface. In this case the constant $U$ curves are circles where $B$ runs from $0$ to $6 \,\pi$.}
\label{sigmafive}
 \iffigs
 \hskip 1cm \unitlength=1.1mm
 \end{center}
  \fi
\end{figure}

We can get an intuitive understa nding of the geometry of these manifolds by realizing them as parameterized surfaces embedded in $\mathbb{R}^3$. Let us consider the metrics (\ref{cricetus5}) and (\ref{cricetus6}). We can write both of them in the form:
\begin{equation}\label{grignolio}
    ds^2 \, = \, p(U) \, \mbox{dU}^2 \, + \, q_{5,6}(U) \, \mbox{dB}^2
\end{equation}
where $p(U) \, = \, \frac{3}{\left( 1\, - \, U^2\right)^2 }$ and where $q_{5,6}(U)$ can be read off from eq.s(\ref{cricetus5}) and (\ref{cricetus6}). In both cases the considered metric can be obtained as the pull-back of the flat Lorentzian  metric in $\mathbb{R}^3$:
\begin{equation}\label{euclidian}
    ds^2_E \, = \, 9 \, \left( dX_1^2 \, + \, dX_2^2 \, - \, dX_3^2 \right)
\end{equation}
on the surface defined by the following parameterization:
\begin{eqnarray}
  X_1 &=& \sqrt{q(U) } \, \cos \left(\ft 13 B\right) \label{x1}\\
  X_2 &=& \sqrt{q(U) } \, \sin \left(\ft 13 B\right) \label{x2} \\
  X_3 &=& g(U) \label{x3}
\end{eqnarray}
where the function $g(U)$ is  the solution of the following differential constraint:
\begin{equation}\label{forbitodicitore}
    \left(\frac{d}{dU}\sqrt{q(U) }\right)^2 \, - \, \left(\frac{d g}{dU}\right)^2 \, = \, \ft 19 \, p(U) \quad \Rightarrow \quad
    g(U) \, = \, \pm \,\int \, \underbrace{\sqrt{\left(\frac{d}{dU}\sqrt{q(U) }\right)^2\, - \, \ft 19 \, p(U)}}_{\sqrt{I(U)}} \, dU
\end{equation}
The important thing is that the radicand $I(U)$ should be a positive definite function in the interval $U\in \left[-1,1\right]$. Focusing on the case $5$, we obtain:
\begin{eqnarray}
g(U) & \equiv & \int \, \sqrt{I(U)} \, dU \nonumber\\
 I(U) &=& \frac{\frac{\left(-3 U^8+9 U^6+4 U^4-40
  U^2+32\right)^2}{\left(U^4-8 U^2+8\right)^3}-2
  \left(U^2-1\right)^2}{6 \left(U^2-1\right)^4}\nonumber\\
   \label{garibaldifuferito}
\end{eqnarray}
The integration defining $g(U)$ does not evaluate to a known special function, so that we cannot provide the analytic form of $g(U)$. Yet we can easily perform a numerical integration and the plots of both the integral and the integrand can be easily displayed. This is done in fig.\ref{Iuplotta}.
Equipped with this numerical result we can finally present the $3D$-plot of the surface $\Sigma_5$ defined by eq.s (\ref{x1}-\ref{x3}). It is displayed in fig.\ref{sigmafive}.
\subsection{Integrable Models from the  $I_7$ series}
In \cite{piesashatwo} two of us studied in detail the K\"ahler manifolds $\Sigma_\gamma $ that occur in the $D$-map image  of the  series of inflaton potentials:
\begin{equation}\label{cosmicolatone}
    V(\phi) \,   \propto \, \left(\cosh \left(\sqrt{3} \gamma \phi \right)\right)^{\ft{2}{\gamma} -2} \quad ; \quad \gamma \, \in \, \mathbb{R} \quad \Rightarrow \quad \mathcal{P}(\phi) \, = \, \left(\cosh \left(\sqrt{3} \gamma \phi \right)\right)^{\ft{1}{\gamma} -1}
\end{equation}
that are classified as $I_7$ in the bestiary of integrable potentials derived in \cite{noicosmoitegr} (see table \ref{tab:families} of this paper). In \cite{piesashatwo} it was shown that for two values of $\gamma$, precisely for $\gamma \, = \, \ft 12$ and for $\gamma \, = \, \ft 13$, the corresponding K\"ahler surface $\Sigma_\gamma$ has constant negative curvature and falls into the scheme discussed in \cite{sergiosashapietroOne}. For instance for the case $\gamma\, = \, \ft 12$ one obtains the following $J(C)$ function (see eq.(7.18) of \cite{piesashatwo}):
\begin{equation}\label{fallicoOne}
    J(C) \, = \, \ft 83 \, C \, - \, \ft 83 \, \log\left( 1 \, - \, e^{2 \, C}\right)
\end{equation}
while for the case $\gamma\, = \, \ft 13$, the result is (see eq.(7.29) of \cite{piesashatwo}):
\begin{equation}\label{fallicoTwo}
    J(C) \, = \, - \, 3 \, C \, - \, \ft 32 \, \log\left( 1 \, - \, e^{2 \, C}\right)
\end{equation}
In both cases the solution of the complex structure equation is provided by:
\begin{equation}\label{complessosoluzia}
    \mathfrak{z} \, = \, \zeta \, \equiv \, \exp\left[C(\phi)\right]\cdot \exp\left[{\rm i} B\right]
\end{equation}
and the interpretation of the isometry group generated by the Killing vector $\vec{k}_B$ is that of a compact $\mathrm{U(1)}$. Furthermore let us recall from \cite{piesashatwo} that for the entire series of potentials (\ref{cosmicolatone}) the VP coordinate equation can be exactly integrated and one obtains:
\begin{equation}\label{fiammingo}
    C(\phi)\, = \, \lambda \,+\,{\rm i} \,\frac{\, {\rm i}^{-1/\gamma }
   B_{(-\mbox{csch}^2\left(\sqrt{
   3} \gamma  \phi
   \right))}\left(\frac{1}{2}
   \left(\frac{1}{\gamma
   }-1\right),\frac{3}{2}-\frac{
   1}{2 \gamma }\right)}{6
   (\gamma -1) \gamma }
\end{equation}
where $\lambda$ is an integration constant and $B_z(a,b)$ is the incomplete Beta-function of classical analysis. In the cases  $\gamma\, = \, \ft 12$ and $\gamma\, = \, \ft 13$ the Beta-function reduces to an elementary transcendental function and can also be inverted.
\par
What one observes from eq.s (\ref{fallicoOne}) and (\ref{fallicoTwo}) is that not only the value of the constant curvature is precisely fixed by the choice of an integrable potential but also the value of the Fayet Iliopoulos constant.
\par
As a counterexample we consider the case $\gamma\, = \, \ft 23$ which yields a pure $\cosh$-potential. This specific integrable cosmological model was discussed at length in section 6.4 of \cite{mariosashapietrocosmo}, where the
properties  of its general integral were analyzed in detail and shown to be particularly nice and elegant. Here we consider the geometric properties of its $D$-map image $\Sigma_{\ft 23}$ which are also specially  nice since for this K\"ahler manifold we can work out the analytic form of a complete set of geodesics. This allows the discussion of their uplifting either to the upper complex plane or to the unit disk, corresponding to the interpretation of the $B$-shift either as a parabolic translation group or as a compact $\mathrm{U(1)}$ symmetry. Which is the right uplifting and the right interpretation of the $B$-shift is once again decided by the asymptotic behavior of the $J(C)$ function. In this case, differently from the attractors of the $I_2$ series, the correct answer
is that $B$ spans  compact $\mathrm{U(1)}$-orbits and that the disk model is the right one. In the same way as it was the case for the $I_2$-attractors the two constant curvature examples together with the $\gamma\, = \, \ft 23$ non-constant curvature one indicate that the compact $\mathrm{U(1)}$ interpretation of the $B$-symmetry appears to be a generic feature of the $I_7$ cosh-model series.
\subsubsection{The K\"ahler manifold in the $D$-image of the $I_2$ model with $\gamma \, = \, \ft 23$}
\begin{figure}[!hbt]
\begin{center}
\iffigs
\includegraphics[height=70mm]{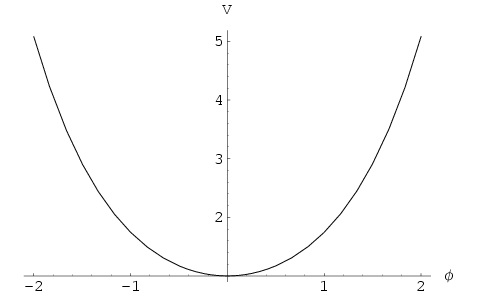}
\else
\end{center}
 \fi
\caption{\it Plot of the integrable inflaton potential $V \, = \, {\mathrm{const}}^2 \times\cosh\left[\frac{2 \sqrt{3}}{3} \phi\right] \, = \,{\mathrm{const}}^2 \times\cosh\left[\sqrt{\frac{2}{3}} \, \hat{\phi}\right] $ which is produced by the choice $\gamma \, = \, \ft 23$ in the integrable series of eq.(\ref{cosmicolatone})}
\label{potenzialone}
 \iffigs
 \hskip 1cm \unitlength=1.1mm
 \end{center}
  \fi
\end{figure}

\begin{figure}[!hbt]
\begin{center}
\iffigs
\includegraphics[height=40mm]{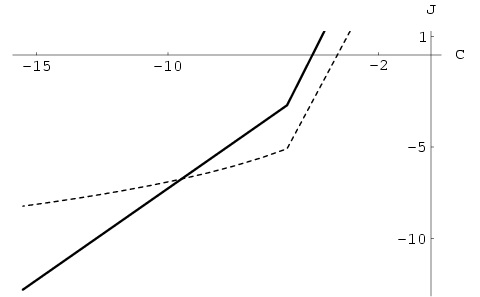}
\includegraphics[height=40mm]{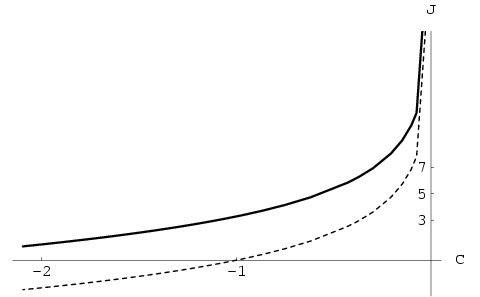}
\else
\end{center}
 \fi
\caption{\it  Plot of the function $J(-C)$ (solid line) in comparison with the function $-\, 3 \, \log(-C)$ (dashed line).The picture on the left shows the behavior of the two functions on a large range of values. The picture on the left is a zoom of the previous plot for small values of $C$. It is quite evident from such a  figure that the function $J(-C)$ has the asymptotic behavior predicted in eq.(\ref{garbarino}).}
\label{bonito}
 \iffigs
 \hskip 1cm \unitlength=1.1mm
 \end{center}
  \fi
\end{figure}

Choosing $\gamma \, = \, \ft 23$ in  eq.(\ref{cosmicolatone}), we obtain $V(\phi) \, \propto \, \cosh\left[\frac{2 \sqrt{3}}{3} \phi\right]$, whose plot is displayed in fig.\ref{potenzialone}. Correspondingly the momentum map and the metric are given by:
\begin{equation}\label{geronimo}
    \mathcal{P}(\phi) \, = \, \mathrm{const} \, \times \, \sqrt{\cosh\left[\frac{2 \sqrt{3}}{3} \phi\right]} \quad \Rightarrow\quad ds^2 \, =\,  \left(\frac{1}{3}
   \sinh \left(\frac{2 \phi
   }{\sqrt{3}}\right) \tanh
   \left(\frac{2 \phi
   }{\sqrt{3}}\right)
   dB^2+d\phi^2\right)
\end{equation}
Instead of the canonical variable it is convenient to utilize the following coordinate $T$ defined by:
\begin{equation}\label{coseccus}
    \phi \, = \, \frac{1}{2}\, \sqrt{3} \, \mbox{ArcCsch}(T) \quad \Rightarrow \quad T \, = \, \mbox{Csch}\left(\frac{2 \phi
   }{\sqrt{3}}\right)
\end{equation}
Upon such a coordinate transformation the K\"ahler metric (\ref{geronimo}) becomes:
\begin{equation}\label{gallus}
    ds^2 \, = \, \frac{9 \, \mbox{dT}^2\, + \, 4 \, \mbox{dB}^2 \sqrt{T^4\, +\, T^2}\, }{12\,\left(T^4\, +\,  T^2 \,\right)}
\end{equation}
which is of the form (\ref{metraxia}) with suitable \footnote{In this case we use the name $T$ for the auxiliary coordinate, since the name $U$ has been reserved for the particular coordinate defined by the position (\ref{trattorialfa}).} $p(T)$ and $q(T)$ and the VP coordinate, that is defined by the integral in eq.(\ref{gomorra}), comes out particularly nice in this case:
\begin{eqnarray}
    C(T) & = & \frac{3}{4} \, (-1)^{3/4} \, B_{(-T^2)}\left(\frac{1}{4},\frac{3}{4}\right) \, + \, \mathrm{const}\nonumber\\
    & = & -3 \sqrt{T} \,
   _2F_1\left(\frac{1}{4},\frac{1}{4};\frac{5}
   {4};-T^2\right)\, + \, \mathrm{const}\label{cfunziona}
\end{eqnarray}
\paragraph{The Curvature and the K\"ahler potential.} Using formulae (\ref{giunone}) and (\ref{gartoccio}) we can calculate the curvature and the K\"ahler potential of the considered manifold, first as functions of $\phi$ and then as functions of
the coordinate $T$. We obtain:
\begin{eqnarray}
  R(T) &=& \frac{1}{3}-\frac{1}{2(T^2+1)} \label{labellocurva}\\
  J(T) &=& - \frac{3}{2} \log
   \left(T\right)\label{labelloJfunz}
\end{eqnarray}
\begin{figure}[!hbt]
\begin{center}
\iffigs
\includegraphics[height=70mm]{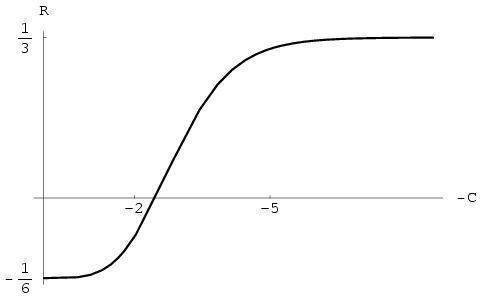}
\else
\end{center}
 \fi
\caption{\it Plot of the curvature of the K\"ahler manifold $\Sigma_{\ft 23}$ as a function of $-C$, the latter being the VP coordinate. The curvature is negative for small negative $C$ and tends to the limit $R_0\, = \, - \ft 16$ as $C\to 0$. Hence at its boundary the manifold becomes a constant curvature manifold $\mathrm{SL(2,R)/O(2)}$. For very large negative values of $C$, namely deep in its bulk, the manifold has instead a positive curvature that tends to the limit $R_{-\infty}\, = \,\ft 13$ for $C\to -\infty$.}
\label{curvatolone}
 \iffigs
 \hskip 1cm \unitlength=1.1mm
 \end{center}
  \fi
\end{figure}
In fig.\ref{curvatolone} we show the behavior of the curvature as function of the VP coordinate $-C$. We see that the curvature is large and positive in the bulk of the manifold, while it tends to a negative value for $C\to 0$. Furthermore the curvature reaches its asymptotic value quite early and it is almost constant when we approach the boundary $C\, = \, 0$. This means that the manifold $\Sigma_{\ft 23}$ is asymptotically a coset $\mathrm{SL(2,R)/O(2)}$ with constant curvature equal to $R_0\, = \, - \, {\nu^2} \, = \, -\,\ft 16$. From this information we can predict the behavior of the K\"ahler potential at the boundary. We should find:
\begin{equation}\label{garbarino}
    J(-C) \, \stackrel{C\to 0}{\simeq}  \, - \, 3 \, \log(-C)
\end{equation}
which is the analogue of eq.(\ref{molinodorino}) and is always true both for the case of a shift-symmetry interpretation or for the case  of a compact $\mathrm{U(1)}$-interpretation. One cannot work out the analytic form of $J(-C)$ since that involves the use of the inverse of the incomplete Beta-function, yet we can easily appreciate the behavior of the function $J(C)$ by means of a parametric plot. This latter is displayed in fig.\ref{bonito}.
\par
\begin{figure}[!hbt]
\begin{center}
\iffigs
\includegraphics[height=55mm]{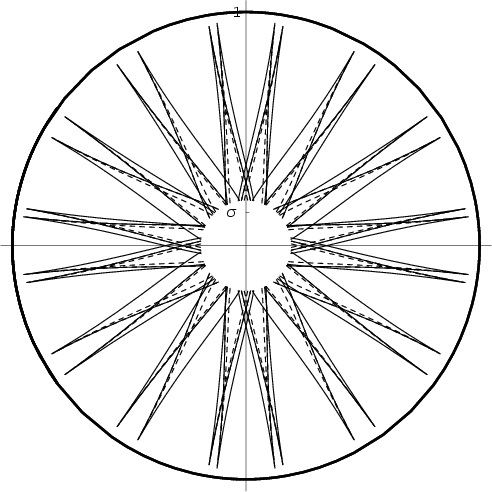}
\includegraphics[height=50mm]{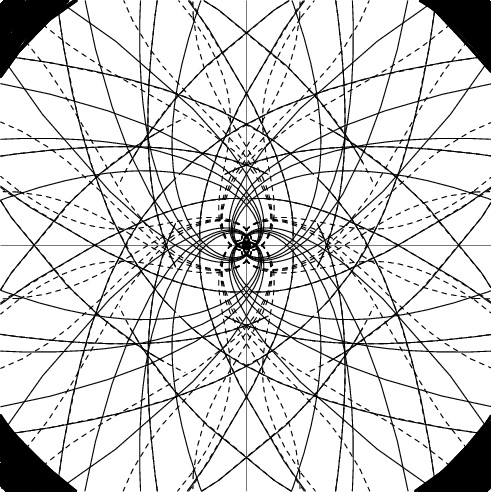}
\includegraphics[height=50mm]{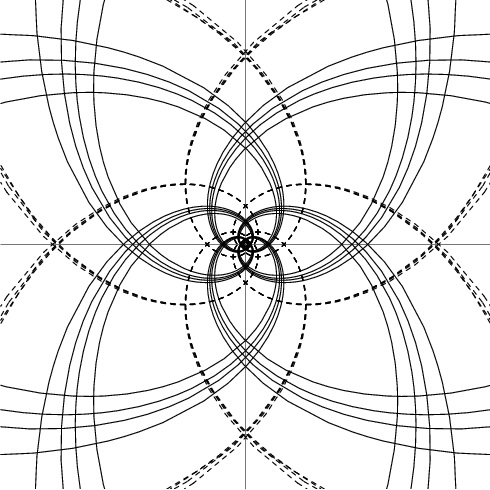}
\else
\end{center}
 \fi
\caption{\it Plots of a set of geodesics of the $\Sigma_{\ft 23}$ manifold uplifted to the interior of the unit disk by means of the choice of the complex structure (\ref{farsesco2}). We have considered various radii $R$ and various centers at $B_0$. Changing the value of $B_0$ simply rotates the shown picture clock-wise or anti-clock-wise.  The figure on the right just shows the full extent of the geodesics and shows their way of approaching the boundary. The small circle at the center has radius $R \, \, \frac{1}{7}$ and it is just empty because the geodesics were not continued to that interior of such a region. This is what is done in the next picture on the right that shows the behavior of the geodesics in the region near $C=\, - \, \infty$.  The last picture on the right is a further zoom of an even smaller region around $C\, = \,-\infty$. The behavior of the geodesics in the region of positive curvature is quite complicated since they wind and cross each other many times. }
\label{bailamme}
 \iffigs
 \hskip 1cm \unitlength=1.1mm
 \end{center}
  \fi
\end{figure}

From this discussion of the behavior of $J(C)$ we cannot decide which complex structure is appropriate to the manifold under consideration:
\begin{eqnarray}
    \mathfrak{z} & = & t \, \equiv \, -\,{\rm i} \, C \, + \, B \label{farsesco1}\\
    \null & \mbox{or} & \null\nonumber\\
    \mathfrak{z} & = & \zeta \, \equiv \, \exp\left[C\right] \,\cdot \,\exp\left[{\rm i} \, B\right] \label{farsesco2}
\end{eqnarray}
We will resolve the question by calculating the asymptotic behavior of $J(C)$ for large values of $C$. Before doing that we consider the structure of the geodesics.
\paragraph{The geodesics.}
As it was shown in eq.(3.33) of \cite{primosashapietro}, a complete system of geodesics for the K\"ahler metric (\ref{metraxia}) (rewritten in the form (\ref{solarium}) upon use of the canonical coordinate $\phi$), is parameterized by  two real constants $R$ and $B_0$ and it is described, in the $C,B$ plane, as it follows:
\begin{equation}\label{gurtisino}
    C(\phi) \, = \, \int \frac{d\phi}{Q(\phi)} \quad ; \quad B(\phi) \,  = \, \pm\,\int \frac {d\phi}{Q(\phi) \, \sqrt{R^2 \, Q^2(\phi)\, - \, 1}}\,  + \, B_0 \, \quad ; \quad Q(\phi) \, \equiv \, \mathcal{P}^\prime(\phi)
\end{equation}
The first of the two integrals advocated in (\ref{gurtisino}) was already calculated above and it was already transformed to the variable $T$ in eq.(\ref{cfunziona}). The second integral $B(\phi)$ is equally  calculable. We get:
\begin{eqnarray}\label{golodoso}
    B(\phi) & = &B_0 \, \pm \, \frac{3}{4} \left(\mbox{Arctanh}\,\left(\frac{2 R^2 \tanh^2\left(\frac{\phi
   }{\sqrt{3}}\right)-3}{\sqrt{9
   \tanh ^4\left(\frac{\phi
   }{\sqrt{3}}\right)+12 R^2
   \tanh ^2\left(\frac{\phi
   }{\sqrt{3}}\right)-9}}\right)\right.\nonumber\\
   &&\left. +\log \left(2 R^2+3 \tanh
   ^2\left(\frac{\phi
   }{\sqrt{3}}\right)+\sqrt{9
   \tanh ^4\left(\frac{\phi
   }{\sqrt{3}}\right)+12 R^2
   \tanh ^2\left(\frac{\phi
   }{\sqrt{3}}\right)-9}\right)\right)
\end{eqnarray}
The transformation to the variable $T$ implies the solution of a second order equation and this leads  to two branches. Therefore we get  the following geodesics:
\begin{eqnarray}
  C(T) &=& \, \frac{3}{4} \, (-1)^{3/4} \, B_{(-T^2)}\left(\frac{1}{4},\frac{3}{4}\right) \nonumber\\
  B_+(T) &=& B_0 \,\pm \, \frac{3}{4} \left(\mbox{Arctan}\left(\frac{2 R^2
   \left(T+\sqrt{T^2+1}\right)^2
   -3}{\sqrt{9
   \left(T+\sqrt{T^2+1}\right)^4
   +12 R^2
   \left(T+\sqrt{T^2+1}\right)^2
   -9}}\right)\right.\nonumber\\
   &&\left.+\log \left(2
   R^2+3
   \left(T+\sqrt{T^2+1}\right)^2 \right.\right.\nonumber\\
   &&\left.\left.+\sqrt{9
   \left(T+\sqrt{T^2+1}\right)^4
   +12 R^2
   \left(T+\sqrt{T^2+1}\right)^2
   -9}\right)\right)\label{branchpiu}\\
  B_-(T)& = & B_+(-T) \label{branchmeno}
  \end{eqnarray}
Since by evaluating the asymptotic behavior of $J(C)$ we are going to show  that the cordinate $B$ is periodic and labels the points inside circular orbits of a  $\mathrm{U(1)}$ compact group  let us anticipate this result and
consider  the choice (\ref{farsesco2}) of the complex structure. This allows to uplift the geodesics (\ref{branchpiu}, \ref{branchmeno}) to the interior of the unit-disk. Such an uplifting is displayed in fig.\ref{bailamme}. In this way the geodesics (\ref{branchpiu}, \ref{branchmeno}) become a family of curves living in the interior of the unit-disk that has the unit circle as the boundary. The geodesics of the Poincar\'e metric are well-known. They are arcs of circles with radius $R$ and center on the unit circle at $B=B_0$. For this reason the two parameters labeling the geodesics have been given such names. In the case of a constant curvature manifold they retrieve the geometric interpretation corresponding to their names. Since the manifold $\Sigma_{\ft 23}$ approaches  Poincar\'e geometry at the boundary, we expect that also the geodesics should approach arcs of circles of radius $R$ with the center on the real axis, when they are located near the boundary, namely for $-C \to 0$. This is precisely what is displayed in fig.\ref{bailamme}. Note that due to the two branches $ B_\pm(U)$ there are two type of geodesics filling the interior of the unit disk. Those coming from $\pm \, B_+(U)$ have been denoted with a solid line, while those associated with $\pm \, B_-(U)$  have been denoted with a dashed line.
\par
The geodesics have the appropriate behavior near the boundary while they differ violently from circles in the deep bulk of the unit disk  where the curvature becomes positive.
In spite of their complicated behavior and multiple winding  in the region of positive curvature, the geodesics approach the shape of geodesics of the Poincar\'e metric in the unit disk, namely arcs of circles with their center on the unit circle.
\par
We conclude that  this interpretation is consistent and viable. Let us show that it is the correct one by examining the asymptotic behavior of the metric for $|C| \to \infty$.
\paragraph{Asymptotic expansion of $J(C)$ for $|C| \to \infty$}
In this case the instrument to calculate the asymptotic expansion of $J(C)$ is based on the use of the classical asymptotic expansions for large values of  the hypergeometric function. Considering the second expression of $C(T)$ in eq.(\ref{cfunziona}) and expanding it around $T=\infty$ we obtain:
\begin{equation}\label{somaro}
  C(T) \, \stackrel{T\to\infty}{\approx}\,   -\,\frac{3}{2}\, \mbox{Log}\left[T\right]\,-\, \frac{3}{8} \left(\pi +\mbox{Log}[64]\right)\, +\,\mathcal{O}\left[\frac{1}{T}\right]^{3/2}
\end{equation}
Comparing  equation (\ref{somaro}) with the expression (\ref{gallus}) for the metric in terms of the variable $T$, we conclude that for large values of $T$ and $C$ we have:
\begin{eqnarray}
  C &\simeq & - \, \ft 32 \, \log \, T \label{asino1}\\
  T^{-2} &\simeq & \exp\left[  \ft 43 \, C\right] \label{asino2}\\
 \ft 12 \, \frac{\mathrm{d}^2J}{\mathrm{d}C^2} &\simeq & \ft 13 \, T^{-2} \, \simeq \, \ft 13 \exp\left[  \ft 43 \, C\right]\label{asino3}
\end{eqnarray}
It follows that we can introduce a complex variable:
\begin{equation}\label{complessavariata}
    \zeta \, = \, \exp \left[\ft 23 \,\left( C(T) +{\rm i} B\right )\right]
\end{equation}
which goes to $1$ when $T \to 0$ (boundary of the manifold) and goes to $0$ when $T\to \infty$ since there  $C\to -\infty$ (origin of the manifold). Near the origin of the manifold the second derivative of the function $J(C) \sim |\zeta|^2$ and this is the necessary and sufficient condition for the metric $g_{\zeta{\bar\zeta}}$ to go to a constant.
In conclusion, the $\mathrm{U(1)}$ interpretation is consistent, as we have already anticipated.
\subsection{Asymptotically flat elliptic models}
\label{dottorsottile}
As announced in the introduction in this section we consider the problem of constructing  K\"ahler surfaces $\Sigma$ with an elliptic isometry whose limiting curvature at the boundary vanishes $R_{\pm\infty}\, = \, 0$. In this case we might be induced to think that the asymptotic behavior of the function $J(C)$ for $C\to \pm\infty$ is fixed. Indeed from our previous results \cite{noicosmoitegr}, we know that for flat K\"ahler manifolds with an elliptic isometry, we have $J(C) \propto \exp\left[\delta \, C\right]$ for some value of $\delta \in \mathbb{R}$. Hence we might naively expect that the function $J(C)$  for surfaces $\Sigma$  with an elliptic isometry and a vanishing limiting curvature should behave has follows:
\begin{equation}\label{gargantua}
    J(C)\, \stackrel{C \to \pm \infty}{\approx} \, \exp\left[ \, \delta_{\pm} \, C \right]\, + \, \mbox{subleading terms}
\end{equation}
There are however two fundamental subtleties that have to be immediately emphasized.
\begin{itemize}
  \item If the topology of the surface $\Sigma$ is the disk topology and $\Sigma$ is simply connected $\pi_1(\Sigma) \, = \, 1$, then one of the two limits $C\to \infty$ has to be interpreted as the interior fixed point, required by Gromov criteria, for elliptic isometries in Hadamard manifolds (and possibily in $\mathrm{CAT}(k)$ manifolds). The other limit corresponds to the unique boundary of disk topology. On the other hand if $\pi_1(\Sigma) \, = \, \mathbb{Z}$ and the K\"ahler surface has the corona topology then there are two boundaries and the limiting curvature can be zero on both boundaries. We will illustrate this with two examples, respectively corresponding to the latter and to the former case.
  \item In the case of non constant curvature, also in the presence of an elliptic isometry, asymptotic flatness can be realized with an asymptotic behavior of type $J(C) \, \stackrel{C\to \infty}{\approx} \, C^2$, not necessarily of $J(C) \, \stackrel{C\to \infty}{\approx} \, \exp\left[\delta \, C\right]$. The only relevant point is that we should have $J(C) \, \stackrel{C\to \, - \, \infty}{\approx} \, \exp\left[\delta \, C\right]$ at the interior fixed point requested by the elliptic character of the isometry. We will illustrate this with the \textit{cigar metric} used for other purposes by Witten in \cite{sigarus}.
\end{itemize}
\subsubsection{The catenoid case with  $\pi_1(\Sigma) \, = \, \mathbb{Z}$}
We begin by considering explicit functions $J(C)$ that have the required asymptotic behavior and we try to work our way backward towards the canonical coordinate $\phi$ and the momentum map $\mathcal{P}(\phi)$. In particular we want to make sure that the considered function $J(C)$ does indeed correspond to a compact isometry. This will certainly be the case if the corresponding metric is the pull-back of the flat three-dimensional euclidian metric on a smooth surface of revolution.
\par
To carry out such a program we consider the  following one-parameter family of $J(C)$ functions:
\begin{equation}\label{fracastoro}
    J_{[\mu]}(C) \, = \,\frac{1}{8} \left(\mu\,
   C^2+\cosh [2\, C]\right)
\end{equation}
which fulfills condition (\ref{gargantua}), by construction.  Many other examples  can be obviously put forward, but this rather simple one is  sufficient to single out the main subtlety that makes many asymptotically flat elliptic models pathological both from the point of view of supergravity applications and from the point of view of Gromov et al classification of isometries.
Using eq.s (\ref{Jmet}) and (\ref{curvatta}) we write the metric and the curvature following from the $J(C)$ function of eq.(\ref{fracastoro}), obtaining
\begin{eqnarray}
  ds^2_\Sigma &=& \frac{1}{16} \left(\, 2 \mu +4 \cosh[2 C]\right)\, \left( dC^2 \, + \, dB^2\right )\label{matrina} \\
  R(C) &=& -\frac{4 \mu  \cosh (C)+1}{(4\,\mu +\cosh [C])^3} \label{curvina}
\end{eqnarray}
From these formulae we draw an important conclusion. In order for $\Sigma$ to be a smooth manifold the curvature should not develop a pole neither in the interior nor on the boundary. This means that $4\mu +\cosh [C]>0$ in the whole range of $C$. This is guaranteed if and only if $\mu > -\, \ft 14$. On the other hand, according to our previous discussions, in the case of an elliptic isometry, there should be, for a finite value of $C$, a zero of the metric coefficient. In mathematical language such a zero  is the fixed point that characterizes elliptic isometries of Hadamard manifolds, while, in physical language, it corresponds to the restoration point where the linearly realized symmetry is unbroken, the mass of the vector field vanishes and the lagrangian of the complex scalar field is the free quadratic one.
Looking at eq.(\ref{matrina}) we see that such a zero exists, if and only if $\mu < -\,\ft 12$. It follows that, at least  in this family of models, there are no smooth manifolds that are asymptotically flat in the elliptic sense and fulfill the physical condition for $\mathrm{U(1)}$-symmetry which corresponds to the Gromov et al identification of elliptic isometries of Hadamard manifolds. At first sight
one should draw  the conclusion that, in the case of the $J(C)$ functions of eq.(\ref{fracastoro}), the isometry is not elliptic. Yet this is somehow strange, since at the boundary, where the curvature goes to zero, the form of $J(C)$ is precisely that which corresponds to elliptic isometries. Furthermore we will shortly show that for every value of $\mu$ the metric in eq.(\ref{matrina}) is just the metric of a smooth revolution surface. Actually for $\mu\, = \,2$ such a revolution surface is the well-known \textbf{catenoid}, firstly constructed by Bernoulli in 1744 as the first example of a minimal surface. Hence we arrive at a puzzle with Gromov et al criteria, whose only resolution can be that the manifolds associated with the $J(C)$ functions of eq.(\ref{fracastoro}) are not Hadamard manifolds. From eq.(\ref{curvina}) we see that, provided $\mu > -\, \ft 14$, the curvature is negative definite and attains its maximal value $R=0$ only on the boundary. Hence in relation with the curvature there is no violation of the properties defining a Hadamard manifold. The violation must be in another item of the definition. Considering the definition \ref{adamardus} of Hadamard manifolds provided in Appendix \ref{mathtopo} we realize that the only way out from the puzzle is that the surfaces corresponding to the $J(C)$ functions of eq.(\ref{fracastoro}) have to be \textbf{non simply connected}.
\begin{figure}[!hbt]
\begin{center}
\iffigs
\includegraphics[height=55mm]{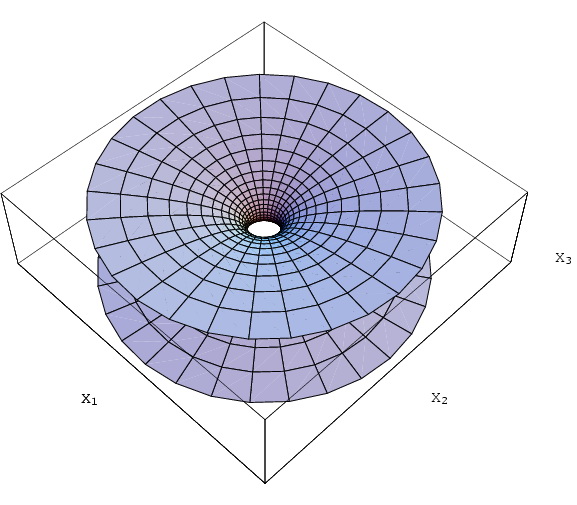}
\includegraphics[height=55mm]{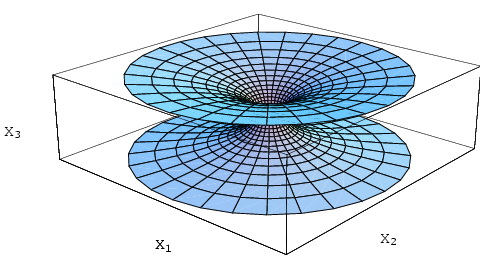}
\else
\end{center}
 \fi
\caption{\it In this picture we present two views of the \textbf{catenoid}, the revolution surface corresponding to $J_{[2]}(C) \, = \,\frac{1}{8} \left(2\,C^2+\cosh [2\, C]\right)$. For large positive or negative values of $C$ one is either in the superior or in inferior plane which is clearly flat with zero curvature. The center of the picture correspond instead to $C\to 0$ and is a sort of strongly negatively curved wormhole that connects the two asymptotic planes. Non simple connectedness is visually spotted. The circles on the surface winding around the throat cannot be contracted to zero and  their homotopy class forms the non trivial element of the first homotopy group $\pi_1(\Sigma)\, = \, \mathbb{Z}$. }
\label{nonconnetto}
 \iffigs
 \hskip 1cm \unitlength=1.1mm
 \end{center}
  \fi
\end{figure}
That this is the case becomes visually obvious when we consider the plot of the surface in three-dimensional space-time
(see fig. \ref{nonconnetto}), yet it is quite clear also analytically. For constant $C$ the orbits of the isometry group spanned by $B\in \left[0,2\,\pi\right]$ are circles of radius:
\begin{equation}\label{radiatore}
    r(C) \, = \, \frac{1}{4} \sqrt{\, 2 \mu +4 \cosh[2 C]}
\end{equation}
The fact that this radius has a minimum different from zero
\begin{equation}\label{formidabile}
    r_{min} \, = \, \frac{1}{4} \sqrt{\, 2 \mu +4} \, > \, 0
\end{equation}
is what spoils simple connectedness and prevents the existence of a fixed point for $\mathrm{U(1)}$. In this way the puzzle is resolved mathematically and we learn an important lesson for Physics. The vanishing of $\pi_1(\Sigma)$ seems to be an essential condition to avoid pathologies in the supergravity lagrangian. This reinforces the idea that  K\"ahler surfaces acceptable in supergravity models should be $\mathrm{CAT}(k)$ manifolds.
\par
Having anticipated this conceptual discussion of their meaning let us work out the details of the models encoded in eq.(\ref{fracastoro}).
Comparing eq.s(\ref{Jmet}) and(\ref{solarium}) we derive the relation between the canonical coordinate $\phi$ and $C$:
\begin{equation}\label{ginnasta}
    \phi \, = \,  \sqrt{2} \, \int \, \sqrt{J_{[\mu]}^{\prime\prime}(C)} \, dC \, = \, \Phi_{[\mu]}(C)\, \equiv \,-\frac{1}{2} \,\mathrm{i} \,\sqrt{\mu +2} \, E\left(\mathrm{i} \,C\left|\frac{4}{\mu +2}\right.\right)
\end{equation}
where $E\left(x\left| m\right.\right)$ denotes the elliptic integral of its arguments.
In the case $\mu \, = \, 2$ which turns out to be that of the \textbf{catenoid},  the function $\Phi_{[\mu]}(C)$ simplifies and it can be easily inverted in terms of elementary functions
\begin{equation}\label{garibaldino}
    \Phi_{[2]}(C) \, = \,\sinh (C) \, \Rightarrow \quad C(\phi) \, = \, \mbox{ArcSinh}(C)
\end{equation}
Substituting into the metric (\ref{matrina}) one finds:
\begin{equation}\label{cagnolinorosso}
\mu \, = \, 2 \quad : \quad     ds^2_\Sigma
\, = \, \frac{\cosh ^2(C)}{2} \left(dC^2 + dB^2\right)
\, = \, \ft 12 \, \left[
 d\phi^2 + \left(\phi ^2+1\right) \, dB^2\right]
\end{equation}
This implies that the derivative of the momentum map is $\mathcal{P}^\prime(\phi)\, = \,\sqrt{\phi ^2+1}$ so that the momentum map and the scalar potential are the following ones:
\begin{equation}\label{sacutillo}
 \mu \, = \, 2 \quad: \quad   \mathcal{P}(\phi) \, = \, \frac{1}{2} \left(\sqrt{\phi^2+1}\, \phi +\mbox{ArcSinh}[\phi \,]\right) \quad \Rightarrow \quad V(\phi) \, \propto\, \left(\sqrt{\phi^2+1}\, \phi +\mbox{ArcSinh}[\phi \,]\right)^2
\end{equation}
The metric (\ref{cagnolinorosso}) can be easily recognized to be the pull-back of the flat three-dimensional euclidian metric:
\begin{equation}\label{sommariva}
    ds^2_{\mathbb{E}^3} \, = \, dX_1^2 \, + \, dX_2^2  + \, dX_3^2
\end{equation}
on the following parametric surface:
\begin{eqnarray}
  X_1 &=& \frac{\cos (B) \cosh
   (C)}{\sqrt{2}}\nonumber \\
  X_2 &=& \frac{\cosh (C) \sin
   (B)}{\sqrt{2}}\nonumber \\
  X_3 &=& \frac{C}{\sqrt{2}} \label{catenoidica}
\end{eqnarray}
which is the classical catenoid. For other values of $\mu$ a similar parametric surface of revolution can be written in terms of appropriate functions of $C$.  As we have already anticipated, although the catenoid is  a rotation surface and its isometry is elliptic, its metric does not satisfy neither Gromov et al criterion nor the Physical consistency requirement that requires the existence of a symmetric point. The reason for this pathology is the non trivial fundamental group $\pi_1(\Sigma)$.
\par
Finally let us appreciate the nature of the same problem from the point of view of complex coordinates. If we introduce the the complex coordinate:
\begin{equation}\label{scarogna}
    \zeta \, = \, \exp\left [ C \, - \, {\rm i} \, B\right] \quad ; \quad \bar{\zeta} \, = \, \exp\left [ C \, + \, {\rm i} \, B\right]
\end{equation}
and we insert it into the expression of (\ref{fracastoro}) of the $J(C)$ function we easily obtain the K\"ahler potential:
\begin{equation}\label{sicumerlo}
    \mathcal{K}(\zeta,\bar{\zeta}) \, = \, 2 \, J(C) \, = \,\frac{1}{16} \mu  \log ^2(\zeta\,\bar{\zeta})+
    \frac{\zeta\,\bar{\zeta}}{8}+\frac{1}{8 \,\zeta\,\bar{\zeta}}
\end{equation}
from which we obtain the metric:
\begin{equation}\label{sicuterat}
    ds^2_\Sigma \, = \, \frac{d\zeta\,{d\bar{\zeta} }
   \left(\zeta  \bar{\zeta }
   \left(\mu +\zeta  \bar{\zeta
   }\right)+1\right)}{8 \, \left(\zeta
   \bar{\zeta }\right)^2} \, \stackrel{\mu \, \to \, 2}{\Longrightarrow} \, \frac{d\zeta\,{d\bar{\zeta} }
   \left(\zeta  \bar{\zeta }
   +1\right)^2}{8 \, \left(\zeta
   \bar{\zeta }\right)^2}
\end{equation}
Examining eq.(\ref{sicuterat}) we see that the metric diverges at the symmetry restoration point $\zeta \, = \, 0$ which now is the boundary of the manifold rather than its interior.
\subsubsection{An asymptotically flat elliptic model with $\pi_1(\Sigma) \, = \, 1$}
\begin{figure}[!hbt]
\begin{center}
\iffigs
\includegraphics[height=70mm]{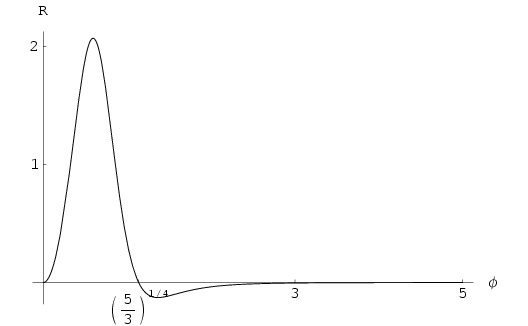}
\else
\end{center}
 \fi
\caption{\it In this picture we present  the plot of the curvature for the elliptic model of eq.(\ref{amomentimappus}). It is limited from above and has three zeros, one at the interior fixed point $\phi \, = \, 0$, a second one at $\phi \, = \,\left( \frac{5}{3}\right)^{1/4 }$ and one on the boundary at $\phi=\infty$
\label{superconnetto}}
 \iffigs
 \hskip 1cm \unitlength=1.1mm
 \end{center}
  \fi
\end{figure}

Let us consider the following momentum map written in terms of the canonical variable $\phi$:
\begin{equation}\label{amomentimappus}
 \mathcal{P}(\phi) \, = \,\phi ^2-\frac{1}{2} \mbox{ArcTan}\left(\phi ^2\right)
\end{equation}
Using the standard formulae (\ref{sodoma}) for the calculation of the VP coordinate we obtain:
\begin{equation}\label{cicciofucca}
  C(\phi) \, = \, \log \left(\frac{\phi }{\sqrt[8]{2
   \phi ^4+1}}\right) \quad \Leftrightarrow \quad \phi \, = \, \left\{\begin{array}{c}
                                                          \pm \sqrt[4]{\sqrt{e^{8\, C}+e^{16 \,C}}+e^{8\, C}} \\
                                                            \pm {\rm i} \,\sqrt[4]{\sqrt{e^{8\, C}+e^{16 \,C}}+e^{8\, C}}\\
                                                           \pm \sqrt[4]{\sqrt{e^{8\, C}\, -\,e^{16 \,C}}+e^{8\, C}}\\
                                                           \pm {\rm i}\sqrt[4]{\sqrt{e^{8\, C}\, -\,e^{16 \,C}}+e^{8\, C}}
                                                         \end{array}
    \right.
\end{equation}
The eighth-root implies the existence of eight branches of the inverse function, that have to considered carefully. Indeed we can accept only those branches where $\phi$ turns out to be  everywhere real. Six branches have to be rejected because of that reason and the only acceptable ones are the first two which are equivalent under the always possible sign revers of $\phi$. In conclusion we have:
\begin{equation}\label{conclusiobranciu}
  \phi \, = \, \sqrt[4]{\sqrt{e^{8\, C}+e^{16 \,C}}+e^{8\, C}}
\end{equation}
Using this branch the infinite interval $\left[-\infty\, , \, \infty\right]$ of the varibale $C$ is mapped into the semi-infinite interval
$\left[ 0 \, , \, \infty\right]$ of the varibale $\phi$. Indeed we have $C(0) \, = \, -\, \infty$, $C(\infty) \, = \,  \infty$.
In the canonical coordinate the form of the metric is:
\begin{equation}\label{cicalucco}
  ds^2_\Sigma \, = \,  d\phi^2 \, + \, f^2(\phi) \, dB^2 \quad ; \quad f^2(\phi) \, = \, \left(\frac{\phi ^5}{\phi
   ^4+1}+\phi \right)^2
\end{equation}
and using eq.(\ref{conclusiobranciu}) we can easily convert it to the VP variable:
\begin{eqnarray}\label{cicaluccoBis}
  ds^2_\Sigma & = &  \ft 12 \, \frac{d^2J}{dC^2} \, \left(dC^2 \, + \, dB^2 \right) \nonumber\\
 & = & \frac{\sqrt{\sqrt{e^{8 C}+e^{16
   C}}+e^{8 C}} \left(2 \sqrt{e^{8
   C}+e^{16 C}}+2 e^{8
   C}+1\right)^2}{\left(\sqrt{e^{8
   C}+e^{16 C}}+e^{8
   C}+1\right)^2} \, \left(dC^2 \, + \, dB^2 \right) \,
\end{eqnarray}
For $C\to \, - \, \infty$ the behavior of the metric coefficient is:
\begin{equation}\label{figliuccio}
  \ft 12 \, \frac{d^2J}{dC^2} \, \stackrel{C\to \, - \, \infty}{\approx} \, e^{2 C}+\frac{5 e^{6 C}}{2} \, + \, \mathcal{O}\left(e^{10 \, C}\right) \quad \Rightarrow \quad J(C) \, \stackrel{C\to \, - \, \infty}{\approx} \, \ft 12 \, e^{2 C}
\end{equation}
while for $C\to \,  \infty$ it is the following:
\begin{equation}\label{spaccanapolino}
 \ft 12 \, \frac{d^2J}{dC^2} \, \stackrel{C\to \, \infty}{\approx}\, 4 \sqrt{2} e^{4\,C} -\frac{3 e^{-4\, C}}{\sqrt{2}}  \, + \, \mathcal{O}\left( e^{-12 C} \right)  \quad \Rightarrow \quad J(C) \, \stackrel{C\to \,  \infty}{\approx} \, \ft 12 \, \ft 1{\sqrt{2}} \, e^{4\,C}
\end{equation}
From previous considerations we see that $C\to \, -\, \infty$ corresponds to $\phi \, = \, 0$ and hence to the fixed point in the interior of the manifold, so that the exponential behavior of $J(C)$ is the expected one for an elliptic isometry. At the same time the exponential behavior on the unique boundary implies that the limiting curvature on the boundary should be zero.
Indeed from the standard formula (\ref{giunone}) for the curvature we obtain:
\begin{equation}\label{lacurbatura}
  R(\phi) \, = \, -\frac{2 \phi ^2 \left(3 \phi ^4-5\right)}{\left(\phi^4+1\right)^2 \left(2 \phi ^4+1\right)} \quad ; \quad R(0)\, = \,0 \quad ; \quad R(\infty)\, = \,0
\end{equation}
whose plot is displayed in fig. \ref{superconnetto}. The vanishing of the limiting curvature is visually evident.
Finally let us make sure that the isometry of this model is indeed elliptic. This we verify by showing that the metric (\ref{cicalucco}) can be retrieved as the pull-back of the flat Lorentz metric in Minkowsian three-dimensional space (\ref{minkiometra}) on the parametric revolution surface (\ref{pianinidivano2}) defined by:
\begin{equation}\label{grugnobello}
  f(\phi)   \, = \, \frac{\phi ^5}{\phi ^4+1}+\phi \quad ; \quad g(\phi) \, \equiv \,\int_0^\phi \, \sqrt{\frac{\sigma ^4
   \left(\sigma ^4+5\right) \left(3
   \sigma ^8+9 \sigma
   ^4+2\right)}{\left(\sigma
   ^4+1\right)^4}} \, d\sigma
\end{equation}
\begin{figure}[!hbt]
\begin{center}
\iffigs
\includegraphics[height=80mm]{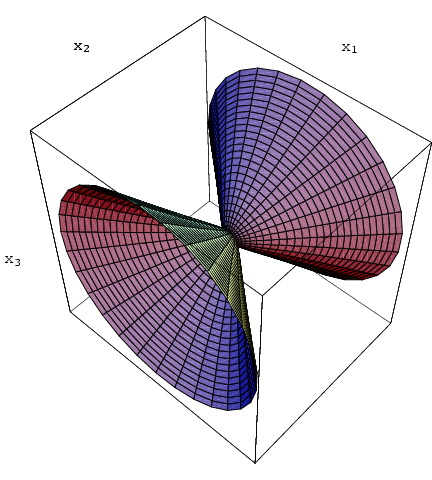}
\includegraphics[height=80mm]{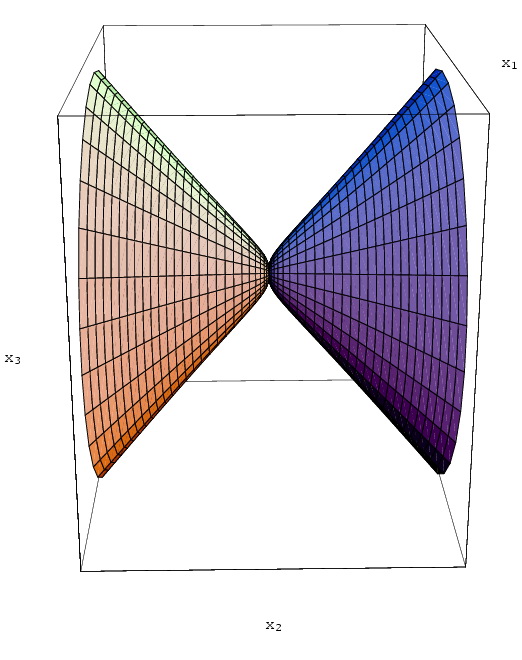}
\else
\end{center}
 \fi
\caption{\it In this picture we present  two views of the revolution surface $\Sigma$ associated with  the elliptic model of eq.(\ref{amomentimappus}).  It is clearly regular and smooth everywhere.}
\label{sbalordo}
 \iffigs
 \hskip 1cm \unitlength=1.1mm
 \end{center}
  \fi
\end{figure}
Two views of this surface are presented in fig.\ref{sbalordo}. It is evident from the picture that this surface is simply connected and that there is in the interior of the manifold a fixed point. It is given by $X_1=X_2=X_3 \, = \, 0$ which lies on the surface and where the radius of the $\mathrm{U(1)}$ orbit shrinks to zero.
\subsubsection{The cigar metric: an asymptotically flat elliptic model with $J(C)\propto C^2$ at the boundary}
\label{barabba}
Inspired by the black-hole model studied in \cite{sigarus} let us consider the following K\"ahler metric written in terms of the complex coordinate $\zeta$:
\begin{equation}\label{fragillo}
    ds^2_{\mathrm{Kahler}} \, = \, \frac{d\zeta \, d\bar{\zeta}}{1 \, + \,\zeta \, \bar{\zeta}}
\end{equation}
It is immediately evident from its form that the metric $\ref{fragillo}$ admits a  $\mathrm{U(1)}$ group of isometries corresponding to the phase transformations: $\zeta \, \to \, \zeta \, e^{{\rm i} \,\theta}$. This is certainly an elliptic isometry. Consider next the transcription of the metric in real variables using the first of the complex structures listed in eq.(\ref{curlandia}), namely $\zeta \, = \, \exp\left[ C \, + \, {\rm i} B\right]$. We obtain:
\begin{equation}
\label{ciurlatamanica}
   ds^2_{\mathrm{Kahler}} \, = \, \frac{\left(\mbox{d}B^2+\mbox{d}C^2\right) e^{2 C}}{1+e^{2 C}} \quad \Rightarrow \quad \ft 12 \, \frac{d^2J}{dC^2} \, = \,\frac{e^{2 C}}{1+e^{2 C}}
\end{equation}
Utilizing next eq.s(\ref{solarium},\ref{sodoma},\ref{Jmet}) in the reverse order we obtain:
\begin{eqnarray}
  \mathcal{P}^\prime(\phi) &=& \tanh(\phi) \quad \Rightarrow \quad \mathcal{P}(\phi) \, = \, \gamma +\log (\cosh (\phi ))\label{momentomimappo}\\
  C(\phi) &=& \log (\sinh (\phi )) \label{toineC}\\
  ds^2_{\mathrm{Kahler}} &=& d\phi^2 \, + \,\tanh^2(\phi) \, \mathrm{d}B^2 \label{canonicsigaret}
\end{eqnarray}
From eq.(\ref{toineC}) we realize that $\phi\, = \, 0$, which is in the interior of the manifold, corresponds to $C\,\to\, - \, \infty$ and provides the necessary fixed point of the elliptic isometry since it is a zero of the $\mathcal{P}^\prime(\phi)$ function. Near this limit the asymptotic behavior of the $J(C)$ function is easily deduced from eq.(\ref{ciurlatamanica}) and it has the required  exponential form:
\begin{equation}\label{gorillusCamel}
    J(C) \, \stackrel{C\,\to\, - \, \infty}{\approx} \, \mbox{const} \, \times \, e^{2 C}
\end{equation}
The asymptotic behavior of $J(C)$ for $C\, \to \, \infty$ is also easily determined from eq.(\ref{ciurlatamanica}) and we get:
\begin{equation}\label{cangerolino}
    J(C) \, \stackrel{C\,\to \, \infty}{\approx} \, \mbox{const} \, \times \, C^2
\end{equation}
This asymptotic behavior is consistent with the vanishing of the curvature at $\phi \, \to \, \infty$ that corresponds to $C\, \to \, \infty$. Indeed using (\ref{momentomimappo}) in eq.(\ref{giunone}) we get:
\begin{equation}\label{carondimonio}
    R(\phi) \, = \, \frac{1}{\cosh^2(\phi)}
\end{equation}
which is strictly positive definite and tends to zero as $|\phi| \, \to \, \infty$. Notwithstanding the elliptic type of the isometry, the asymptotic behavior of $J(C)$ at the boundary realizes the flat metric in the style which is usually associated with the case of a parabolic isometry. This apparent mystery is completely resolved by looking at the form of the parametric surface of which (\ref{canonicsigaret}) is the curvature.
\par
\begin{figure}[!hbt]
\begin{center}
\iffigs
\includegraphics[height=80mm]{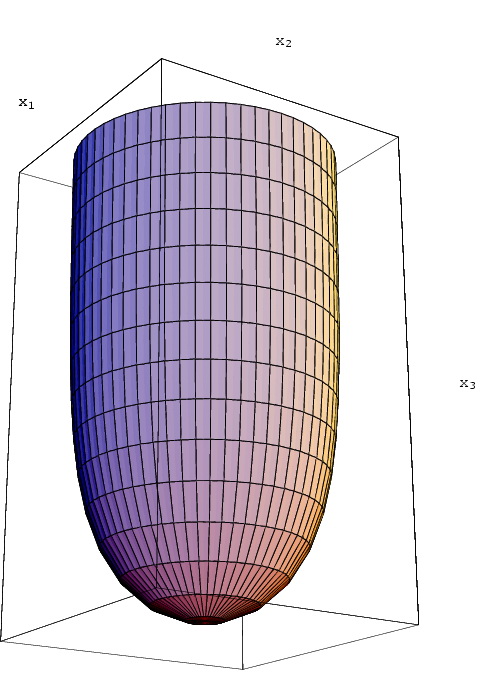}
\else
\end{center}
 \fi
\caption{\it The cigar shaped euclidian surface whose K\"ahler metric is displayed in eq.(\ref{fragillo}). It is a revolution surface which becomes asymptotically cylindrical.
\label{tobacconista}}
 \iffigs
 \hskip 1cm \unitlength=1.1mm
 \end{center}
  \fi
\end{figure}

Defining the following revolution surface $\Sigma_{cigar}$ in the euclidian three-space $\mathbb{E}^3$ :
\begin{eqnarray}
  X_1 &=& \tanh(\phi) \, \cos (B)\nonumber\\
  X_2 &=& \tanh(\phi) \, \sin (B) \nonumber\\
  X_3 &=& g(\phi) \label{leoncinogonfiabile}
\end{eqnarray}
where
\begin{equation}\label{sinardo}
 g(\phi) \, = \,   \log \left(\sqrt{2} \cosh (\phi)+\sqrt{\cosh (2 \phi)+3}\right)
 -\frac{\sqrt{\cosh (2 \phi )+3} \mbox{sech}(\phi)}{\sqrt{2}}
\end{equation}
we easily verify that the pull-back on  $\Sigma_{cigar}$ of the euclidian metric $dX_1^2+dX_2^2+dX_3^2$ is the metric (\ref{canonicsigaret}) under consideration. A picture of this revolution surface is displayed in fig.\ref{tobacconista}.
Looking at the picture we realize that for small $\phi$, close to $C=-\infty$ we are the near the round shaped vertex of the surface, the fixed point of elliptic isometries. As we go to large $\phi$ the surface becomes asymptotically cylindrical, namely flat, and in $C,B$ coordinates it reads:
\begin{equation}\label{rundulitus}
    ds^2_{Kahler} \, \approx \, dC^2 \, + \, dB^2
\end{equation}
like the metric in the plane, yet the coordinate $B$ remains periodic because we are on the cylinder and not on the plane.
\section{Examples of non maximally symmetric K\"ahler manifolds with an  isometry group of the parabolic type.}
\label{parabolica}
In this section we present four examples of non maximally symmetric surfaces with parabolic isometry that lead to cosmological potentials of interest by means of their use in minimal supergravity where the parabolic isometry is gauged:
\begin{enumerate}
  \item The D-map image of the simplest attractor models based on a momentum map of the form $\mathcal{P}(\hat{\phi}) \, = \,\tanh^n\left( \frac{\phi}{\sqrt{6}}\right)$. These models are not integrable in the sense explained in Appendix \ref{integralnymodely} but are quite interesting from the phenomenological point of view. We show that the K\"ahler surface in the D-map image of these potentials  is regular and has a finite curvature only if $n \le 2$.
  \item The D-map image of the $I_6$ integrable potential that, as we have already explained, is also an  $\alpha$-attractor  with a value of $\alpha \, = \, \ft 49$ within the permissible range defined by the authors of \cite{alfatrattori}. We present some detailed information about this remarkable model.
  \item The D-map image of the $I_2$ integrable potentials whose class includes the best fit model studied by Sagnotti et al in the context of the climbing scalar set up. Here we show how the integrable representation of the Appel function $F_1$ allows to determine the asymptotic behavior of the $J(C)$ function in a precise way.
   \item A counterexample of a surface with a parabolic isometry where the asymptotic behavior of the $J(C)$ function and of the metric coefficient in one of the boundaries are not those of eq.(\ref{polentaconcia}) rather those of eq.(\ref{oseletti}). The limiting curvature is zero in this case.
\end{enumerate}
\subsection{The simplest attractors}
\label{simplicio}
 Considering the simplest attractor defined in \cite{alfatrattori} as $\mathfrak{P}(U)\, = \, \lambda U^n$ we are now going to show that the corresponding surface $\Sigma$ is non singular if and only if $n\le 2$. For $n>2$ the curvature of $\Sigma$ always develops a rather violent singularity and $\Sigma$ is not a smooth manifold.
\par
\begin{figure}[!hbt]
\begin{center}
\iffigs
\includegraphics[height=80mm]{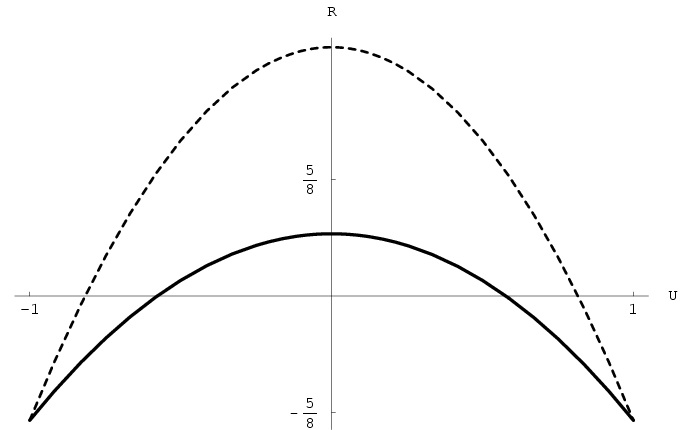}
\caption{\it   Plot of the curvature $R_{(n)}(U)$ for the simplest attractor models $n=1,2$. The dashed line corresponds to $n=1$, while the solid line corresponds to $n=2$. As we see the value of the curvature at the boundary $U\, = \, \pm 1$ is always negative but it increases while we go to the interior of the bulk and it reaches a positive valued maximum at $U=0$.}
\label{curvan12}
 \iffigs
 \hskip 1cm \unitlength=1.1mm
 \end{center}
  \fi
\end{figure}
\begin{figure}[!hbt]
\begin{center}
\iffigs
\includegraphics[height=80mm]{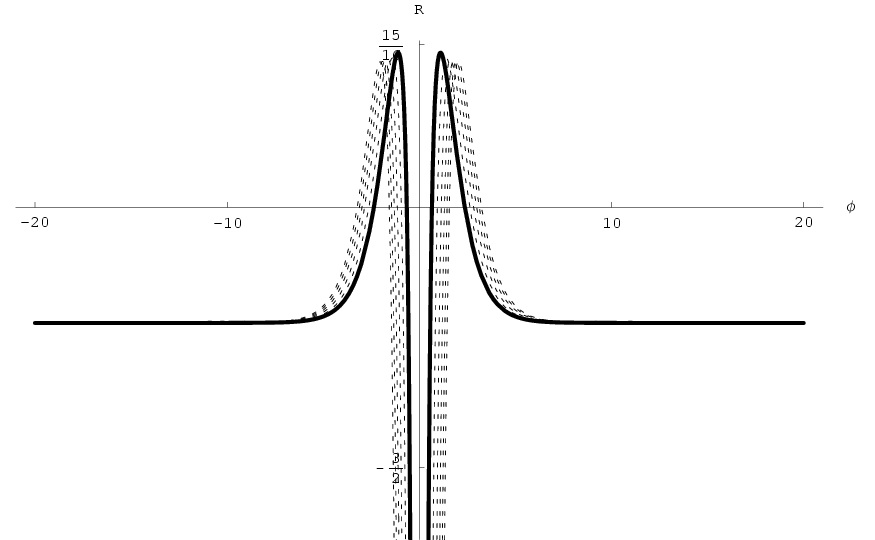}
\caption{\it   Plot of the curvature $R_{(n)}(\phi)$ for the simplest attractor models $n=3,4,5,6,7,8$. The solid line corresponds to $n=3$, while the dashed lines corresponds to $n=4,5,6,7,8$. As we see the value of the curvature at the boundary $U\, = \, \pm 1$ is always negative and has a universal value independent from $n$.  In the interior of the manifold it reaches a positive maximum but then drops to $-\infty$ at $\phi \, = \,0$ which is a true singularity.}
\label{curvapolla}
 \iffigs
 \hskip 1cm \unitlength=1.1mm
 \end{center}
  \fi
\end{figure}

In this case the function $\mathfrak{P}(U)$ is a simple power as we already said: $\mathfrak{P}_{(n)}(U)\, = \, \lambda U^n$. Hence the momentum map is:
\begin{equation}\label{celesteimpero}
    \mathcal{P}_{(n)}(\phi)\, = \, \lambda  \tanh ^n\left(\frac{\phi }{\sqrt{3}}\right)
\end{equation}
and, applying the formula (\ref{giunone}) for the curvature of the associated K\"ahler surface, we find:
\begin{equation}\label{jupiter}
   R_{(n)}(\phi) \, = \, -\frac{1}{3} \left(2 n^2-6 \cosh \left(\frac{2 \phi
   }{\sqrt{3}}\right) n+\cosh \left(\frac{4 \phi
   }{\sqrt{3}}\right)+3\right) \mbox{csch}^2\left(\frac{2
   \phi }{\sqrt{3}}\right)
\end{equation}
Converting (\ref{jupiter}) to the $U$-coordinate we find:
\begin{equation}\label{Uciocio}
    R_{(n)}(U) \, = \,-\frac{(n+1) (n+2)\,U^4-2 n^2\, U^2+(n-3) n+2}{6 U^2}
\end{equation}
From the above explicit expression we see that the curvature has always a double pole at $U=0$ unless the constant term in the numerator \textit{i.e.} $(n-3) n+2$ vanishes. This happens only for $n=1,2$ that are the roots of such a quadratic polynomial in $n$.  Therefore the K\"ahler surface associated with such simplest models is a smooth manifold in the full range $U\in\left[-1,1\right]$ only for $n=1,2$. The plot of the curvature for the two admissible cases is given in fig.\ref{curvan12}.
In the other figure \ref{curvapolla} we display instead the behavior of the curvature for $n\ge 3$. In this case, due to the presence of the pole at $U=0 \,\Leftrightarrow \,\phi =0$, it is visually better to plot the curvature against the variable $\phi$, the boundary $U\, = \, \pm 1$ corresponding to $\phi \, = \, \pm \, \infty$.
Inspired by the above observation one can introduce a mixed quadratic attractor as described in the next subsection.
\subsubsection{Mixed quadratic attractor.}
We just name \textit{mixed quadratic attractor} the model where the function $\mathfrak{P}(U)$ advocated in eq.(\ref{trattorialfa}) is a generic quadratic polynomial:
\begin{equation}\label{gonadirotte}
    \mathfrak{P}_{(a,b,c)}(U) \, = \, a \, U^2+b \, U+c \quad \Rightarrow \quad \mathcal{P}(\phi) \, = \, a \tanh ^2\left(\frac{\phi }{\sqrt{3}}\right)+b \tanh
   \left(\frac{\phi }{\sqrt{3}}\right)+c
\end{equation}
The corresponding curvature is easily calculated as function of $U$ and we find:
\begin{equation}\label{culiscon}
    R_{(a,b)}(U) \, = \, \frac{-12\, a \, U^3-3\, b \, U^2+8\, a \,U+b}{3\, b+6\, a\, U}
\end{equation}
Note that the curvature is independent from the value of $c$, namely from the constant shift in the momentum map that deserves the name of Fayet Iliopoulos term.
\begin{figure}[!hbt]
\begin{center}
\iffigs
\includegraphics[height=80mm]{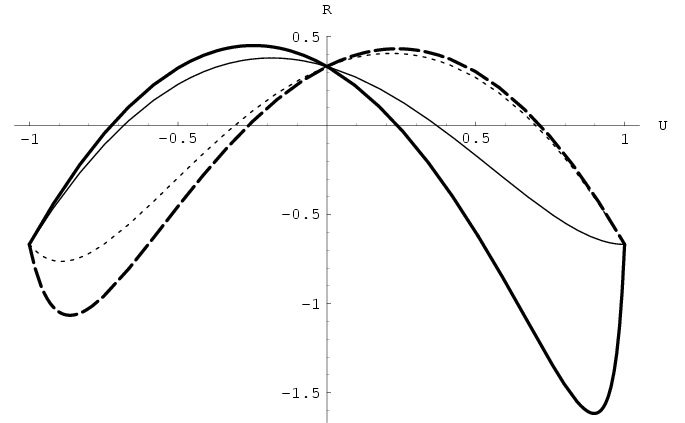}
\caption{\it   Plot of the curvature $R_{(a,b)}(U)$ for the mixed attractor model where $\mathfrak{P}(U)$ is just a quadratic polynomial. The solid thick line corresponds to parameters $(a,b) \, = \, (\ft 13\, , \, -\ft{7}{10})$. The dashed thick line corresponds to parameters $(a,b) \, = \, (\ft 13\, , \, \ft{4}{5})$. The solid thin line corresponds to parameters $(a,b) \, = \, (\ft 14\, , \, -1)$. The thin dashed line corresponds to parameters $(a,b) \, = \, (-\ft 13\, , \, -1)$}
\label{curvabiscia}
 \iffigs
 \hskip 1cm \unitlength=1.1mm
 \end{center}
  \fi
\end{figure}
The curvature is finite over the whole range $[-1,1]$ of definition of the variable $U$  if the denominator has no zero in that range. This occurs
\begin{equation}\label{ginotappo}
  \mbox{if} \quad  |\,b\,| \ge |2 \, a| \quad \mbox{for  $b,a \ne 0$} \quad \mbox{or if $a \,=\, 0$}\quad \mbox{or if $b \,=\, 0$}.
\end{equation}
In fig.\ref{curvabiscia} we have displayed the behavior of the curvature for a few cases of parameters $(a,b)$ satisfying the restrictions imposed by non singularity. Furthermore let us observe that independently from the parameters $a,b$ the curvature has a universal value at $U\, = \, \pm 1$, namely:
\begin{equation}\label{puliscitibocchino}
    R_{(a,b)}(\pm 1) \, = \, - \frac{2}{3}
\end{equation}
\par
In this case it is quite easy to calculate the integral (\ref{sodoma}) which defines the VP coordinate, of which we can provide the analytic form both as a function $\phi$ and as a function of $U$. We find:
\begin{eqnarray}
  C_{(a,b)}(\phi) &=& \frac{1}{4
   \left(b^2-4 a^2\right)^2} \, \times \left(96 \log \left(b \cosh \left(\frac{\phi
   }{\sqrt{3}}\right)+2 a \sinh \left(\frac{\phi
   }{\sqrt{3}}\right)\right) a^3 \right.\nonumber\\
   &&\left. +6 \left(4 a^2-b^2\right)
   \cosh \left(\frac{2 \phi }{\sqrt{3}}\right) a+2 \sqrt{3}
   b \left(b^2-12 a^2\right) \phi +3 b \left(b^2-4
   a^2\right) \sinh \left(\frac{2 \phi }{\sqrt{3}}\right)\right) \nonumber \\
 C_{(a,b)}(U) &=& \frac{1}{4 \left(b^2-4 a^2\right)^2
   \left(U^2-1\right)}\times \left( 3 \left((U-1) (U+1) \left(-16 \left(\log
   \left(1-U^2\right)-2 \log (b+2 a U)\right) a^3\right.\right.\right.\nonumber\\
   &&\left.\left.\left.+\left(12
   a^2 b-b^3\right) \log (1-U)+\left(b^3-12 a^2 b\right)
   \log (U+1)\right)\right.\right.\nonumber\\
   &&\left.\left.-2 (2 a-b) (2 a+b) \left(a U^2-b\,
   U+a\right)\right)\right)
\end{eqnarray}
\begin{figure}[!hbt]
\begin{center}
\iffigs
\includegraphics[height=80mm]{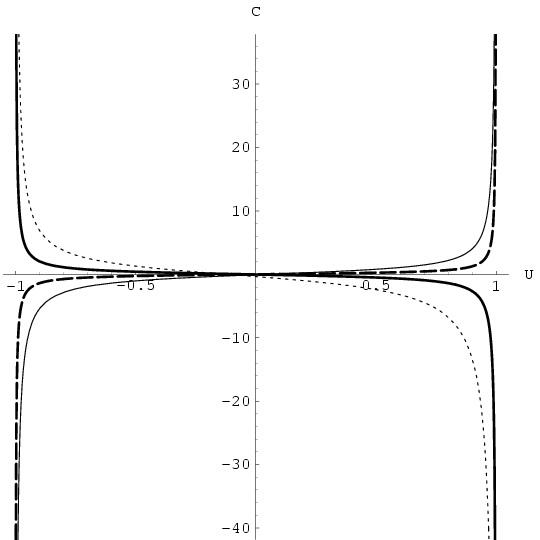}
\caption{\it   Plot of the VP coordinate $C_{(a,b)}(U)$ as function of $U$ in the mixed quadratic attractor models.  The solid thick line corresponds to parameters $(a,b) \, = \, (\ft 13\, , \, -\, 4)$. The dashed thick line corresponds to parameters $(a,b) \, = \, (\ft 14\, , \, 7)$. The solid thin line corresponds to parameters $(a,b) \, = \, (\ft 15\, , \, 2)$. The thin dashed line corresponds to parameters $(a,b) \, = \, (\ft 15\, , \, -1)$}
\label{CUmixquadr}
 \iffigs
 \hskip 1cm \unitlength=1.1mm
 \end{center}
  \fi
\end{figure}
Plots of $C_{(a,b)}(U)$ for a few different choices of the parameters are displayed in fig.\ref{CUmixquadr}. As we easily realize from such plots the VP coordinate goes always from $-\infty$ to $+\infty$ as the coordinate $U$ spans its own range
$[-1,1]$. The direction of growth can be inverted into a descent direction depending on the values of $(a,b)$. In any case the function $C_{(a,b)}(U)$ is monotonic.
\par
Let us now consider the behavior of the metric coefficient. Utilizing the $U$ coordinate the line-element on the K\"ahler surfaces associated with the mixed quadratic attractor is:
\begin{eqnarray}\label{giornalaio}
    ds^2_\Sigma & = & \frac{3 \,\mbox{dU}^2}{\left(1-U^2\right)^2}+\frac{1}{3}
    \,(b+2 \, a \, U)^2 \left(U^2-1\right)^2 \, \mbox{dB}^2\nonumber\\
   &\Downarrow & \nonumber\\
   p(U) & = & \frac{3 }{\left(1-U^2\right)^2} \quad ; \quad q(U) \, = \, (b+2 \,a\,U)^2 \left(U^2-1\right)^2
\end{eqnarray}
 Since we have:
\begin{equation}\label{colaianno}
   \ft 12 \, \frac{d^2J}{dC^2} \, = \, q(U)
\end{equation}
we see that the metric coefficient tends to zero while $U \to \pm 1$, but unless $b=0$ there are no  zeros of $q(U)$ in the interior of the manifold, namely for $|U|<1$. This means that unless $b=0$ the elliptic interpretation of the gauged isometry is forbidden and we have to interpret it as parabolic.
\par
Let us consider the asymptotic behavior in the two boundaries $U \to \pm 1$ and verify that it agrees with that predicted by the limiting value of the curvature.
In one of the two boundaries the $C$-function goes to $-\infty$, while in the other it goes to so $+\infty$.
\par
Let us set $U=1-\xi$ and expand $q(1-\xi)$ in series of $\xi$. We have:
\begin{eqnarray}\label{corneliogracco}
    q(1-\xi) & = & \frac{4 a^2 \xi ^6}{3}+\left(-8 a^2-\frac{4 b a}{3}\right)
   \xi ^5+\left(\frac{52 a^2}{3}+\frac{20 b
   a}{3}+\frac{b^2}{3}\right) \xi ^4+\left(-16 a^2-\frac{32
   b a}{3}-\frac{4 b^2}{3}\right) \xi ^3\nonumber\\
   &&+\left(\frac{16
   a^2}{3}+\frac{16 b a}{3}+\frac{4 b^2}{3}\right) \xi ^2
\end{eqnarray}
 Consider however the behavior of $C(1-\xi)$, we have:
\begin{eqnarray}\label{bestialdo}
    C_{(a,b)}(1-\xi) & = & \frac{3 \left(16 a^3-8 b a^2-4 b^2 a+2 b^3\right)}{8
   \left(b^2-4 a^2\right)^2 \xi }\nonumber\\
   &&+\frac{1}{4 \left(b^2-4 a^2\right)^2} \times \left(3 \left(\frac{1}{4}
   \left(16 a^3-8 b a^2-4 b^2 a+2 b^3\right)\right.\right.\nonumber\\
   &&\left.\left.+\frac{1}{2}
   \left(64 \log (2 a+b) a^3-32 \log (\xi ) a^3-32 \log (2)
   a^3-16 a^3+8 b a^2\right.\right.\right.\nonumber\\
   &&\left.\left.\left.+24 b \log (\xi ) a^2-24 b \log (2)
   a^2+4 b^2 a-2 b^3-2 b^3 \log (\xi )+2 b^3 \log
   (2)\right)\right)\right)+\mathcal{O}\left(\xi
   ^1\right)\nonumber\\
\end{eqnarray}
Hence the leading term going to infinity in $C(1-\xi)$ is $\frac{1}{\xi}$ which always wins on $\log(\xi)$.
\par
Consider next   the leading term in the $\xi$-expansion of $C$ provided by eq.(\ref{bestialdo}). We have:
\begin{equation}\label{Casinone}
  C_{(a,b)} \, \stackrel{\xi \, \to \, 0}{\approx} \, \frac{3}{4( 2\,a \,+\, b)}\times \frac{1}{\xi}
\end{equation}
At the same time from eq.(\ref{corneliogracco}) we derive the asymptotic behavior of the square of the momentum map derivative or metric coefficient
 \begin{equation}\label{cosmonauta}
\ft 12 \frac{\mathrm{d}^2J}{\mathrm{d}C^2} \, \equiv \, q(1-\xi)  \, = \, \left(\mathcal{P}^\prime(\phi)\right)^2 \, \stackrel{\xi \, \to \, 0}{\approx} \, \frac{2(2\, a \, + \, b)}{\sqrt{3}} \, \times \, \xi^2
\end{equation}
from which we deduce:
\begin{equation}\label{floristilio}
  \ft 12 \frac{\mathrm{d}^2J}{\mathrm{d}C^2} \, \stackrel{\xi \, \to \, 0}{\approx} \, \frac{3}{4}\,  \times \, \frac{1}{C^2}
\end{equation}
This implies that:
\begin{equation}\label{fluoromangio}
  J(C) \, \stackrel{C \, \to \, - \, \infty}{\approx}  \, - \, \ft 32 \, \log\left[\, C \right] \, = \, \frac{1}{R_{(a,b)}(\infty) } \,  \log\left[\, C \right]
\end{equation}
As predicted by general considerations the function $J(C)$ has a logarithmic divergence both on both boundaries of the manifold ($C\, \to \, -\, \infty$) and ($C\, \to \, \infty$). Furthermore the coefficient of the $\log$ is just the inverse of the limiting value of the curvature, which in this case is identical in both extrema.
\par
Hence, relying on such conclusions and the parabolic interpretation of the gauged isometry we identify the correct complex variable as the plane-one:
\begin{equation}\label{grugnotto}
    t \, = \, {\rm i} \, C \, - \, B
\end{equation}
\par
\begin{figure}[!hbt]
\begin{center}
\iffigs
\includegraphics[height=80mm]{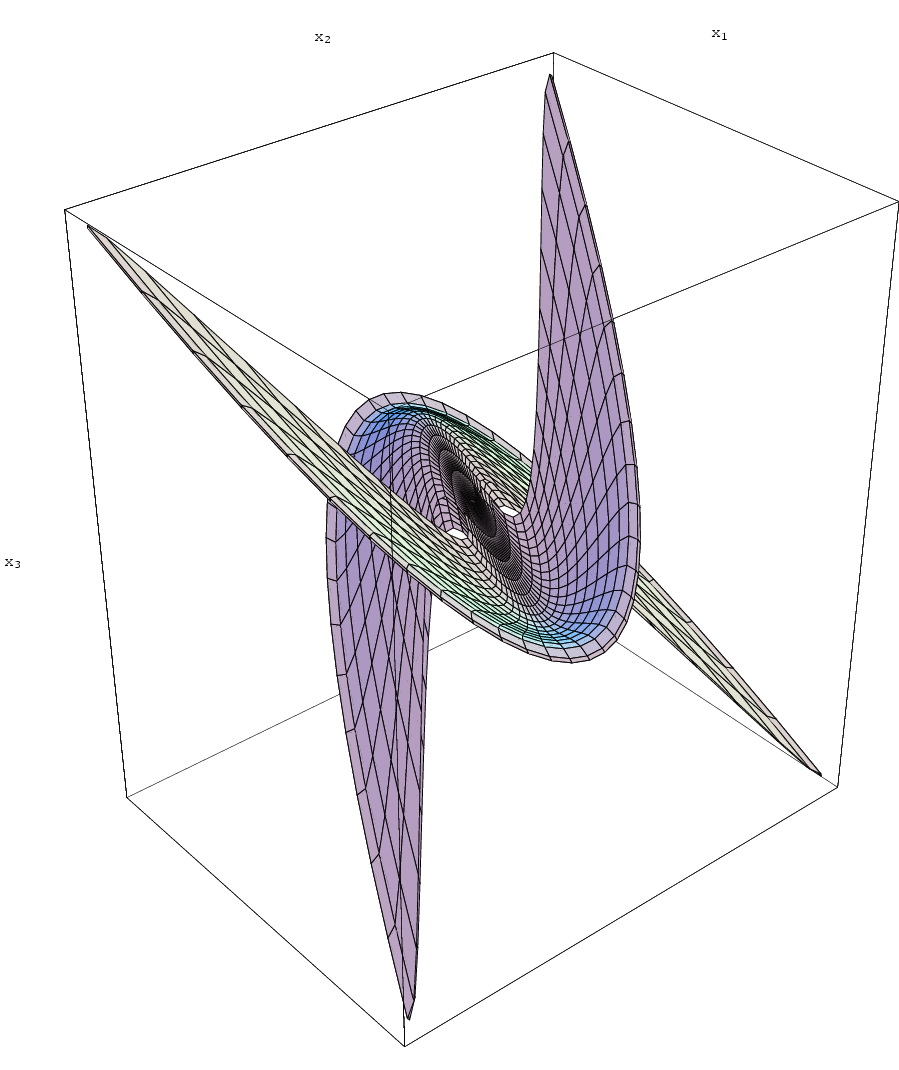}
\includegraphics[height=80mm]{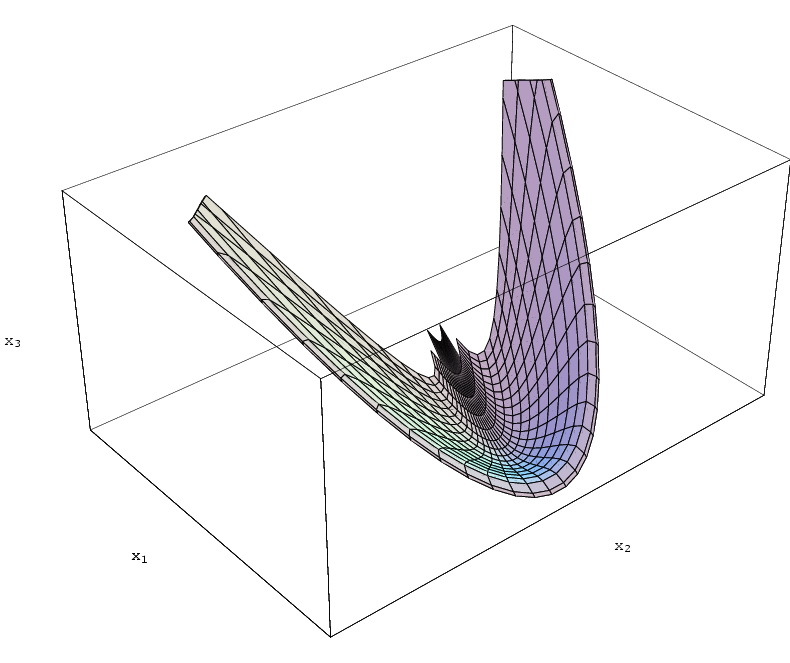}
\caption{\it   In this figure we present the $3D$-appearance of the K\"ahler surface $\Sigma$ associated with the mixed quadratic attractor of parameters $a=\sqrt{3}$ and $b=\ft 52 \, \sqrt{3}$. The surface is composed of two branches associated with the two choices of the signs $\left\{\pm X_1, X_2, \pm X_3\right \}$. The picture on the left displays the two branches together. The picture on the right displays only one branch.}
\label{sigmagorgona}
 \iffigs
 \hskip 1cm \unitlength=1.1mm
 \end{center}
  \fi
\end{figure}

Furthermore we can use the embedding (\ref{pianinidivanoparab}-\ref{granolatoparab}) appropriate to the parabolic case   in order to obtain a three-dimensional model of the surface $\Sigma$ in which the orbits of the parabolic translation group are parabolae drawn on the surface. Identifying $f(\phi) \, = \, \mathcal{P}^\prime(\phi)$, as it is appropriate for the metric (\ref{solarium})  we can always reproduce this latter by choosing in (\ref{pianinidivanoparab}) a $g(\phi)$ function that satisfies the differential equation:
\begin{equation}\label{granolato}
    f'(\phi ) \, g'(\phi ) \, = \, 1 \quad \Rightarrow \quad g(\phi) \, = \,
    \int \frac{1}{\mathcal{P}^{\prime\prime}(\phi)} \, d\phi
\end{equation}
As an example we have chosen the case of parameters $a=\sqrt{3}$ and $b=\ft 52 \, \sqrt{3}$. So doing we obtain:
 \begin{eqnarray}
   f(\phi) &=&\frac{1}{2} \mbox{sech}^3\left(\frac{\phi }{\sqrt{3}}\right)
   \left(5 \cosh \left(\frac{\phi }{\sqrt{3}}\right)+4 \sinh
   \left(\frac{\phi }{\sqrt{3}}\right)\right)\nonumber\\
   g(\phi) &=& \frac{1}{23652} \times \left(-37376 \sqrt{3} \phi -9855 \cosh \left(\frac{2 \phi
   }{\sqrt{3}}\right)\right.\nonumber\\
   &&\left.+50370 \log \left|4 \cosh \left(\frac{2
   \phi }{\sqrt{3}}\right)+5 \sinh \left(\frac{2 \phi
   }{\sqrt{3}}\right)-8\right|\right.\nonumber\\
   &&\left.-7062 \sqrt{73} \left(\log
   \left|-12 \sqrt{73} \tanh \left(\frac{\phi
   }{\sqrt{3}}\right)-5 \sqrt{73}+73\right|\right.\right.\nonumber\\
   &&\left.\left.-\log \left|12
   \sqrt{73} \tanh \left(\frac{\phi }{\sqrt{3}}\right)+5
   \sqrt{73}+73\right|\right)+7884 \sinh \left(\frac{2 \phi
   }{\sqrt{3}}\right)\right)
 \end{eqnarray}
 In the above formula $\log\left|\dots\right|$ denotes the logarithm of the absolute value.
The complete surface $\Sigma$ is composed of two branches obtained by adding to the parameterized surface (\ref{pianinidivanoparab}) the other one  obtained by  inverting the signs of $X_1$ and $X_3$.
In fig.\ref{sigmagorgona} we have displayed the $3D$-dimensional image of $\Sigma$.
\subsubsection{The quadratic attractor: an exception}
\label{bububu}
Let us now consider the particular case $a=1,b=0$ of the mixed attractor, namely the pure quadratic one. In this case, written in the coordinate $U$ the metric takes the form:
\begin{eqnarray}\label{giornalistone}
    ds^2_\Sigma & = & \frac{3 \,\mbox{dU}^2}{\left(1-U^2\right)^2}+\frac{4}{3}
    \, U^2 \left(U^2-1\right)^2 \, \mbox{dB}^2\nonumber\\
   &\Downarrow & \nonumber\\
   p(U) & = & \frac{3 }{\left(1-U^2\right)^2} \quad ; \quad q(U) \, = \, \frac{4}{3}\,U^2 \left(U^2-1\right)^2
\end{eqnarray}
and we realize the presence of a zero of $q(U)$ in the interior of the $U$ definition interval, namely for $U=0$. This suggests the existence of a fixed point in the interior of the manifold implying an elliptic interpretation of the gauged isometry. In order to transform this suggestion into a certainty we just have to verify that the VP coordinate $C(U)$ goes to $-\infty$ for $U\to 0$ and that in the same limit  $q(U)$ approaches zero  as $\exp[\delta \, C(U)]$, for some positive $\delta$. These conditions are easily verified. Indeed we have:
\begin{equation}\label{cufunzi}
    C(U) \, = \, \frac{3 \left(-8 \left(U^2+1\right)-16 (U-1) (U+1)
   \left(\log \left(1-U^2\right)-2 \log (2
   U)\right)\right)}{64 \left(U^2-1\right)}
\end{equation}
which tends to $-\, \infty$ for $U\to 0$. Furthermore we have:
 \begin{eqnarray}
   q(U) &\stackrel{U\to 0}{\approx}& \frac{4 U^2}{3} \, + \, \mathcal{O}(U^4)\nonumber \\
   \exp\left[\ft 43 \, C(U)\right] &\stackrel{U\to 0}{\approx}& 4 \sqrt{e} U^2 \, + \, \mathcal{O}(U^4)
 \end{eqnarray}
so that we conclude:
\begin{equation}\label{ciuralafigno}
    \ft 12 \, \frac{\mathrm{d}^2J}{\mathrm{d}C^2} \, \stackrel{C\, \to \, - \,\infty}{\approx} \, \frac{1}{3\, \sqrt{e} } \, \exp\left[\ft 43 \, C\right]
\end{equation}
which confirms that the isometry is indeed elliptic.
\begin{figure}[!hbt]
\begin{center}
\iffigs
\includegraphics[height=90mm]{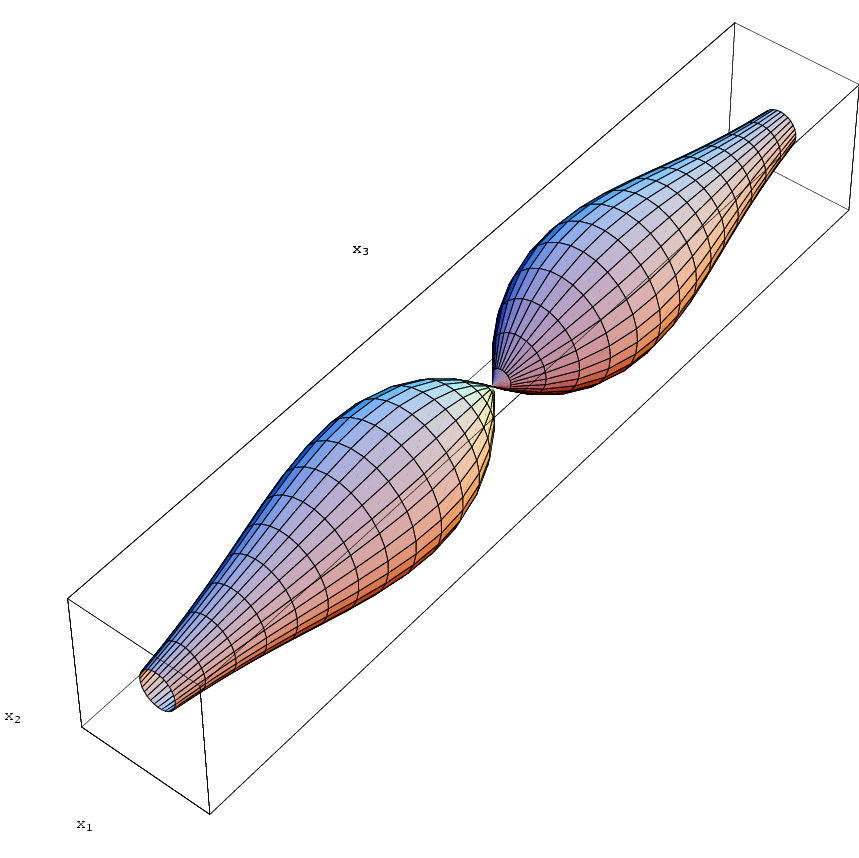}
\caption{\it   The revolution surface in euclidian three-dimensional space that corresponds to the simplest
quadratic $\alpha$-attractor. Differently from the case of the linear and mixed attractors, the gauged isometry in the quadratic case is elliptic and the K\"ahler metric corresponding to this inflaton potential is just the pull back of the euclidian three-dimensional metric on the revolution surface displayed in this figure. The fixed point is clearly visible in the drawing of this surface that is shaped like a doubled baseball bat.}
\label{mazzata}
 \iffigs
 \hskip 1cm \unitlength=1.1mm
 \end{center}
  \fi
\end{figure}
In terms of the canonical variable $\phi$, the metric (\ref{giornalistone}) takes the following explicit expression:
\begin{equation}\label{edicolante}
    ds^2_\Sigma \, = \,  \mathrm{d}\phi^2 \, + \, \frac{4}{3} \mbox{sech}^4\left(\frac{\phi
   }{\sqrt{3}}\right) \tanh ^2\left(\frac{\phi
   }{\sqrt{3}}\right) \, \mathrm{d}B^2
\end{equation}
We can easily check that (\ref{edicolante}) is the pull-back of the euclidian metric
\begin{equation}
ds^2_{\mathbb{E}^3}\, = \, dX_1^2\,+\,dX_2^2\,+\, dX_3^2
\end{equation}
on the following parametric surface:
\begin{eqnarray}
  X_1 &=& f(\phi) \, \cos(B) \nonumber\\
  X_2 &=& f(\phi) \, \sin(B)\nonumber\\
  X_3 &=& g(\phi)
\end{eqnarray}
where:
\begin{eqnarray}
\label{basaballata}
  f(\phi) &=& \frac{2 \mbox{sech}^2\left(\frac{\phi }{\sqrt{3}}\right)
   \tanh \left(\frac{\phi }{\sqrt{3}}\right)}{\sqrt{3}}\\
  g(\phi) &=& \int_0^\phi \, \sqrt{1-\frac{4}{9} \left(\cosh \left(\frac{2
   t}{\sqrt{3}}\right)-2\right)^2
   \mbox{sech}^8\left(\frac{t}{\sqrt{3}}\right)} \, \mathrm{d}t
\end{eqnarray}
A visualization of the surface (\ref{basaballata}) is provided in fig.\ref{mazzata}.
From its doubled baseball--bat shape  we easily see that this is a revolution surface and we also spot the fixed point, invariant under rotations, which sits in the middle.
\subsubsection{The simplest linear attractor}
In this section,  as a further example, we consider the very simplest linear attractor where the function $\mathfrak{P}(U) \, = \, U$ is just linear.
The momentum map and the metric in this case are:
\begin{equation}\label{guru}
    \mathcal{P}(\phi) \, = \, \sqrt{3} \, \tanh \left(\frac{\phi}{\sqrt{3}} \right) \quad \Rightarrow \, ds^2 \, =\, d\phi^2 \, + \,\mbox{sech}^4\left(\frac{\phi }{\sqrt{3}}\right)\, \mathrm{dB}^2
\end{equation}
The VP coordinate is very easily calculated and has a very simple form:
\begin{equation}\label{VPClinear}
    C(\phi) \, = \, \frac{\phi }{2}+\frac{1}{4} \sqrt{3} \sinh \left(\frac{2
   \phi }{\sqrt{3}}\right)
\end{equation}
$C(\phi)$  is a monotonic increasing function of $\phi$ going from  $-\infty$ to $\infty$. The metric coefficient $\mbox{sech}^4\left(\frac{\phi }{\sqrt{3}}\right)$ goes to zero at $\phi=\pm\infty$ so that we should check whether it goes as fast to zero as $\exp\left[\mp 2\,\delta\, C(\phi)\right]$. Without repeating the intermediate steps as in the previous case we can immediately see that the test for $\mathrm{U(1)}$-symmetry fails since $\mbox{sech}^4\left(\frac{\phi }{\sqrt{3}}\right)$ goes to zero as a negative exponential of $\phi$, while $\exp\left[\mp 2\,\delta\, C(\phi)\right]$ goes to zero as a negative exponential of an exponential. Therefore the limit (\ref{sirtaki}) is always infinite and the correct interpretation of the $B$-shift is a parabolic translation symmetry.
\begin{figure}[!hbt]
\begin{center}
\iffigs
\includegraphics[height=90mm]{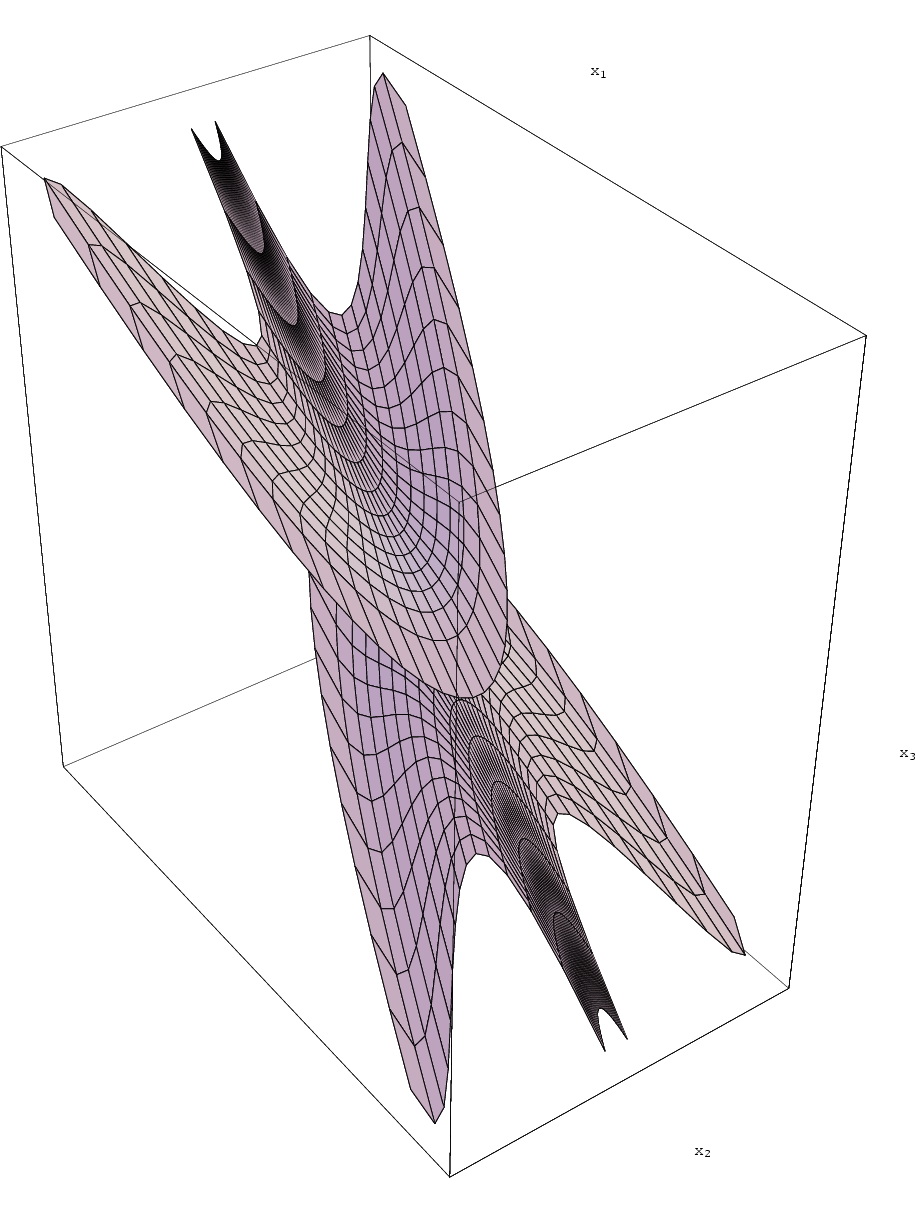}
\includegraphics[height=85mm]{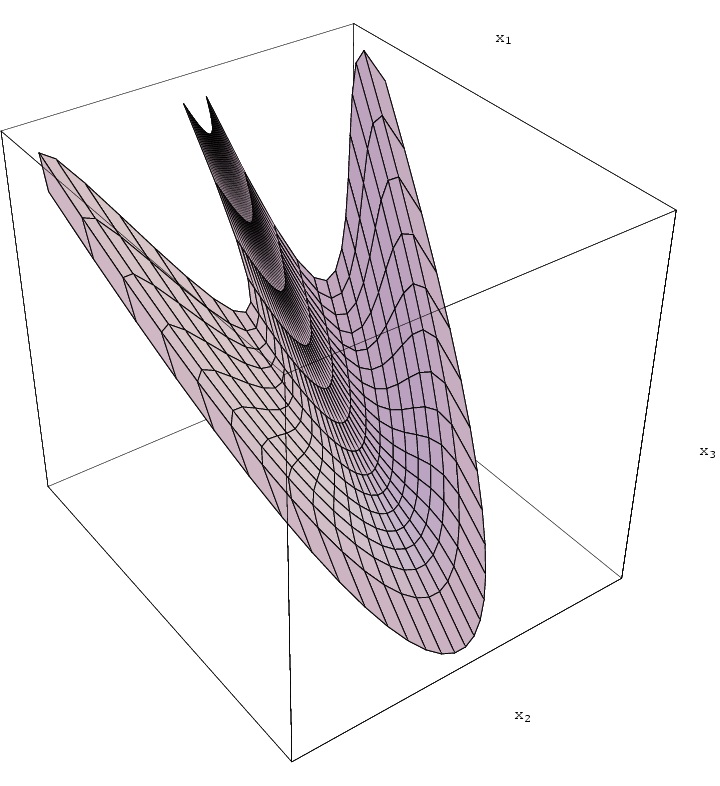}
\caption{\it   In this figure we present the $3D$-plot of the surface $\Sigma$ associated with the simplest linear attractor. As in the mixed quadratic case, also here the surface has two branches due to the choice of signs in front  of $X_1$ and $X_3$. The left picture presents the two branches together while the right picture presents only one branch.}
\label{sigmarazza}
 \iffigs
 \hskip 1cm \unitlength=1.1mm
 \end{center}
  \fi
\end{figure}
In accord with this interpretation we can construct a three-dimensional model of the surface $\Sigma$ associated with the simplest linear attractor by using the parabolic parametrization (\ref{pianinidivanoparab}). In this case the relevant functions
$f(\phi)$ and $g(\phi)$ are the following ones:
\begin{eqnarray}
\label{falaisa}
  f(\phi) &=& \mbox{sech}^2\left(\frac{\phi }{\sqrt{3}}\right) \\
  g(\phi) &=& -\frac{3}{8} \left(\cosh \left(\frac{2 \phi
   }{\sqrt{3}}\right)+4 \log \left(\sinh \left(\frac{\phi
   }{\sqrt{3}}\right)\right)\right)
\end{eqnarray}
The $3D$-appearance of the surface $\Sigma$ produced by the parabolic parameterization (\ref{pianinidivanoparab}) with the functions (\ref{falaisa}) is displayed in fig.\ref{sigmarazza}.
\subsection{The integrable attractor $I_6$}
\label{arcotanno}
The integrable model $I_6$ (see table \ref{tab:families}) is very remarkable. On one hand it is an instance of an $\alpha$-attractor with a value of $\alpha$ in the range determined by the authors of \cite{alfatrattori}, on the other it is integrable in the sense explained in appendix \ref{integralnymodely} and belongs to the bestiary compiled in \cite{noicosmoitegr}. For this reason we dwell a little bit on the properties of the K\"ahler surface in the D-map of this interesting potential and we also summarize the explicit cosmological solution of the corresponding Friedman equations that  can be displayed in closed analytic form thanks to integrability.
\par
\begin{figure}[!hbt]
\begin{center}
\iffigs
\includegraphics[height=70mm]{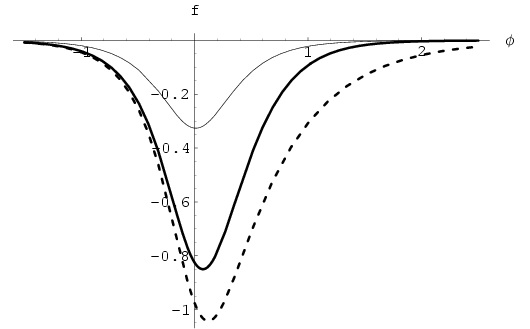}
\caption{\it   In this figure we present the plots  of the function $f_\beta(\phi)$ for the $\Sigma_\beta$ surfaces in the image of the $D$-map of the $I_6$ integrable potentials. The thick solid line corresponds to $\beta \, = \,  \ft {1}{10}$ . The thick dashed line corresponds to $\beta \, = \, 0$. The thin solid line corresponds to $\beta \, = \, 2$. }
\label{I6trolleybus}
 \iffigs
 \hskip 1cm \unitlength=1.1mm
 \end{center}
  \fi
\end{figure}
\begin{figure}[!hbt]
\begin{center}
\iffigs
\includegraphics[height=120mm]{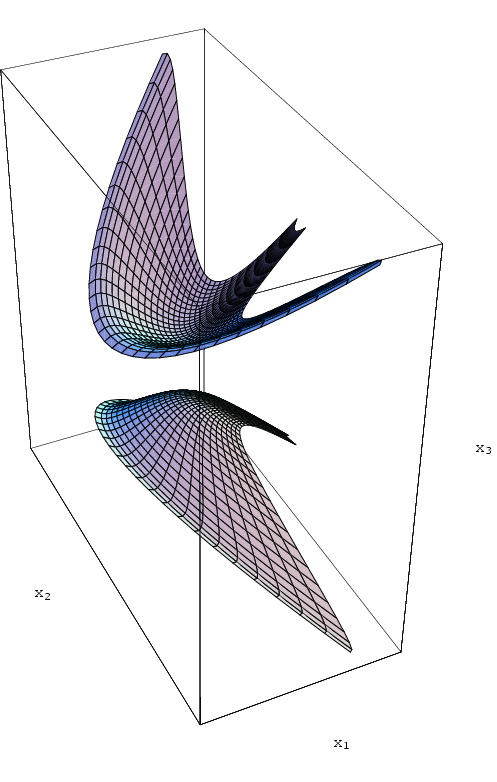}
\caption{\it   In this figure we display an example of the parametric parabolic surfaces realizing the $\Sigma_\beta$ K\"ahler manifolds in the $D$-map image of the $I_6$ integrable potentials. The chosen value of $\beta$ is $\beta \, = \, \ft 12$. }
\label{arcotannosuperflua}
 \iffigs
 \hskip 1cm \unitlength=1.1mm
 \end{center}
  \fi
\end{figure}

To discuss this model we consider the following momentum map and potential:
\begin{equation}\label{arcotannoMomMap}
V(\phi) \,\propto \, \left(\mathcal{ P}(\phi)\right)^2\quad ; \quad \mathcal{ P}(\phi) \, = \,\sqrt{\pi  \beta +\mbox{ArcTan}\left(e^{-2 \sqrt{3} \phi }\right)}
\end{equation}
\begin{figure}[!hbt]
\begin{center}
\iffigs
\includegraphics[height=70mm]{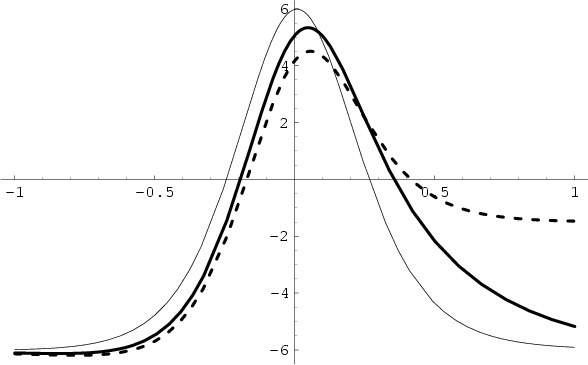}
\caption{\it   In this figure we present the plots  of the curvature for the $\Sigma_\beta$ surfaces in the image of the $D$-map of the $I_6$ integrable potentials. The thick solid line corresponds to $\beta \, = \,  \ft {1}{10}$ . The thick dashed line corresponds to $\beta \, = \, 0$. The thin solid line corresponds to $\beta \, = \, 2$. }
\label{I6tramurbano}
 \iffigs
 \hskip 1cm \unitlength=1.1mm
 \end{center}
  \fi
\end{figure}
which lead to the following expression for the curvature:
\begin{eqnarray}\label{sputacaso}
R_\beta(\phi) & = & \frac{N(\phi)}{D(\phi)} \nonumber\\
N(\phi) & = &
  3 \left(-4 \left(1-6 e^{4
   \sqrt{3} \phi }+e^{8 \sqrt{3}
   \phi }\right) \mbox{Arctan}\left(e^{-2 \sqrt{3} \phi
   }\right)^2\right.\nonumber\\
   &&\left.-4 e^{4 \sqrt{3} \phi }
   \left(4 \pi  \beta  \left(\cosh
   \left(4 \sqrt{3} \phi
   \right)-3\right)-3 \sinh \left(2
   \sqrt{3} \phi \right)\right) \mbox{Arctan}\left(e^{-2 \sqrt{3} \phi
   }\right)\right.\nonumber\\
   &&\left.+e^{4 \sqrt{3} \phi }
   \left(24 \pi ^2 \beta ^2+4 \pi
   \left(3 \sinh \left(2 \sqrt{3}
   \phi \right)-2 \pi  \beta  \cosh
   \left(4 \sqrt{3} \phi
   \right)\right) \beta
   -3\right)\right) \nonumber\\
   D(\phi) & = &2 \left(1+e^{4
   \sqrt{3} \phi }\right)^2
   \left(\pi  \beta +\mbox{Arctan}\left(e^{-2 \sqrt{3} \phi
   }\right)\right)^2
\end{eqnarray}
As it is evident from the plot of this function, displayed for a few values of $\beta$ in fig.\ref{I6tramurbano}, the curvature of the K\"ahler surfaces $\Sigma_\beta$ is always finite in the whole range of the $\phi$-coordinate. It tends to two different negative values $R_{\pm\infty}$ at $\phi\, = \, \pm \infty$ and becomes positive inside the bulk. This means that $\Sigma_\beta$ are not Hadamard manifolds according to the definition recalled in the mathematical appendix \ref{mathtopo}. Hence strictly speaking  the rigorous proof provided in  appendix  \ref{mathtopo} of the criteria that discriminate among elliptic, parabolic and hyperbolic isometries does not apply to this case. Yet the fact that the metric coefficient is regular and finite everywhere and that it  has just a zero at $\phi \, = \, \pm \infty$ leads to the conclusion that the isometry has only one fixed point at the boundary and it is therefore parabolic. The explicit form of the metric is:
\begin{equation}\label{dsqbeta}
  ds^2_{\beta} \, = \, d\phi^2 \, + \, f_\beta^2(\phi) \, dB^2
\end{equation}
where the coefficient $f_\beta(\phi)$ is given by:
\begin{equation}\label{stupidario}
  f_\beta(\phi) \, = \, -\frac{\sqrt{3} e^{2 \sqrt{3} \phi}}{\left(1+e^{4 \sqrt{3} \phi}\right) \sqrt{\pi  \beta
   +\mbox{Arctan}\left(e^{-2 \sqrt{3} \phi}\right)}}
\end{equation}
its plot being displayed for a few values of $\beta$ in fig.\ref{I6trolleybus}.
Adopting as geometrical model of $\Sigma_\beta$ the parabolic   surface in three-dimensional Minkowski encoded in eq.s (\ref{pianinidivanoparab}) we have to calculate the function $g(\phi)$. This is defined by the integral (\ref{granolatoparab}) which, in the pesent case, takes the following explicit appearance:
\begin{equation}\label{sumachka}
g_\beta(\phi) \, = \, \frac{1}{3} \int \frac{e^{-4
   \sqrt{3} \phi } \left(1+e^{4
   \sqrt{3} \phi }\right)^2
   \left(\pi  \beta +\mbox{Arctan}\left(e^{-2 \sqrt{3} \phi
   }\right)\right)^{3/2}}{4
   \left(\pi  \beta +\tan
   ^{-1}\left(e^{-2 \sqrt{3} \phi
   }\right)\right) \sinh \left(2
   \sqrt{3} \phi \right)-1} \, d\phi
\end{equation}
The integral (\ref{sumachka}) does not evaluate to any known special function, yet it can be evaluated numerically to any desired order of precision which, inserted into (\ref{pianinidivanoparab}) allows to draw the corresponding surface. This is done in fig.\ref{arcotannosuperflua}
\subsubsection{Asymptotic expansion of the function $J(C)$}
Next it is convenient to revert to the coordinate $U$ defined by
 \begin{equation}\label{phiUphiU}
   \phi   \, = \, -\frac{\mbox{Arctan}(U)}{2 \sqrt{3}}
 \end{equation}
In terms of $U$ the metric (\ref{dsqbeta}) becomes:
\begin{equation}\label{giorgione}
  ds^2_\beta \, = \, \frac{1}{12}
   \left(\frac{\mbox{dU}^2}{\left(1 \, - \, U^2\right)^2}+\frac{9 \mbox{dB}^2
   \left(1 \, - \, U^2\right)}{\pi  \beta
   +\mbox{Arctan}\left(e^{\mbox{ArcTanh}(U)}\right)}\right)
\end{equation}
while the VP coordinate is defined by the following integral:
\begin{equation}\label{giumenta}
C(U) \, = \,  -\frac{1}{6} \int \frac{e^{-\mbox{ArcTanh}(U)} \left(1+e^{2 \mbox{ArcTanh}(U)}\right) \sqrt{\pi  \beta
   +\mbox{Arctan}\left(e^{\mbox{ArcTanh}(U)}\right)}}{U^2-1} \, dU
\end{equation}
which does not evaluate to a  known special function. Notwithstanding this, the asymptotic behavior of $C(U)$ for $U$ very close to both boundaris $U\, = \, \pm 1$ can be easily evaluated expanding the integrand of (\ref{giumenta}) in power series. For instance, close to $U=1$ we can substitute $U=1-\xi$ and expand in powers of $\xi$. We obtain:
\begin{eqnarray}\label{fisciolo}
  C & =& \int \, d\xi \, \left[ \frac{\sqrt{2 \pi  \beta +\pi }}{12
   \xi ^{3/2}}-\frac{1}{24 \sqrt{\pi
    \beta +\frac{\pi }{2}} \xi
   }+\frac{3 (2 \pi  \beta +\pi
   )^2-1}{48 (2 \pi  \beta +\pi
   )^{3/2} \sqrt{\xi }}\right.\nonumber\\
   &&\left. -\frac{10 (2
   \pi  \beta +\pi )^2+3}{144
   \left(\sqrt{2} (2 \pi  \beta +\pi
   )^{5/2}\right)}+\frac{\left(45 (2
   \pi  \beta +\pi )^4-22 (2 \pi
   \beta +\pi )^2-15\right)
   \sqrt{\xi }}{1152 (2 \pi  \beta
   +\pi )^{7/2}}\right.\nonumber\\
   &&\left.-\frac{\left(88 (2
   \pi  \beta +\pi )^4+40 (2 \pi
   \beta +\pi )^2+35\right) \xi
   }{1920 \left(\sqrt{2} (2 \pi
   \beta +\pi
   )^{9/2}\right)}+O\left(\xi
   ^{3/2}\right)
  \right]
\end{eqnarray}
Performing the integration we get the expansion of $C$ in powers of $\xi \equiv (1-U)$:
\begin{eqnarray}\label{gromovaspasso}
  C & = & -\frac{\sqrt{2 \pi  \beta +\pi }}{6
   \sqrt{\xi }}-\frac{\log (\xi
   )}{24 \sqrt{\pi  \beta +\frac{\pi
   }{2}}}+\frac{\left(3 (2 \pi
   \beta +\pi )^2-1\right) \sqrt{\xi
   }}{24 (2 \pi  \beta +\pi
   )^{3/2}}\nonumber\\
   &&-\frac{\left(10 (2 \pi
   \beta +\pi )^2+3\right) \xi }{144
   \left(\sqrt{2} (2 \pi  \beta +\pi
   )^{5/2}\right)}+\frac{\left(45 (2
   \pi  \beta +\pi )^4-22 (2 \pi
   \beta +\pi )^2-15\right) \xi
   ^{3/2}}{1728 (2 \pi  \beta +\pi
   )^{7/2}}\nonumber\\
   &&-\frac{\left(88 (2 \pi
   \beta +\pi )^4+40 (2 \pi  \beta
   +\pi )^2+35\right) \xi ^2}{3840
   \left(\sqrt{2} (2 \pi  \beta +\pi
   )^{9/2}\right)}+O\left(\xi
   ^{5/2}\right)
\end{eqnarray}
which uniquely fixes the leading term:
\begin{equation}\label{forlizia}
  C \,\simeq \,  -\frac{\sqrt{2 \pi  \beta +\pi }}{6
   \sqrt{\xi }}
\end{equation}
Expanding in powers of $\xi$ the coefficient of $dB^2$ in the metric (\ref{giorgione}) we obtain instead:
\begin{eqnarray}
  \ft 12 \, \frac{d^2J}{dC^2}&=& -\frac{3 \left(U^2-1\right)}{4 \left(\pi  \beta +\mbox{Arctan}\left(e^{\tanh^{-1}(U)}\right)\right)}\nonumber \\
   &=& \frac{3 \xi }{2 \pi  \beta +\pi
   }+\frac{3 \sqrt{2} \xi ^{3/2}}{(2
   \pi  \beta +\pi )^2}-\frac{3
   \left((2 \pi  \beta +\pi
   )^2-4\right) \xi ^2}{2 (2 \pi
   \beta +\pi )^3}\nonumber\\
   &&+\frac{\left(24-5
   (2 \pi  \beta +\pi )^2\right) \xi
   ^{5/2}}{2 \sqrt{2} (2 \pi  \beta
   +\pi )^4}+O\left(\xi ^3\right)
\end{eqnarray}
which determines the leading term:
\begin{equation}\label{leaderBalla}
  \ft 12 \, \frac{d^2J}{dC^2} \, \simeq \, \frac{3 \xi }{2 \pi  \beta +\pi}
\end{equation}
Comparing eq. (\ref{leaderBalla}) with eq.(\ref{forlizia}) we work out:
\begin{equation}\label{signorefeudale}
   \ft 12 \, \frac{d^2J}{dC^2} \, \stackrel{C \to  \infty}{\simeq} \, \frac{1}{12}
   \left(\frac{1}{C}\right)^2+\mathcal{O}\left(\left(\frac{1}{C}\right)^3\right   ) \quad\Leftrightarrow \quad
   J(C) \, \stackrel{C \to \infty}{\simeq} \,
  \frac{1}{6} \log(C)+\mathcal{O}\left(\frac{1}{C}\right)
   \end{equation}
   The coefficient in front of the logarithm in the second half of equation (\ref{signorefeudale}) has a precise meaning. It is just the reciprocal of the asymptotic value of the curvature:
   \begin{equation}\label{isgmundo}
     \lim_{C \to \infty} R_\beta \, = \, - \, 6
   \end{equation}
What we have explicitly confirmed by the above token is that at both extrema of the $C$-range, respectively corresponding to the boundary and to the deep interior of the manifold, we have:
\begin{equation}\label{finocchione}
  J(C) \, \stackrel{C \to  \infty}{\simeq} \, - \, \frac{1}{R_\infty} \, \log(C) \quad ; \quad J(C) \, \stackrel{C \to  0}{\simeq} \, - \, \frac{1}{R_0} \, \log(C)
\end{equation}
This is the expected behavior in case od a parabolic symmetry.
\subsubsection{The analytic solution of Friedman equations}
In view of the special relevance of the $I_6$-model, in this subsection we present the exact analytic solution of the Friedman equations corresponding to this attractor integrable potential. Following the strategy introduced in \cite{noicosmoitegr},\cite{mariosashapietrocosmo} and briefly summarized in appendix \ref{integralnymodely}, the effective lagrangian encoding the dynamics of the scale factor and the scalar field is the following one:
\begin{equation}\label{effelagra}
  \mathcal{L} \, = \, e^{3 \,{A}(t)-\,{\mathcal{B}}(t)}
   \left(-\frac{3}{2}
   \, {A}'(t)^2+\frac{1}{2} \phi
   '(t)^2-e^{2 \, {\mathcal{B}}(t)}
   \left(\pi  \beta +\mbox{Arctan}\left(e^{-2 \sqrt{3} \phi(t)}\right)\right)\right)
\end{equation}
By definition $A(t) \, = \, \log a(t)$ is the logarithm of the scale factor, $\mathcal{B}(t)$ is a lagrangian multiplier tht imposes the vanishing of the hamiltonian (first Friedman equation) and that can be chosen at will, its choice corresponding to a choice of the dependence of the cosmic time on the parametric time (see eq.(\ref{piatttosa})).
\par
The explicit integration of the field equations of (\ref{effelagra}) can be performed by means of the following change of variables:
\begin{eqnarray}
  A(t)&=& \frac{1}{6} \log \left[X(t ) Y(t )\right] \nonumber \\
  B(t) &=& -\frac{1}{2} \log \left[X(t ) Y(t  )\right] \nonumber\\
  \phi(t) &=& \frac{\log \left[\frac{X(t
   )}{Y(t )}\right]}{2 \sqrt{3}} \label{romanzorosa}
\end{eqnarray}
where $X(t)$ and $Y(t)$ are two new fields. By means of (\ref{romanzorosa}) the lagrangian (\ref{effelagra}) is transformed into the following one:
\begin{equation}\label{giuridico}
\mathcal{L }\, = \,   -\pi  \beta -\mbox{Arctan}\left(\frac{Y(t )}{X(t )}\right)-\frac{1}{6}\, X'(t )\,Y'(t )
\end{equation}
whose field equations read as follows:
\begin{eqnarray}
  0 &=& -\frac{Y(t )}{X(t )^2+Y(t
   )^2}-\frac{Y''(t )}{6} \nonumber \\
  0&=& \frac{X(t )}{X(t )^2+Y(t
   )^2}-\frac{X''(t )}{6} \label{scimitarra}
\end{eqnarray}
Introducing the complex field $Z(t) \, = \, X(t) \, + \, {\rm i}\, Y(t)$ the two equations (\ref{scimitarra}) combine into the following holomorphic equation:
\begin{equation}\label{rominapovera}
  \frac{1}{Z(t )}-\frac{Z''(t)}{6} \, = \, 0
\end{equation}
The general integral of the above equation can be expressed in terms of the classical error function defined by:
\begin{equation}\label{errorafunzia}
  \mbox{erf} (z) \, \equiv \, \frac{2}{\sqrt{\pi}} \, \int_0^z \, \exp\left[- \, t^2\right] \, dt
\end{equation}
and of its imaginary argument continuation:
\begin{equation}\label{erfidefi}
 \mbox{Erfi} (z) \, \equiv \, \frac{1}{\rm i} \, \mbox{erf} \left({\rm i} \, z \right)
\end{equation}
Indeed the general solution of (\ref{rominapovera}) is the following
\begin{equation}\label{gnugno}
Z(t) \, = \,  \frac{2 \, \exp\left[\mbox{Inverse}[\mbox{Erfi}]\left(\Delta \, t\right)^2\right] \sqrt{\frac{3}{\pi}}}{\Delta }
\quad\Leftrightarrow \quad \Delta^2 \, t^2 \, = \, \mbox{Erfi} \left( \log \left[ \sqrt{\frac{\pi}{3}} \, \frac{\Delta}{2} \,Z(t)\right] \right)
\end{equation}
where $\Delta$ is the relevant integration constant. The other integration constant is simply a shift of the parametric time $t$ and can be disposed off without loss of generality.
\par
Taking the real and imaginary parts of the function $Z(t)$ defined by eq.(\ref{gnugno}) one obtains the functions $X(t)$ and $Y(t)$, which substituted back into eq.(\ref{romanzorosa}) yield the solution of Friedman equations.
\subsection{Integrable Models from the $I_2$ series}
Among the integrable potentials of the series $I_2$ of table \ref{tab:families} there is the best fit potential $\gamma \, = \, -\, \ft 76$, that has been indicated by Sagnotti and collaborators \cite{SagnottiDubna}, as a good modeling of the CMB low angular momentum spectrum within the general panorama of climbing scalars \cite{Sagnotti:2013ica},\cite{Dudas:2012vv},\cite{Dudasprimo}.
Choosing  $I_2$ with $C_{1}/C_{2} >0$, $C_1 >0$ and any $\gamma$ and setting:
\begin{equation}\label{follicolare}
 \phi \, = \, \sqrt{3} \, \mbox{Arctanh} \, (U)
\end{equation}
we obtain the following expression of the momentum map in terms of the $U$-variable.
  \begin{equation}\label{I2seria}
    \mathfrak{P}_\gamma(U) \, = \,  \frac{\sqrt{\frac{2}{3}}
   \sqrt{\left(\frac{U+1}{1-U}\right)^{3 \gamma
   }+\left(\frac{2}{U+1}-1\right)^{-\frac{3}{2}
   (\gamma +1)}}}{\gamma ^2}
  \end{equation}
  Note that the above function for $|\gamma| > 1$ and $\gamma >0$ has a zero at $U=-1$ and increases monotonically to $\infty$ as $U\to 1$. For $|\gamma| > 1$ and $\gamma < 0$, $\mathfrak{P}(U)$ has a zero at $U=1$ and increases monotonically to $\infty$ as $U\to -1$. In other words the behavior is the same exchanging $U \leftrightarrow - U$ which simply corresponds to change the irrelevant sign of the original field ${\hat \phi}$.  In the case $|\gamma| < 1$,  the function $\mathfrak{P}(U)$ has a zero  only for $\gamma < 0$ while for $\gamma >0$ it has no zeros. The best fit model of \cite{SagnottiDubna} has $\gamma \, = \, - \ft 76$ and fits in the first pattern, namely it has a zero at $U=1$ and increases monotonically to $\infty$ as $U\to -1$. In no case the models of this series are $\alpha$-attractors and can be used to describe the exit from inflation. Yet as shown by Sagnotti et al they might be quite appropriate to describe the very early stage of inflation and they might connect to the description in terms of other potentials at later times.
  \par
  In this section we consider the inclusion of these potentials in minimal supergravity models and we explore the properties of the K\"ahler surface $\Sigma$ in the image of their $D$-map.
    Many properties of this series have already been discussed in \cite{primosashapietro} and \cite{piesashatwo} where it has been pointed out the remarkable property that for all values of $\gamma$ the integral defining the VP coordinate yields a simple analytic expression in terms of the Appel function $F1$. Let us briefly recall this result. First it was observed that the only relevant choice concerning the coefficients $C_{1,2}$ is their relative sign. In order to have a positive definite potential they have to be both positive. In this case by means of a constant shift of $\phi$ they can be always equalized and reabsorbed into an overall constant in front of the potential. Hence following the normalizations of  \cite{piesashatwo} the momentum map can be chosen the following one:
\begin{equation}\label{firitone}
  \mathcal{P}_\gamma \, = \,  -\frac{\sqrt{e^{2 \sqrt{3}
   \gamma  \phi }+e^{\sqrt{3}
   (\gamma +1) \phi }} }{3 \gamma ^2}
\end{equation}
and performing the integral (\ref{sodoma}) one obtains the following result:
\begin{equation}\label{fattuccio}
  C_\gamma(\phi) \, = \, e^{-\sqrt{3} \gamma  \phi }\, F_1\left(\frac{\gamma }{\gamma-1};-\frac{1}{2},1;2+\frac{1}{\gamma -1};-e^{-\sqrt{3} (\gamma -1) \, \phi },-\frac{e^{-\sqrt{3} (\gamma -1) \phi } (\gamma +1)}{2 \, \gamma }\right)
\end{equation}
where $F_1(\dots)$ denotes an Appel function F1 which is defined by the following series development in two variables:
\begin{equation}\label{appellofunzio}
    F_1(a,b_1,b_2,c;x,y) \, = \, \sum_{m,n=0}^\infty \, \frac{(a)_{m+n} \, (b_1)_m \, (b_2)_{n}}{(c)_{m+n} \, m! \, n!} \, x^m \,y^n
\end{equation}
having denoted:
\begin{equation}\label{risingfactorial}
    (r)_n \, \equiv \, \frac{\Gamma(r+n)}{\Gamma(n)}
\end{equation}
The Appel series has convergence radius $1$ in both  $x$ and $y$. Yet the function can be analytically prolonged to larger values of both variables by means of the integral representation:
\begin{equation}\label{gorchiza}
 F_1(a,b_1,b_2,c;x,y) \, = \,  \frac{\Gamma (c)}{\Gamma (a) \Gamma
   (c-a)} \,  \int \, (1-t)^{-a+c-1} \,t^{a-1} \,(1-t x)^{-b_1} \,(1-t y)^{-b_2}\, \mathrm{dt}
\end{equation}
which  now will be very useful to determine the asymptotic behavior of the metric in $C$ and resolve the question about the correct interpretation of the $B$-isometry.
Introducing the variable:
\begin{equation}\label{Tvaria}
    T \, = \, \exp\left[ - \, \sqrt{3} \, \gamma \, \phi \right] \, = \, \left(\frac{\sqrt{U+1}}{\sqrt{1-U}}\right)^{-3 \gamma }
\end{equation}
which covers the interval $[0,\infty]$ while $U$ ranges from $-1$ to $1$, the VP coordinate can be rewritten as follows:
\begin{equation}\label{gupka}
    C_{\gamma}(T) \, = \, T \, F_1\left(\frac{\gamma }{\gamma
   -1};-\frac{1}{2},1;2+\frac{1}{\gamma
   -1};-T^{\frac{\gamma -1}{\gamma
   }},-\frac{T^{\frac{\gamma -1}{\gamma
   }} (\gamma +1)}{2 \gamma }\right)
\end{equation}
In particular for the case of the best fit model considered by Sagnotti et al \cite{SagnottiDubna}, namely for $\gamma \, = \, -\ft 76$ we obtain:
\begin{equation}\label{formicolio}
  C_{-\ft 76}(T) \,=\, T^{7/13}
   F_1\left(\frac{7}{13};-\frac{1}{2},1;\frac{20}{13};-T,
   -\frac{T}{14}\right) \label{C76funzia}
\end{equation}
Expressed in terms of the finite range variable $U$, the curvature of the $\Sigma_\gamma$ surfaces associated with the models of the $I_2$ series is the following one\footnote{We do not repeat the intermediate steps for the calculation of the curvature that should by now be standard. Furthermore note that, with respect to \cite{piesashatwo}, we have changed the normalization of the curvature, dividing it by a factor $4$, in order to agree with the notations of other papers of the inflationary literature.}:
\begin{eqnarray}\label{curvareI2}
    R_{\gamma}(U) & = & - \frac{\mathrm{N}_\gamma(U)}{\mathrm{D}_\gamma(U)}\nonumber\\
    \mathrm{N}_\gamma(U)& = & 3 \left(8 \gamma ^3\left(\frac{U+1}{1-U}\right)^{9\gamma }
    +2 (\gamma  (\gamma  (8\gamma-3)+6)+1)\left(\frac{U+1}{1-U}\right)^{6\gamma+3}
    +\left(\frac{2}{U+1}-1\right)^{-\frac{9}{2} (\gamma +1)} (\gamma+1)^3\right.\nonumber\\
    &&\left.+4
   \left(\frac{2}{U+1}-1\right)^{-\frac{3}{2} (5 \gamma +1)} \left(5 \gamma^3+1\right)\right)\nonumber\\
    \mathrm{D}_\gamma(U) & = & 8 \left(\left(\frac{U+1}{1-U}\right)^{3\gamma}+\left(\frac{2}{U+1}-1\right)^{-\frac{3}{2} (\gamma +1)}\right)^2 \left(2\gamma
   \left(\frac{U+1}{1-U}\right)^{3\gamma}+\left(\frac{2}{U+1}-1\right)^{-\frac{3}{2} (\gamma +1)} (\gamma+1)\right)\nonumber\\
\end{eqnarray}
\begin{figure}[!hbt]
\begin{center}
\iffigs
\includegraphics[height=90mm]{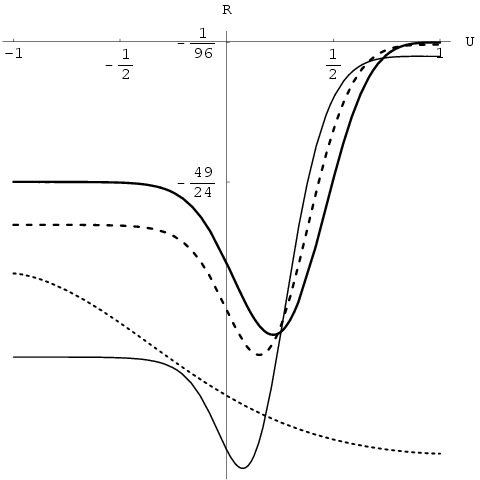}
\caption{\it   In this figure we present the plots  of the curvature for some of the $\Sigma_\gamma$ surfaces in the image of the $D$-map of the $I_2$ integrable potentials. The thick solid line corresponds to $\gamma \, = \, - \, \ft 76$ (the best fit model of Sagnotti et al \cite{SagnottiDubna}). The thick dashed line corresponds to $\gamma \, = \, - \, \ft 43$. The thin solid line corresponds to $\gamma \, = \, - \, \ft 74$. Finally the thin dashed line correspond to $\gamma \, = \, 2$.}
\label{I2curvafischia}
 \iffigs
 \hskip 1cm \unitlength=1.1mm
 \end{center}
  \fi
\end{figure}
As one sees from the plots displayed in fig.\ref{I2curvafischia} the curvature is mostly negative and either descends or climbs from one value at the boundary $U\, = \, -1$ to another value at the other boundary $U\, = \, 1$. When $\gamma >0$ the function is monotonic, while for $\gamma < 0$ the function has a minimum.
\par
\begin{figure}[!hbt]
\begin{center}
\iffigs
\includegraphics[height=70mm]{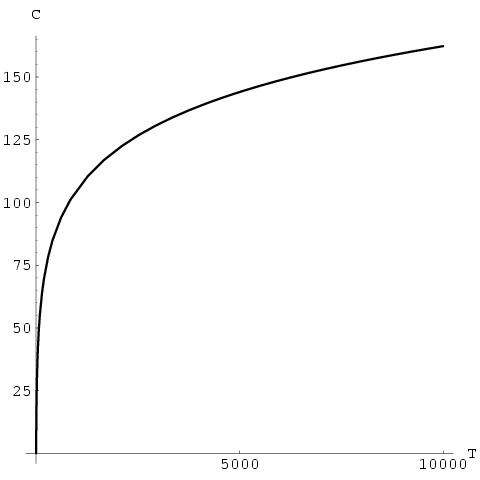}
\includegraphics[height=70mm]{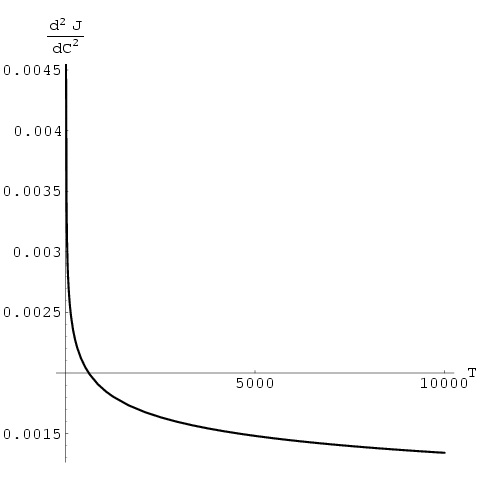}
\caption{\it   In this figure, in relation with the best fit integrable attractor model $\gamma \, = \, -\ft 76$,  we present the plots of the function $C_{-\ft 76}(T)$ and of the metric $\frac{\mathrm{d}^2}{\mathrm{d}C^2} \, J_{-\ft 76}(C)$ against the variable $T$. From these plots we get a visual demonstration that the metric goes to zero for very large value of the VP coordinate $C$.}
\label{CTplot}
 \iffigs
 \hskip 1cm \unitlength=1.1mm
 \end{center}
  \fi
\end{figure}

In \cite{piesashatwo}, focusing on the case of the best fit model ($\gamma \, = \, - \, \ft 76$) two of us studied the asymptotic expansion of the K\"ahler function $J(C)$ for small values of the field $C$. We obtained:
\begin{equation}\label{molinodorino}
    J_{-\ft 76}(C) \, \stackrel{|C| \ll 1}{=} \, - \, \frac{12}{49} \, \log[C] \, + \, 1 + \, \sum_{k=1}^\infty c_k \, C^k
\end{equation}
where the first $7$ values of the $c_k$ are displayed in eq.(8.40) of \cite{piesashatwo}. From the point of view of geometry the expansion (\ref{molinodorino}) refers to the boundary of the manifold $\Sigma_{\gamma}$ and displays a logarithmic singularity which is universal apart from the numerical coefficient in front of the $log$ related with the  boundary value of the curvature. Hence from the expansion (\ref{molinodorino}) we cannot distinguish whether the appropriate interpretation of the $B$-symmetry is that of a compact $\mathrm{U(1)}$ or of a parabolic translation. To resolve this dichotomy we need the expansion of the function $J_{-\ft 76}(C)$ or better of the metric $ \frac{\mathrm{d}^2}{\mathrm{d}C^2} \, J_{-\ft 76}(C)$ for very large values of $C$. The plot of $C_{-\ft 76}(T)$ is displayed in fig.\ref{CTplot} which displays also the plot of the metric $ \frac{\mathrm{d}^2}{\mathrm{d}C^2} \, J_{-\ft 76}(C)$ against the variable $T$.
Indeed by means of a straightforward substitution in $\frac{\mathrm{d}^2}{\mathrm{d}C^2} \, J_{-\ft 76}(C) \, = \,\left(\, \mathcal{P}_{-\ft 76}(\phi)\right)^2$ one obtains:
\begin{eqnarray}\label{superjet}
    \frac{\mathrm{d}^2}{\mathrm{d}C^2} \, J_{-\ft 76}(C) & = &\frac{12
   \left(T^{13/7}+14\right)^2}{2401
   \left(T^{27/7}+T^2\right)} \nonumber\\
   &\stackrel{T\to \infty}{\approx}&\frac{12
   \sqrt[7]{\frac{1}{T}}}{2401}+\frac{32
   4\left(\frac{1}{T}\right)^2}{2401}+\frac{2028
   \left(\frac{1}{T}\right)^{27/7}}{2401
   }+\mathcal{O}\left(\left(\frac{1}{T}\right)^{29
   /7}\right)
\end{eqnarray}
If we were able to estimate the asymptotic behavior for large $T$ of the $C_{-\ft 76}(T)$-function, than by comparison of the leading order of $C$ with the leading order of (\ref{superjet}) we might establish the asymptotic behavior of the metric with respect to $C$. The desired expansion of $C_{-\ft 76}(T)$ can be obtained from the integral representation (\ref{gorchiza}). Inserting the value $\gamma \, = \, - \, \ft 76$ and that of the arguments in (\ref{gorchiza}) we get:
\begin{eqnarray}
  C_{-\ft 76}(T) &=& \int_0^1 \, dt \, K(t,T) \\
  K(t,T) &=& \frac{98 \, T \, \sqrt{t\, T^{13/7}\, +\, 1}}{t^{6/13}
   \left(13 \, t \, T^{13/7}\, +\, 182\right)}
\end{eqnarray}
For large values of $T$ the kernel $K(t,T)$ of the integral representation behaves as follows:
\begin{equation}\label{fursenkoBis}
   K(t,T) \,\stackrel{T\to \infty}{\approx} \, \frac{98 \sqrt[14]{T}}{13 t^{25/26}}\, + \,\mathcal{O}(T^{-\ft{25}{14}})
\end{equation}
and performing the integral we obtain:
\begin{equation}\label{fursenko}
   C_{-\ft 76}(T) \,\stackrel{T\to \infty}{\approx} \, 196 \sqrt[14]{T}\, + \,\mathcal{O}(T^{-\ft{25}{14}})
\end{equation}
Combining eq.(\ref{fursenko}) with eq.(\ref{superjet}) we obtain:
\begin{equation}\label{gromolado}
  \ft 12 \,  \frac{\mathrm{d}^2}{\mathrm{d}C^2} \, J_{-\ft 76}(C) \,\stackrel{C\to \infty}{\approx} \,\frac{192}{C^2} \, + \, \mathcal{O} \left(C^{-3}\right)
\end{equation}
This result is extremely beautiful and can be compared with the result (\ref{molinodorino}) for the small $C$ behavior of the $J(C)$ function. Combining the two informations we can conclude that:
\begin{eqnarray}
  J(C) &\stackrel{C\to 0}{\approx} & - \, \frac{1}{R_{-\ft 76}[0]} \, \log (C) \label{piccolo}  \\
  J(C) &\stackrel{C\to \infty}{\approx}& -  \, \frac{1}{ R_{-\ft 76}[\infty]} \, \log (C) \label{grande}
\end{eqnarray}
Indeed as it is evident from fig.\ref{I2curvafischia} the limiting values of the curvature at $C=0 \, \Leftrightarrow \, U=-1$ and at $C=\infty \, \Leftrightarrow \, U=1$ are:
\begin{equation}\label{cosnomallo}
    R_{-\ft 76}[0] \, = \, \lim_{U\to-1} \, R_{-\ft 76}(U) \, = \, - \, \frac{49}{24} \quad ; \quad R_{-\ft 76}[\infty] \, = \, \lim_{U\to 1} \, R_{-\ft 76}(U) \, = \, - \, \frac{1}{96}
\end{equation}
\begin{figure}[!hbt]
\begin{center}
\iffigs
\includegraphics[height=90mm]{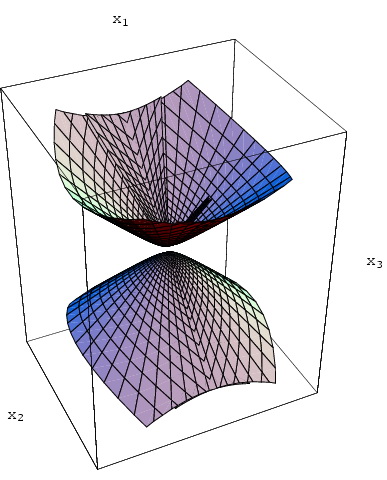}
\includegraphics[height=85mm]{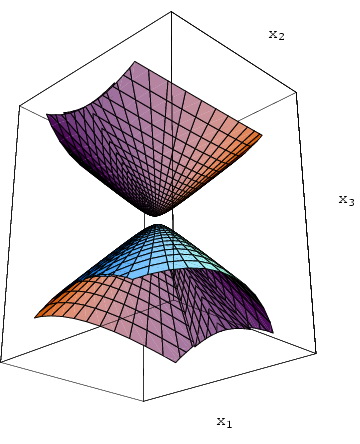}
\caption{\it   In this figure we present the $3D$-plot of the surface $\Sigma$ associated with the $I2$ integrable attractor with $\gamma= -\ft 76$. This is the best fit model of Sagnotti et al \cite{SagnottiDubna}. The left and right pictures present the same surface from two different viewpoints.}
\label{I2surface}
 \iffigs
 \hskip 1cm \unitlength=1.1mm
 \end{center}
  \fi
\end{figure}
This result shows that at its two boundaries the surface $\Sigma_{-\ft 76}$ approaches a constant negative curvature manifold in the plane presentation and that therefore the correct interpretation of the $B$-isometry is that of a parabolic translation group. Indeed the behavior $\frac{\mathrm{d}^2}{\mathrm{d}C^2}J(C) \sim C^{-2}$ is contradictory with the interpretation of $C$ as the logarithm of the modulus of the complex coordinate.
\par
Relying on these conclusions we can now realize a three-dimensional model of the surface $\Sigma_{-\,\ft 76}$ utilizing the parabolic parameterization (\ref{pianinidivanoparab}). The picture of $\Sigma_{-\,\ft 76}$ is displayed in fig.\ref{I2surface}.
In this case the two functions are:
\begin{eqnarray}
  f_{-\ft 76}(\phi) &=& -\frac{2 \sqrt{3} e^{-\frac{7 \phi
   }{\sqrt{3}}} \left(14+e^{\frac{13
   \phi }{2 \sqrt{3}}}\right)}{49
   \sqrt{e^{-\frac{7 \phi
   }{\sqrt{3}}}+e^{-\frac{\phi }{2
   \sqrt{3}}}}} \label{df}\\
   g_{-\ft 76}(\phi) &=& \int_0^\phi \, dx \, \frac{98 e^{\frac{7 x}{2 \sqrt{3}}}
   \left(1+e^{\frac{13 x}{2
   \sqrt{3}}}\right)^{3/2}}{196+366
   e^{\frac{13 x}{2
   \sqrt{3}}}+e^{\frac{13 x}{\sqrt{3}}}} \label{gf}
\end{eqnarray}
The integral (\ref{gf}) does not evaluate to any known special function but can be computed numerically to any desired precision and this is sufficient to obtain the three-dimensional plots displayed in fig.\ref{I2surface}. Note also that as in the previous case the complete surface is the union of four branches related to the choices of signs of $X_1$ and $X_3$.
\par
As a final remark let us emphasize that the conclusions reached for the best-fit model can be naturally generalized, with a little bit of work to all the values of  $\gamma$ (at least the negative ones) so that the parabolic interpretation of the $B$-symmetry seems a generic feature of the entire $I_2$-series of integrable potential. Note also that the surface in fig. \ref{I2surface} is smooth and we do not detect the presence of any singular point.
\subsection{A non maximally symmetric symmetric K\"ahler manifold with parabolic isometry and zero curvature at one boundary}
\label{piattocsquare}
As a final example we consider a parabolic model where the curvature at one of the two boundaries goes to zero so that the asymptotic behavior of the $J(C)$-function on that boundary becomes exceptional.
\par
Let the momentum map be the following one:
\begin{equation}\label{curchato}
  \mathcal{P}(\phi) \, = \, \exp\left[ \nu \, \phi \right] \, + \, \mu \, \phi
\end{equation}
The corresponding $f(\phi)$-function is:
\begin{equation}\label{ffunz}
  f(\phi) \, = \, \mathcal{P}^\prime(\phi) \, = \, \nu \, \exp\left[ \nu \, \phi \right] \, + \, \mu
\end{equation}
which has no zeros for finite $\phi$ if $\mu$ and $\nu$ have the same sign. If the two parameters have opposite signs there is such a zero and this creates a fixed point of the isometry $B\to B+c$ at finite $\phi$ which implies that the isometry is elliptic. Yet in case of opposite signs the curvature has a singularity so that any smooth K\"ahler manifold with a momentum map of type (\ref{curchato}) has a parabolic isometry group. Indeed
using equation (\ref{giunone}) we can immediately calculate the curvature and we find:
\begin{equation}\label{Rcontro}
  R(\phi) \, = \, -\frac{e^{\nu  \phi } \nu
   ^3}{2 \left(\mu +e^{\nu
   \phi } \nu \right)}
\end{equation}
This shows what we just said. The manifold is smooth and singularity-free if and only if $\mu$ and $\nu$ have the same sign so that at no value of $\phi$ the denominator can develop a zero. Without loss of generality we can assume that $\nu >0$ since the sign of $\phi$ can be flipped without changing its kinetic term. With this understanding it follows that also $\mu >0$ for regularity.
\par
Consider next the integral defining the VP coordinate $C$. We immediately obtain:
\begin{equation}\label{golomma}
  C(\phi) \, = \, \int \, \frac{1}{ \mathcal{P}^\prime(\phi)} \, d\phi \, = \, \frac{\phi }{\mu }-\frac{\log
   \left(\mu +e^{\nu  \phi }
   \nu \right)}{\mu  \nu }
\end{equation}
The range of $C$ is now easily determined considering the limits of the above function for $\phi \, = \, \pm \infty$. When $\mu >0\, ,\,\nu>0$ we have:
\begin{equation}\label{mnogobullo}
  C(-\infty) \, = \, - \, \infty \quad ; \quad C(\infty) \, = \,  - \, \frac{\log[\nu]}{\mu \, \nu}
\end{equation}
Hence  $C \, \in \, \left[-\infty\, , \, - \, \frac{\log[\nu]}{\mu \, \nu}\right]$. The VP coordinate  is always negative and it spans a semininfinite interval. Keeping this range in mind we can invert the relation (\ref{golomma})  between $\phi$ and $C$ obtaining:
\begin{equation}\label{cirimellusbonus}
 \phi \, = \, -\frac{\log
   \left(\frac{e^{-C \mu  \nu }}{\mu }-\frac{\nu }{\mu}\right)}{\nu }
\end{equation}
The $J$-function is easily calculated from eq.(\ref{gartoccio}) and we find:
\begin{eqnarray}\label{polilogga}
  \mathcal{J}(\phi) & = &\frac{\nu ^2 \phi ^2+(2-2 \nu \phi ) \log \left(\frac{e^{\nu  \phi } \nu }{\mu }+1\right)
  -2\mbox{Li}_2\left(-\frac{e^{ \nu  \phi } \nu }{\mu}\right)}{\nu ^2}
\end{eqnarray}
where $\mbox{Li}_n(z)$ is the polylogarithmic function. Introducing in (\ref{polilogga}) the relation between $\phi$ and $C$, we get an explicit analytic expression for the $J(C)$ function, namely:
\begin{eqnarray}\label{casoraro}
  J(C) & = & \frac{\log ^2\left(\frac{e^{-C \mu  \nu }-\nu }{\mu  }\right)
  +2 \left(\log \left(\frac{e^{-C \mu  \nu   }-\nu }{\mu}\right)+1\right) \log \left(\frac{1}{1-e^{C \mu \nu } \nu }\right)-2
   \mbox{Li}_2\left(1+\frac{1}  {e^{C \mu  \nu } \nu  -1}\right)}{ \nu ^2}
\end{eqnarray}
As for the metric, having the explicit expression (\ref{casoraro}), we easily calculate its second derivative and we find:
\begin{equation}\label{giustino}
  ds^2 \, = \, \ft 12 \, \frac{\mathrm{d}^2J}{\mathrm{d}C^2} \, \left(\mathrm{d}C^2 \, + \, \mathrm{d}B^2\right) \, = \, \frac{\mu ^2}{\left(e^{C \mu
    \nu } \nu -1\right)^2} \, \left(\mathrm{d}C^2 \, + \, \mathrm{d}B^2\right)
\end{equation}
For $C\, \to \, - \, \infty$ the metric coefficient $\ft 12 \, \frac{\mathrm{d}^2J}{\mathrm{d}C^2}$ tends to a costant:
\begin{equation}\label{grimaldi}
 \ft 12 \, \frac{\mathrm{d}^2J}{\mathrm{d}C^2} \, \stackrel{C \to  - \, \infty}{\approx} \, {\mu^2}  \quad \Rightarrow \quad J(C) \,  \stackrel{C \to  - \, \infty}{\approx} \, \frac{\mu^2}{2} \, C^2
\end{equation}
This asymptotic behavior differs from the usual logarithmic behavior of $J(C)$ at the boundary because at $C\, = \, -\infty$ and hence at $\phi \, = \, -\, \infty$ the curvature goes to zero.
\par
In the other extremum of the $C$-range, namely for $C\to\,  - \, \frac{\log[\nu]}{\mu \, \nu}$ the metric coefficient diverges and we have the standard logarithmic singularity. To see this, set $C \, = \, -  \, \frac{\log[\nu]}{\mu \, \nu} \, - \, \xi$ and substitute it  into the expression of the metric coefficient. We obtain:
\begin{eqnarray}\label{saracinescu}
   \ft 12 \, \frac{\mathrm{d}^2J}{\mathrm{d}C^2} & = & \frac{\mu ^2}{ \left(e^{\mu
   \nu  \left(-\xi -\frac{\log
   (\nu )}{\mu  \nu }\right)}
   \nu -1\right)^2} \nonumber\\
   &\stackrel{\xi\to 0}{\approx} &\frac{1}{\nu ^2 \xi
   ^2}+\frac{\mu }{\nu  \xi
   }+\frac{5 \mu
   ^2}{12}+\frac{1}{12} \mu ^3
   \nu  \xi +\mathcal{O}\left(\xi
   ^2\right)
\end{eqnarray}
and we conclude that, naming $C_0 \, = \,  - \, \frac{\log[\nu]}{\mu \, \nu}$, we have:
\begin{equation}\label{gilgamesh}
  J(C) \, \stackrel{C\to C_0}{\approx} \, \frac{2}{\nu^2} \log\left[C_0 \, - \, C\right]
\end{equation}
This is the standard logarithmic singularity and the coefficient in front of the logarithm is indeed the inverse of the limiting curvature: $R_{C_0} \, = \, \ft 12 \, \nu^2$.
\par
This result confirms once again the relation between the asymptotic behavior of the $J(C)$ function and the character  of the isometry group. For a parabolic isometry the asymtotic behavior is just that anticipated in eq.s(\ref{polentaconcia},\ref{oseletti}). For a vanishing limiting curvature the correct asymptotic is (\ref{oseletti}).
\par
The present example is  very paedagical in order to avoid possible misconceptions. If we looked at the expression (\ref{giustino}) and we forgot the precisely defined range of the variable $C$ which is determined by the integration of the complex structure equation, we might be tempted to consider the same metric also for positive values of $C$. We would conclude that when $C\to \infty$ the metric coefficient goes to zero as $\exp[-\nu \,C]$. Then we would dispute that the last mentioned behavior  indicates an elliptic interpretation of the isometry and advocate that there is a clash with our a priori knowledge that the isometry  is instead parabolic. In fact there is no clash since the positive range of $C$ is excluded and it is not to be considered. At the extrema of the $C$-interval,  the function $J(C)$ displays the expected asymptotic behavior foreseen for the parabolic case.
\section{An example of a non maximally symmetric K\"ahler manifold with an isometry group of the hyperbolic type}
\label{iperbarico}
\begin{figure}[!hbt]
\begin{center}
\iffigs
\includegraphics[height=50mm]{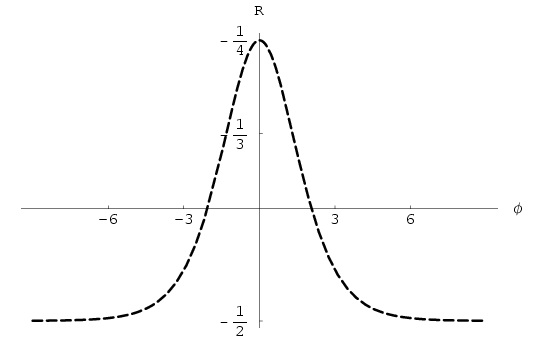}
\includegraphics[height=50mm]{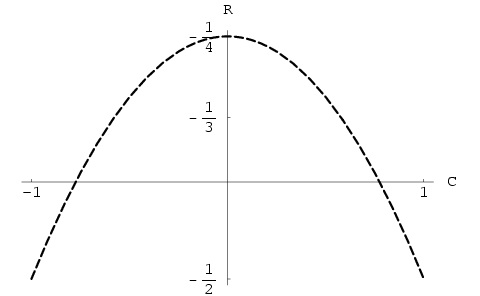}
\caption{\it   In this figure we present the plot of the curvature of the surface $\Sigma$ defined by eq.(\ref{yaku}) that has a hyperbolic isometry. The picture on the left displays the dependence of the curvature on the canonical coordinate $\phi$, while the picture on the right displays its dependence on the VP coordinate $C$.}
\label{cicciocurvo}
 \iffigs
 \hskip 1cm \unitlength=1.1mm
 \end{center}
  \fi
\end{figure}

In order to exhibit an example of a surface with non constant curvature that has a hyperbolic isometry
we consider the following momentum map and potential:
\begin{equation}\label{ugrovico}
    V(\phi) \, = \, \left[\mathcal{P}(\phi)\right]^2 \quad ; \quad \mathcal{P}(\phi)\, = \, \phi +\sinh (\phi )
\end{equation}
which yields:
\begin{equation}\label{yaku}
    \mathcal{P}^\prime\left(\phi\right)\, = \, 1\, +\, \cosh(\phi) \quad ; \quad \mathrm{d}s^2_\Sigma \, = \, \mathrm{d}\phi^2 \, + \, \left(1\, +\, \cosh(\phi)\right)^2 \, \mathrm{d}B^2
\end{equation}
According to the mathematical classification discussed in appendix \ref{mathtopo} the metric (\ref{yaku}) has a hyperbolic type of isometry due to the two fixed points on the boundary of the manifold corresponding to the two singularities $\phi \, = \, \pm \infty$.
The curvature of this manifold is finite but not constant. Indeed, applying eq.(\ref{giunone}) we obtain:
\begin{equation}\label{ghirlandadifiori}
  R(\phi) \, = \, -\frac{\cosh (\phi )}{2 (\cosh
   (\phi )+1)}
\end{equation}
whose plot is presented in fig.\ref{cicciocurvo}.
In this case it is very simple to integrate the complex structure equation which defines the VP coordinate. We obtain:
\begin{equation}\label{solimato}
 C(\phi) \, = \, \tanh \left(\frac{\phi   }{2}\right) \quad ; \quad \phi \, = \,  2 \, \mbox{ArcTanh}(C)
 \end{equation}
 and we observe that in line with our general criteria for hyperbolic symmetry, the range of the VP coordinate is in this case finite:
 \begin{equation}\label{corallonero}
   C \, \in \, \left[-1\, , \, 1 \right ]
 \end{equation}
 From the integration of eq.(\ref{gartoccio}) that defines the $J$-function and the K\"ahler potential we obtain:
 \begin{equation}\label{granulatodiferro}
 J(\phi) \, = \,  2 \,  \phi  \tanh \left(\frac{\phi}{2}\right) \, = \, J(C) \, = \, 4 C \,\mbox{ArcTanh}(C)
 \end{equation}
 Calculating the metric coefficient from (\ref{granulatodiferro})  we get:
 \begin{equation}\label{sinuheBis}
 \ft 12 \, \frac{d^2J}{dC^2} \, = \, \frac{4}{\left(C^2-1\right)^2} \, \quad ; \quad ds^2 \, = \, \frac{4}{\left(C^2-1\right)^2}  \, \left( dC^2 + dB^2 \right)
 \end{equation}
 displaying a polar singularity at both extrema of the $C$-range, namely at $C=\pm 1$.
 \par
\begin{figure}[!hbt]
\begin{center}
\iffigs
\includegraphics[height=50mm]{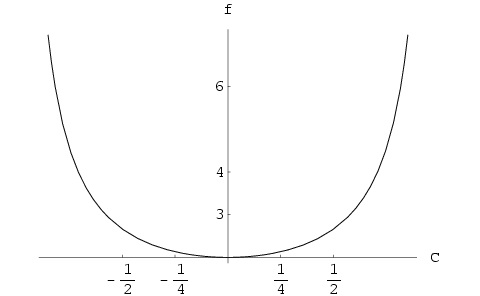}
\includegraphics[height=50mm]{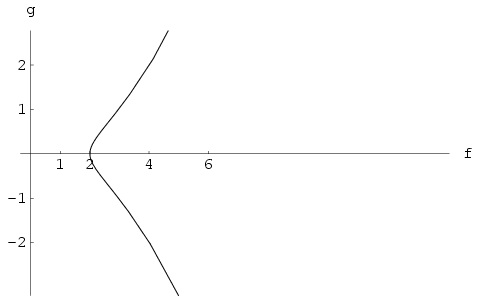}
\caption{\it  In this picture we present the plots of the functions $f(C)$, $g(C)$ that define the realization of the K\"ahler manifold $\Sigma$ associated with the potential (\ref{ugrovico}) as a parametric surface in flat Minkowski three-dimensional space. The geometrical model is that appropriate to the hyperbolic character of the isometry $B\,\to\, B+c$. The first two pictures display the plot of $g$ and $f$ as functions of the VP coordinate $C$. The last plot is the parametric plot of the curve in the plane $f,g$. Geometrically this is the curve cut out by the surface $\Sigma$ in any plane orthogonal to the axis $X_2$}
\label{veryuseful}
 \iffigs
 \hskip 1cm \unitlength=1.1mm
 \end{center}
  \fi
\end{figure}
\begin{figure}[!hbt]
\begin{center}
\iffigs
\includegraphics[height=110mm]{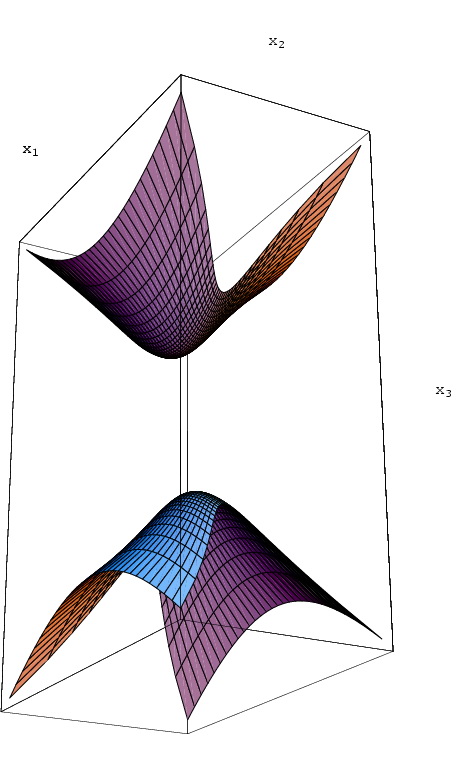}
\caption{\it   In this figure we present the $3D$-plot of the surface $\Sigma$ associated with the potential \ref{ugrovico}. The correct interpretation of the isometry in this case is that of a hyperbolic group. Indeed the hyperbolic embedding (\ref{pianinidivano3}) in three-dimensional Minkowski space works beautifully and we have the smooth surface displayed here.}
\label{HypSigmasurface}
 \iffigs
 \hskip 1cm \unitlength=1.1mm
 \end{center}
  \fi
\end{figure}

 In order to present a geometrical model of this K\"ahler manifold, we resort to the hyperbolic parametric surface encoded in formulae (\ref{pianinidivano3}) and we calculate the relevant functions $f(\phi)$ and $g(\phi)$. In this case it is more convenient to express them in terms of the finite range VP coordinate $C$.
 We have:
 \begin{equation}\label{frullatodipecorino}
   f(\phi) \, = \, \cosh (\phi )+1 \, = \,\frac{2}{1\, - \, C^2}
 \end{equation}
 and inserting the result into eq.(\ref{granolatoHyp}) we get \footnote{Note that in the integral  of eq.(\ref{granolatoHyp}) we consistently trnsformed the integrtion variable from $\phi$ to $C$.}:
 \begin{equation}\label{legiziano}
 g(C) \, = \,  \frac{1}{8} \left(\frac{2 C\, \left(C^2-3\right)}{\left(C^2-1\right)^2}\,+\,
 \log (C-1)-\log (C+1)\right)
 \end{equation}
 The plots of the these functions is presented in fig.\ref{veryuseful}.
In fig. \ref{HypSigmasurface} we display the three dimensional shape of the parametric surface $\Sigma$ realizing the desired K\"ahler manifold.
\section{Conclusions}
\label{concludo}
As we announced in the introduction the main goal of the present paper was to establish the proper geometrical interpretation of
the gauged isometry group $\mathcal{G}_\Sigma$ for the one-complex dimensional K\"ahler manifolds $\Sigma$ that sustain the inclusion into Minimal Supergravity Models of positive definite inflaton potentials.
\par
We were able to show that the classification of isometries into elliptic, hyperbolic and parabolic which is proper to the case of the coset manifold $\mathrm{SL(2,\mathbb{R})/O(2)}$ that, in particular, gives rise to the Starobinsky model, can be extended to non-constant curvature manifolds certainly if they are Hadamard manifolds (see Appendix \ref{mathtopo}) and probably also if they are $\mathrm{CAT}(k)$ manifolds with $k>0$. The mathematical classification of isometries of Gromov et al \cite{Gromov1985} is mirrored, using more physical terms, into the available asymptotic expansions of the $J(C)$ function on the boundary, or in the origin of the manifold $\Sigma$ which, by definition, is a fixed point of $\mathcal{G}_\Sigma$ when this latter is a compact $\mathrm{U(1)}$. We illustrated these concepts with several examples for each type of isometry both integrable and not integrable.
\par
A few side results seem quite relevant:
\begin{enumerate}
  \item Inflaton potentials able to describe inflation in a way qualitatively consistent with experimental data are mostly based on the gauging of a parabolic isometry, Starobinsky model being the prototype of all such cases. Yet it should be stressed that there is at least one example of $\alpha$-attractor (the quadratic one discussed in section \ref{bububu}) that emerges from the gauging of an elliptic isometry.
  \item Among the parabolic models denominated $\alpha$-attractors by Kallosh et al, which seem to provide a viable description of PLANCK data, there is just one integrable model corresponding to the case $I_6$ of the bestiary compiled in \cite{noicosmoitegr}. This model provides the unique possibility of writing an analytically explicit Sasaki-Mukhanov equation for a background compatible with experimental data.
   \item Models based on the gauging of a $\mathrm{U(1)}$-symmetry have, by definition, a fixed point which typically corresponds, in cosmic evolution, to a stable de Sitter attractor. Possibly such models can be useful to describe the late re-accelerated phase  of the Universe.
   \item On the contrary other parabolic integrable models that are not attractors might be relevant for the description of the early pre-inflationary phase just after the Big Bang \cite{Dudasprimo,Dudas:2012vv,Sagnotti:2013ica,SagnottiDubna}.
    \item The requirement of smoothness and curvature-finiteness of the K\"ahler surface $\Sigma$ poses constraints on the viable positive definite inflaton potentials. In view of what we mentioned above about the classification of isometries it is quite attractive to conjecture that an appropriate geometrical constraint on the available potentials is provided by the assumption that the D-map image of the potential should be a $\mathrm{CAT}(k)$ manifold.
\end{enumerate}
The last point in the above list deserves a further comment. Since the notion of $\mathrm{CAT}(k)$ manifolds applies to manifold of any dimension, one might argue that the $\mathrm{CAT}(k)$ condition might apply to  multi-field  K\"ahler manifolds in supergravity theories with several multiplets. The conjectured possibility of  classifying the isometries of such manifolds as elliptic, hyperbolic or parabolic, might be the clue to select the inflaton in a multifield space.
\section*{Acknowledgments}
One of us (S.F.) would like to aknowledge enlightening discussions with R. Kallosh, A. Linde and M. Porrati on related work.
\par
S.F. is supported by ERC Advanced Investigator Grant n. 226455
\emph{Supersymmetry, Quantum Gravity and Gauge Fields (Superfelds)}.
The work of A.S. was supported in part by the RFBR Grants No. 11-02-01335-a, No. 13-02-91330-NNIO-a and No. 13-02-90602-Arm-a.
\newpage
\appendix
\section{Integrable Models}
\label{integralnymodely}
Since we frequently refer to them in the main text, for  reader's convenience in this appendix we recall the classification of integrable cosmological potentials that was derived in \cite{noicosmoitegr} and further reviewed and developed in \cite{mariosashapietrocosmo},\cite{piesashatwo}. First of all let us explain what is meant by integrable potentials. Considering four-dimensional gravity  coupled to a canonically normalized scalar field $\phi$ that self interacts via  a potential $V(\phi)$, the coupled solutions of the Einstein-Klein-Gordon system where the metric is of the spatially flat form:
\begin{equation}\label{piatttosa}
        ds^2 \ = \ - \ e^{\,2\,{\cal B}(t)} \, \mathrm{d}t^2 \ + \ a^2(t) \ \mathrm{d}\mathbf{x}\cdot \mathrm{d}\mathbf{x}
        \ ,
      \end{equation}
${\cal B}(t)$ being an auxiliary function of the parametric time $t$ are all encoded in the solutions of the following equivalent dynamical system
 \begin{equation}\label{unopratino}
    {\cal L}_{eff} \, = \, \mbox{const} \, \times \, e^{\,{3 \, A} \ - \ {\cal B} } \ \left[ \ - \, \frac{3}{2} \dot{ A}^{\,2} \ + \ \frac{1}{2} \, \dot{\phi}^2 \ - \  e^{\,2\,{\cal B}} \  {{V}}(\phi) \right]
\end{equation}
that contains three degrees of freedom ${\cal B}(t)$, $At)\, \equiv \, \log \, a(t)$ and $\phi(t)$. The function $B(t)$ is just a Lagrangian multiplier that imposes a hamiltonian constraint corresponding to the first of Friedman equations. The true degrees of freedom are two, $A(t)$ and $\phi(t)$. By definition the potential $V(\phi)$ is named \textbf{integrable} if, for a suitable choice of the gauge function $B(t)$, the dynamical system (\ref{unopratino}) is Liouville integrable admitting, besides the standard quadratic hamiltonian a second functionally independent conserved hamiltonian that Poisson commutes with the standard one. Thanks to the additional hamiltonian the field equations of (\ref{unopratino}) can always be reduced to quadratures and in most cases one can explicitly derive the general integral in terms of classical special functions.
\begin{table}[h!]
\centering
{\small
\begin{tabular}{|l|c|}
\hline
\null & Infinite families of Integrable Potentials \\
\hline
\hline
\null&\null\\
$I_1$ & $\! C_{11} \, e^{\,\sqrt{\ft 32} \, {\hat \phi}} \, + \, 2\, C_{12} \, + \, C_{22} \, e^{\, - \sqrt{\ft 32} \, {\hat \phi}}$  \\
\null&\null\\
\hline
\null&\null\\
$I_2$ & $\! C_1 \, e^{,\gamma \,\sqrt{6} \, {\hat \phi}}\, +\, C_2e^{\,(\gamma+1)\, \sqrt{\ft 32} \, {\hat \phi}} \ \ \ $  \\
\null&\null\\
\hline
\null&\null\\
$I_3$ & $\! C_1 \, e^{ \sqrt{6} \, {\hat \phi}} \ + \ C_2$    \\
\null&\null\\
\hline
\null&\null\\
$I_4$ & $\! C_1 \,  {\hat \phi} \,  e^{ \sqrt{6} \, {\hat \phi}} $    \\
\null&\null\\
\hline
\null&\null\\
$I_5$ & $\! C_1 \, \log \left( \coth [\sqrt{\ft 32} \, {\hat \phi}]\right) \, + \,C_2 $    \\
\null&\null\\
\hline
\null&\null\\
$I_6$ & $\! C_1 \arctan\left(e^{- \,\sqrt{6} \, {\hat \phi}}\right)\, + \, C_2 $    \\
\null&\null\\
\hline
\null&\null\\
$I_7$ & $\! C_1 \, \Big(\cosh\,\gamma\,\sqrt{\ft 32} \, {\hat \phi} \Big)^{\frac{2}{\gamma} \, - \, 2}\, + \, C_2 \Big( \sinh\,\gamma\,\sqrt{\ft 32} \, {\hat \phi} \Big)^{\frac{2}{\gamma} \, - \, 2}$  \\
\null&\null\\
\hline
\null&\null\\
$I_8$ & $\! C_1 \left(\cosh [ \gamma \, \sqrt{6} \, {\hat \phi}] \right)^{\ft 1 \gamma -1}\,\cos\left[\left(\ft 1 \gamma -1\right)\, \arccos\left(\tanh[\gamma\, \sqrt{6} \, {\hat \phi}]\,+\,C_2\right)\right]$ \\
\null&\null\\
\hline
\null&\null\\
$I_9$ & $\! C_1 \ e^{\gamma\,\sqrt{6} \, {\hat \phi}} \  + \ C_2 \
e^{\frac{1}{\gamma}\,\sqrt{6} \, {\hat \phi}}\ \ $ \\
\null&\null\\
\hline
\null&\null\\
$I_{10}$ & $\! C_1 \ e^{\gamma\,\sqrt{6} \, {\hat \phi}} \ \cos\left( \sqrt{6} \, {\hat \phi} \, \sqrt{1-\gamma^2} \ + \, C_2 \right) \ \ $ \\
\null&\null\\
\hline
\hline
\end{tabular}
}
\caption{The families of integrable potentials classified in \cite{noicosmoitegr}. In all   cases  $C_i$ should be real parameters and $\gamma \in \mathbb{Q}$ should just be a rational number. Furthermore the field  ${\hat \phi}$ admits a canonical normalization in $D=4$ and corresponds to the standard normalization $M_P^2 \, = \, 1$. }
\label{tab:families}
\end{table}
The list of integrable potentials is composed of $10$ families each depending on one or more parameters and $28$ sporadic cases which are parameterless apart from an overall scale in front of the potential. The $10$ families are listed in table (\ref{tab:families}) where we already present them in the normalizations used in the current literature on inflationary models.
\begin{table}[h!]
\centering
{\small
\begin{tabular}{|lc|}
\hline
\null & \null \\
\null & Sporadic Integrable Potentials \\
\null & \null \\
 \null & $\begin{array}{lcr}\mathcal{V}_{Ia}( {\hat \phi})
   & = & \frac{\lambda}{4} \left[(a+b)
   \cosh\left(\frac{3}{5}\sqrt{6} \, {\hat \phi}\right)+(3 a-b)
   \cosh\left(\frac{1}{5}\sqrt{6} \, {\hat \phi}\right)\right] \end{array}$ \\
   \null & \null \\
   \null & $\begin{array}{lcr}\mathcal{V}_{Ib}( {\hat \phi}) & = & \frac{\lambda}{4} \left[(a+b)
   \sinh\left(\frac{3}{5}\sqrt{6} \, {\hat \phi}\right)-(3 a-b)
   \sinh\left(\frac{1}{5}\sqrt{6} \, {\hat \phi}\right)\right] \end{array} $\\
   \null & \null \\
   where & $ \left\{a,b\right\} \, = \, \left\{
\begin{array}{cc}
 1  & -3  \\
 1  & -\frac{1}{2} \\
 1  & -\frac{3}{16}
\end{array}
\right\} $\\
\null & \null \\
\hline
\null & \null \\
\null & $\begin{array}{lcr}
\mathcal{V}_{II}( {\hat \phi})
   & = & \lambda  \left(a \cosh ^4\left(\frac{\hat{\phi} }{\sqrt{6}}\right)+c \sinh
   ^2\left(\frac{\hat{\phi} }{\sqrt{6}}\right) \cosh ^2\left(\frac{\hat{\phi}
   }{\sqrt{6}}\right)+b \sinh ^4\left(\frac{\hat{\phi} }{\sqrt{6}}\right)\right)\nonumber\ ,
\end{array}$\\
\null & \null \\
where  & $\left\{a,b,c\right\} \, = \, \left\{
\begin{array}{ccc}
 1  & 1  & -2
   \\
 1  & 1  & -6
   \\
 1  & 8  & -6
   \\
 1  & 16  & -12
    \\
 1  & \frac{1}{8} &
   -\frac{3}{4} \\
 1  & \frac{1}{16} &
   -\frac{3}{4}
\end{array}
\right\} $\\
\null & \null \\
\hline
\null & \null \\
\null & $\begin{array}{lcr}
\mathcal{V}_{IIIa}({\hat \phi}) & = & \frac{\lambda}{16} \left[\left(1-\frac{1}{3
   \sqrt{3}}\right) e^{-3 \sqrt{6} \, {\hat \phi}
   /5}+\left(7+\frac{1}{\sqrt{3}}\right)
   e^{-\, \sqrt{6} \, {\hat \phi}
   /5} \right. \\
   && \left. +\left(7-\frac{1}{\sqrt{3}}\right)
   e^{ \sqrt{6} \, {\hat \phi} /5}+\left(1+\frac{1}{3
   \sqrt{3}}\right) e^{3 \sqrt{6} \, {\hat \phi}
   /5}\right]\ .
   \end{array}$\\
\null&\null\\
\hline
\null & \null \\
\null &$\begin{array}{lcr}
  \mathcal{V}_{IIIb}( {\hat \phi}) &=& \frac{\lambda}{16} \left[\left(2-18
   \sqrt{3}\right) e^{-3 \sqrt{6} \, {\hat \phi}
   /5}+\left(6+30 \sqrt{3}\right) e^{- \,
   \sqrt{6} \, {\hat \phi} /5}\right.\\
   &&\left. +\left(6-30
   \sqrt{3}\right) e^{  \sqrt{6} \, {\hat \phi}
   /5}+\left(2+18 \sqrt{3}\right) e^{3
   \sqrt{6} \, {\hat \phi} /5}\right]
   \end{array}
 $ \\
 \null&\null\\
\hline
\end{tabular}
}
\caption{In this Table of the $28$ sporadic integrable potentials classified in \cite{noicosmoitegr} we have selected those that are pure linear combinations of exponentials. As for the previous cases they are written in the normalizations where $M_P^2 \,= \, 1$ and the kinetic term of ${\hat \phi}$ is canonical.}
\label{Sporadic}
\end{table}
A sublist of the $28$ sporadic cases that are relevant to the discussion of this paper is displayed in table \ref{Sporadic}.
\begin{table}[h!]
\begin{center}
{\scriptsize
\begin{tabular}{|c|c|c|c|c|c|}
\hline
Curv. & Gauge Group & $V(\sqrt{\ft 12} \, {\hat \phi})$ & Values of $\nu$  & Values of $\mu$ & Mother series \\
\hline
\hline
\null & \null & \null & \null & \null & \null \\
$-\, 8 \,{\hat \nu}^2$ & $\mathrm{U(1)}$ & $\left( \cosh\left({\hat \nu} \, {\hat \phi} \right) \, + \, \mu\right)^2 $ & ${\hat \nu} \, = \, \frac{1}{2}\,\sqrt{\frac{3}{2}}$ & $\mu \, = \,0$ & $I_1$ or $I_7$ with $\gamma \, = \, \ft 12$  \\
\null & \null & \null & \null & \null & \null \\
\hline
\null & \null & \null & \null & \null & \null \\
$-\, 8 \,{\hat \nu}^2$ & $\mathrm{U(1)}$ & $\left( \cosh\left({\hat \nu} \, {\hat \phi} \right) \, + \, \mu\right)^2 $ & ${\hat \nu} \, = \, \frac{1}{\sqrt{6}}$ & $\mu \, = \,1$ &   $I_7$ with $\gamma \, = \, \ft 13$  \\
\null & \null & \null & \null & \null & \null \\
\hline
\null & \null & \null & \null & \null & \null \\
$-\, 8 \,{\hat \nu}^2$ & $\mathrm{U(1)}$ & $\left( \cosh\left({\hat \nu} \, {\hat \phi} \right) \, + \, \mu\right)^2 $ & ${\hat \nu} \, = \, \frac{1}{\sqrt{6}}$ & $\mu \, = \, - \,1$ &   $I_7$ with $\gamma \, = \, \ft 13$  \\
\null & \null & \null & \null & \null & \null \\

\hline
\null & \null & \null & \null & \null & \null \\
$-\, 8 \,{\hat \nu}^2$ & $\mathrm{SO(1,1)}$ & $\left( \sinh\left({\hat \nu} \, {\hat \phi} \right) \, + \, \mu\right)^2 $ & ${\hat \nu} \, = \, \frac{1}{2}\,\sqrt{\frac{3}{2}}$ & $\mu \, = \,0$ & $I_1$ or $I_7$ with $\gamma \, = \, \ft 12$  \\
\null & \null & \null & \null & \null & \null \\
\hline
\null & \null & \null & \null & \null & \null \\
$-\, 8 \,{\hat \nu}^2$ & $\mathrm{parabolic}$ & $\left( \exp\left({\hat \nu} \, {\hat \phi} \right) \, + \, \mu\right)^2 $ & ${\hat \nu} \, = \, \mbox{any}$ & $\mu \, = \,0$ & all pure exp are integ.  \\
\null & \null & \null & \null & \null & \null \\
\hline
\end{tabular}
}
\end{center}
\caption{In this table we mention which particular values of the curvature and of the Fayet Iliopoulos constant yield cosmological potentials that are both associated to constant curvature and integrable according to the classification of \cite{noicosmoitegr}}
\label{integpotenziallini}
\end{table}
Finally let us recall from \cite{sergiosashapietroOne} the list of constant curvature integrable potentials. They are listed in table \ref{integpotenziallini}.
\newpage
\section{A Mathematical Appendix on the Topology of Isometries}
\label{mathtopo}
In this appendix we provide a mathematically more rigorous illustration of the criteria discriminitating among elliptic, parabolic and hyperbolic isometries of a two dimensional manifold whose metric is written in the standard form utilized throughout this paper namely:
\begin{equation} \label{metric-Killing}
  ds^2 \, = \,  d\phi^2 \, + \, f(\phi)^2 \, dB^2,
  \end{equation}
In relation with minimal supergravity models the function $f(\phi)$ is obviously the first derivative $\mathcal{P}^\prime(\phi)$
with respect to the canonical coordinate $\phi$ of the momentum map $\mathcal{P}(\phi)$, yet the question we address here is of a much more general nature and it is purely mathematical. Considering the metric (\ref{metric-Killing}) as god-given, it obviously admits the one dimensional group of isometries $B\, \to \, B+c$ for any choice of the smooth function $f(\phi)$ parameterizing the metric coefficient and the question is what is the topology of such a group, is it compact or non-compact, and in the second case is it parabolic or hyperbolic. When we deal with a constant negative curvature manifold, namely with the coset $\mathrm{SL(2,\mathbb{R})}/\mathrm{O(2)}$ these questions have a precise answer within Lie algebra theory, since the considered one-dimensional group of isometries $\mathcal{G}_{iso}$ is necesarily a subgroup of $\mathrm{SL(2,\mathbb{R})}$ and as such its generator $\mathfrak{g}\in \slal(2,\mathbb{R})$ can be of three types:
\begin{description}
  \item[a)] $\mathfrak{g}$ is compact, which means that, as a matrix, in whatever representation of the Lie algebra $\slal(2,\mathbb{R})$ it is diagonalizable and its eigenvalues are purely imaginary. In this case the one-dimensional subrgroup is topologically a circle $\mathbb{S}^1$ and isomorphic to $\mathrm{U(1)}$. We name \textit{elliptic} the isometry group $\mathcal{G}_{iso}$ generated by such a $\mathfrak{g}$.
 \item[b)] $\mathfrak{g}$ is non-compact and semisimple, which means that, as a matrix, in whatever representation of the Lie algebra $\slal(2,\mathbb{R})$, it is diagonalizable and its eigenvalues are real and non vanishing. In this case the one-dimensional subgroup is topologically a line $\mathbb{R}$ and it is isomorphic to $\mathrm{SO(1,1)}$. We name \textit{hyperbolic} the isometry group $\mathcal{G}_{iso}$ generated by such a $\mathfrak{g}$.
  \item[c)] $\mathfrak{g}$ is non-compact and nilpotent, which means that, as a matrix, in whatever representation of the Lie algebra $\slal(2,\mathbb{R})$, it is nilpotent and its eigenvalues are zero. In this case the one-dimensional subrgroup is topologically a line $\mathbb{R}$. We name \textit{parabolic} the isometry group $\mathcal{G}_{iso}$ generated by such a $\mathfrak{g}$.
\end{description}
The interesting question is whether the characterization of an isometry as \textit{elliptic,parabolic} or \textit{hyperbolic} can be reformulated in pure geometrical terms and applied to cases where there is no ambient Lie algebra for the the unique one-dimensional continuous isometry $\mathcal{G}_{iso}$. In this respect it is useful to remark that a metric of type (\ref{metric-Killing}) implies a  fibre-bundle structure of the underlying two-dimensional manifold $\Sigma$:
\begin{equation}\label{fibrillazione}
  \Sigma \, = \, \mathrm{P}(\mathbb{R},\mathcal{F},\mathcal{G}_{iso}) \, \rightarrow \, \mathbb{R}
\end{equation}
where the base manifold is the real line $\mathbb{R}$ spanned by the coordinate $\phi \, \in \, \left[-\infty\, , \, +\infty\right]$, the structural group is the one-dimensional isometry group $\mathcal{G}_{iso}$ and the standard fibre $\mathcal{F}$ is a one dimensional space on which $\mathcal{G}_{iso}$ has a transitive action. In other words the manifold $\Sigma$ is fibered into orbits of the isometry group. An explicit geometrical realization of this fibration in the three cases was already provided in the main text by means of the three types of parametric surfaces encoded in:
\begin{enumerate}
  \item Eq.s (\ref{pianinidivano2}) which realize a surface in three-dimensional Minkowski space which is fibered in circles $S^1$ representating the orbits of an elliptic isometry group $\mathcal{G}_{iso}$.
       \item Eq.s (\ref{pianinidivano3}) which realize a surface in three-dimensional Minkowski space which is fibered in hyperbolae representating the orbits of a hyperbolic isometry group $\mathcal{G}_{iso}$.
  \item Eq.s (\ref{pianinidivanoparab}) which realize a surface in three-dimensional Minkowski space which is fibered in parabolae representating the orbits of a parabolic isometry group $\mathcal{G}_{iso}$.
 \end{enumerate}
As we argued in the main text, providing also some counterexamples, the subtle point is that the explicit geometric construction as a parametric surface fibered in circles, parabolae or hyperbolae, which a priori seems always possible, should lead to a smooth manifold having no singularity and being simply connected.
\par
In more abstract terms the question was formulated by mathematicians for a single isometry $\Gamma$, even belonging to a discrete isometry group, not necessarily continuous and Lie, which can be characterized unambiguously as elliptic, parabolic, or hyperbolic, for Riemannian manifolds also of higher dimension than two, provided they are Hadamard manifolds.
\begin{definizione}
\label{adamardus}
A Hadamard manifold is a simply connected, geodesically complete Riemannian manifold $\mathcal{H}\, = \, \left(\mathcal{M},g\right)$ whose scalar curvature $R(x)$ is {\bf nonpositive definite and finite}, namely $-\infty \,<\, R(x) \leq\, 0, \quad \forall x \, \in \, \mathcal{M}$.
\end{definizione}
The virtue of Hadamard manifolds is that they allow for what is usually not available in generic Riemannian manifolds, namely the definition of a bilocal distance function  $d(x,y)$ providing the absolute distance  between any two points $x,\,y\,\in \,\mathcal{H}$. As we teach our students when introducing (pseudo)-Riemanian geometry and General Relativity, the concept of absolute space-(time) distance is lost in Differential Geometry and we can only define the lenght of any curve  $\beta^{\mu}(t)$ ($t\,\in\,[0\,, 1]$), which starts at the point $x^{\mu}\,=\,\beta^{\mu}(0)$ and ends at the point $y^{\mu}\,=\,\beta^{\mu}(1)$. Given the metric $g_{\mu \nu}(x)$ we introduce the  length functional which provides such a length:
\begin{equation} \label{length}
\ell(\beta) \, = \, \int_0^1 \sqrt{g_{\mu \nu} \frac{d \beta^{\mu}}{d t} \frac{d \beta^{\nu}}{d t}} dt
  \end{equation}
The curves corresponding to extrema of the length functional are the geodesics, but in a generic Riemannian manifold there is no guarantee that for any two-points $x,y\,\in \, \mathcal{M}$ there is an arc of geodesic connecting them that is an absolute minimum of the length functional and that such minimum is unique and non-degenerate. Instead the hypotheses characterizing Hadamard manifolds guarantee precisely this (see, e.g. \cite{Gromov1985} and references therein) and one can define the distance function:
\begin{equation} \label{distance}
\forall \, x,y \, \in \, \mathcal{H}\quad : \quad d(x,y) \, = \, \mbox{infimum} \left[\ell(\beta)\right]
  \end{equation}
Hence restricting one's attention to Hadamard manifolds one can introduce a very useful geometrical concept that allows for a geometrical classification of isometries $\Gamma$:
\begin{equation}\label{figliodimamma}
\Gamma \quad : \quad \mathcal{M }\, \to \, \mathcal{M } \quad\quad ; \quad\quad  \Gamma_\star \,\left[ ds^2_g \right] \, = \, ds^2_g
\end{equation}
where $\Gamma_\star$ denotes the pull-back of $\Gamma$.  The geometrical concept which provides the clue for such a classification is the displacement function defined below for any isometry $\Gamma$:
\begin{equation}\label{dispiazzo}
d_{\Gamma}(x)\,\equiv \,d(x,\Gamma x)
\end{equation}
\subsection{Classification of isometries of Hadamard manifolds $\mathcal{H}\, = \, \left(\mathcal{M},g\right)$}
The isometries of a Hadamerd manifold belong to the following types (see, e.g. \cite{Gromov1985} and references therein):
\begin{description}
  \item[a)] \textbf{elliptic}, if $d_{\Gamma}(x)$ attains an absolute minimum of vanishing displacement $\min_{x\in \mathcal{M}}d_{\Gamma}(x)\,=\,0$, or, to say it in other words, if and only if $\Gamma$ has a fixed point $x_0 \, \in \, \mathcal{M}$ in the interior of the manifold  for which $d\left(x_0, \, \Gamma x_0\right) \, = \, 0$.
  \item[b)] \textbf{hyperbolic}, if $d_{\Gamma}(x)$ attains an absolute minimum larger than zero $\min_{x\in \mathcal{M}}d_{\Gamma}(x)\,> \,0$, or equivalently if $\Gamma$ has two distinct fixed points on the boundary $\partial \mathcal{M}$ of $\mathcal{M}$
  \item[c)] \textbf{strictly parabolic}, if $d_{\Gamma}(x)$ never attains its infimum which is zero $\inf_{x\in \mathcal{M}}d_{\Gamma}(x)\,=\,0$, or equivalently if $\Gamma$ has just one fixed point on the boundary $\partial \mathcal{M}$ of $\mathcal{M}$;
  \item[d)]  \textbf{mixed}, if $d_{\Gamma}(x)$ does not attain its the infimum which is larger than zero: $\inf_{x\in \mathcal{H}}d_{\Gamma}(x)\,>\,0$.
\end{description}
The above classification of  isometries is a generalisation
to a nonconstant curvature case of the classification of isometries of the very particular constant curvature case, namely the Poincar\'e-Lobachevsky plane $\frac{\mathrm{SL(2,\mathbb{R})}}{\mathrm{O(2)}}$, where only the isometries (a), (b) and (c) are realized.
\subsection{Application to the K\"ahler surfaces relevant to Minimal Supergravity Models}
Not all K\"ahler surfaces $\Sigma$ in the image of the $D$-map are Hadarmard since the curvature sometimes becomes positive in the interior of the manifold but most of them are such  and moreover the limiting curvature of the boundary is negative for all models. Therefore it makes sense to utilize the above geometric classification of isometries and verify that it just agrees with the criteria based on asymptoti expansions of the funtion $J(C)$ utilized in the main text in order to discriminate among elliptic, parabolic and hyperbolic groups. Negative curvature guarantees the existence of a distance function, but probably in all considered examples such a distance function is well defined in spite of the existence of positive curvature domains in the deep interior of the manifold.
\par
Hence with reference to the metric (\ref{metric-Killing}) let us consider the isometry $\Gamma$ corresponding to $B$-shifts:
\begin{equation}\label{isom1}
 B \ \rightarrow \, \Gamma\,B\,=\,B\,+\,\delta \,,
  \end{equation}
where $\delta$ is a constant parameter, let us assume that the curvature
 \begin{equation} \label{curvature1}
  R \, = \,  -\, \ft 12 \, \frac{\frac{d^2}{d \phi^2}f(\phi)}{f(\phi)},
  \end{equation}
 fulfills the  Hadamard condition: $-\infty \,<\, R\leq\, 0$ and let us apply  the classification scheme introduced above.
\par
The first observation is the following. If the function $f(\phi)$ has neither a singularity nor a  zero (i.e., if $f(\phi) \, \neq \, \pm \infty \,\, and \,f(\phi) \, \neq \, 0$) both in the  range of the coordinates $\{\phi,\,B\}$ corresponding to the interior of the manifold $\mathcal{M}$ and for those limiting values corresponding to the boundary $\{\phi,\,B\}\,\in\,\partial \mathcal{M}$ then the metric (\ref{metric-Killing}) has no  coordinate singularity and  the isometry (\ref{isom1}) admits only one fixed point $B\,=\,\infty\,\in\,\partial \mathcal{M}$ on the boundary of the manifold. In this case the isometry $\Gamma$ is strictly parabolic, according to item (c) of the the above classification.
\par
On the other hand, if the function $f(\phi)$ possesses a coordinate singularity at some value of $\phi \,=\,\phi_0\,\in\,\mathcal{M}$ in the interior of $\mathcal{M}$, then in order to establish which is the type of the isometry $\Gamma$ one has to introduce a new coordinate system $\{\phi,\,B\}\,\rightarrow \,\{\widetilde \phi,\, \widetilde B\}$  such that the metric expressed in terms of the new coordinates is non-singular in the vicinity of the former coordinate singularity. The existence of such a coordinate system is guaranteed by the non-singularity of the curvature and by the smoothness of the manifold. If in the newly constructed coordinate system the isometry has  a fixed point   corresponding to the former coordinate singularity  then, according to item a) of the above classification, it is elliptic. Since this happens for alll elements of the isometry group $\mathcal{G}_{iso}$, this latter is a compact $\mathrm{U(1)}$ and the appropriate complex structure is $\mathfrak{z} \, = \, \zeta \, = \, \exp\left[ \delta \left (c \, - \,{\rm i} B\right)\right]$. Otherwise the isometry is certainly not elliptic and non-compact.
\par
Summarizing, the necessary condition for the isometry $\Gamma$ to be elliptic is that the function $f(\phi)$ has a zero or a pole in the interior of $\mathcal{M}$ at some $\phi\,=\,\phi_0\,\equiv\,-\,\frac{a_1}{a_2}$, where $a_1$ and $a_2 \,>\,0$ are arbitrary constant parameters. In case such a singularity is  power--like, we conclude that in a neighborhood  $ \mathbb{U}_{\phi_0}$ of $\phi_0$ we have:
\begin{equation} \label{singularity0}
 f(\phi)|_{\phi\,\in\, \mathbb{U}_{\phi_0}} \, = \,  (a_2\,\phi\,+\,a_1)^n
  \end{equation}
  where $n$ is a positive or negative integer. Comparing eq. (\ref{curvature1}) we see that the condition of a regular and finite curvature is fulfilled if and only if $n\,=\,1$. In other words  the function $f(\phi)$ has the following behavior at $\phi \, = \, \phi_0$:
\begin{equation} \label{singularity}
 f(\phi)|_{\phi\,\in\, \mathbb{U}_{\phi_0}} \, = \,  a_2\,\phi\,+\,a_1 + \mathcal{O}\left[\left (\phi \, - \, \phi_0\right)^3\right]
  \end{equation}
Correspondingly the curvature is zero at leading order:
\begin{equation} \label{singularityCurv}
 R|_{\phi\,\in\, \mathbb{U}_{\phi_0}}\,=\,0\, + \, \mathcal{O}\left[\left (\phi \, - \, \phi_0\right)^3\right]
  \end{equation}
In the new coordinate system $\{x,\, y\}$, $\{\phi,\,B\}\,\rightarrow \,\{x,\, y\}$, defined by
\begin{equation}\label{newcoord}
  x \, = \,   (\phi \,+\,\frac{a_1}{a_2})\,\cos (a_2\,B)\,, \quad   y \, = \,   (\phi \,+\,\frac{a_1}{a_2})\,\sin (a_2\,B)\,,
  \end{equation}
the metric (\ref{metric-Killing}) becomes
\begin{eqnarray} \label{flat0}
  ds^2|_{\phi\,\in\, \mathbb{U}_{\phi_0}} & \simeq &d\phi^2 \, + \, (a_2\,\phi \,+\, a_1)^2 \, dB^2 \nonumber\\
   & = &  dx^2 \, + \, dy^2,
  \end{eqnarray}
and the isometry transformations (\ref{isom1}) takes the following form:
\begin{equation}\label{isom2}
 \{x,\,y\} \ \rightarrow \,  \{x \, \cos \delta\,+\,y\,\sin\delta\,,\, -\, x \, \sin \delta\,+\,y\,\cos\delta\}\,,
  \end{equation}
 The original coordinate singularity has disappeared, but in the new coordinates (\ref{newcoord}) the isometry (\ref{isom2}) acquires the fixed point $\{x_0\,=\,0\,,\,y_0\,=\,0\}$, $\{0,\,0\}\,\rightarrow \,\{0,\, 0\}$, in the interior of $\mathcal{M}$. Hence if the above situation is verified  according to titem a) of the above classification the isometry group is elliptic.
\par
Consider next the behavior of the VP coordinate, defined by eq.(\ref{sodoma}), in the neighborhood of $\phi_0$. To leading order we have
\begin{equation}\label{singularity1}
 \phi \,\rightarrow\, C\,\simeq\, \frac{1}{a_2}\,\ln\,(a_2\,\phi\,+\,a_1)\, + \, \mathcal{O}\left[\left (\phi \, - \, \phi_0\right)^{-1}\right] \quad \quad \Rightarrow \quad
  \phi_0 \,\Leftrightarrow\, C_0\,=\,-\infty\
\end{equation}
so that the  metric (\ref{metric-Killing}) becomes
\begin{equation}\label{singularity2}
  ds^2|_{C\,\in\, \mathbb{U}_{C_0}} \, \simeq \,e^{2 \, a_2 \, C}\, \left(\mathrm{d}B^2\,+\,\mathrm{d}C^2\right)
\end{equation}
in the $C_0$--neighborhood $C\,\in\, \mathbb{U}_{C_0}$. Inspection of the latter formula shows that it reproduces the criterion to decide that the isometry is elliptic advocated in eq.(\ref{pagnocorto}).
\begin{equation}\label{singularity3}
   \ft 12 \,   \frac{d^2}{d C^2}J(C) |_{C\,\in\, \mathbb{U}_{C_0}} \, = \,e^{2 \, a_2 \, C}|_{C\,\in\, \mathbb{U}_{C_0}}\,\rightarrow\, 0
\end{equation}
Let us stress that the fixed point in the interior of the manifold required for an elliptic interpretation of the isometry group is just, in the phsical language the origin of the manifold where the K\"ahler metric becomes approximately the flat one, required for a quadratic limit of the $\sigma$-model and a linear representation of the symmetry on the complex field.
\par
Let us now turn to the case where the singularity of the metric coefficient  is of the exponential type, namely for $\phi_0\,=\, \infty$ and for  $\phi\,\in\, \mathbb{U}_{\phi_0}$, we have
\begin{equation} \label{singularityExp1}
 f(\phi)|_{\phi\,\in\, \mathbb{U}_{\phi_0}} \, = \,a_1\, e^{a_2\,\phi}\,, \quad a_2\,>\,0
  \end{equation}
this behavior is also consistent with the regularity of the curvature $R$ (see eq.(\ref{curvature1})), which, in this case takes a finite negative value in the leading order approximation:
\begin{equation} \label{singularityCurv1}
 R|_{\phi\,\in\, \mathbb{U}_{\phi_0}}\,\simeq\,-\,a_2^2 \, + \, \mbox{subleading terms}
  \end{equation}
The  metric (\ref{metric-Killing}) reproduces locally the metric of the hyperbolic (Poincar\'e -Lobachevsky)
plane
\begin{equation} \label{Lobachevsky1}
  ds^2|_{\phi\,\in\, \mathbb{U}_{\phi_0}} \, \approx \,  d\phi^2 \, + \, \,a^2_1\, e^{2 a_2\,\phi} \, dB^2
  \end{equation}
for which it is well known that the value of $\phi_0\,=\infty $ corresponds to the boundary $\partial \mathcal{M}$. If the function $f(\phi)$ does not have other singularities of the exponential type, but (\ref{singularityExp1}), then one can immediately conclude that the isometry (\ref{isom1}) is the strictly parabolic  according to item c) of the above classification, since it possesses jus a single fixed point $B\,=\,\infty$ on the boundary $\partial \mathcal{M}$.
\par
If besides the singularity (\ref{singularityExp1}) the function $f(\phi)$ possesses a second   exponential singularity at $\widetilde \phi_0\, =\, -\,\infty $ for $\phi\,\in\, \mathbb{U}_{\widetilde \phi_0}$, namely
\begin{equation} \label{singularityExp2}
 f(\phi)|_{\phi\,\in\, \mathbb{U}_{\widetilde \phi_0}} \, = \,\widetilde a_1\, e^{-\widetilde a_2\,\phi}\,, \quad \widetilde a_2\,>\,0\,,
  \end{equation}
then by the same token as above we come to the conclusion  that the point $\widetilde \phi_0$ belongs  to the boundary of  another hyperbolic plane locally isomorphic to the neighborhood $U_{\tilde{\phi}_0} \subset \mathcal{H}$ and that isometry (\ref{isom1}) possesses a second fixed point on such a boundary. Hence the  isometry is  hyperbolic according to item (b) of the above classification and since this applies to all elements of the isometry group $G_{iso}$ this latter is hyperbolic and isomorphic to $\mathrm{SO(1,1)}$.
\par
One can not exclude  the existence of more sophisticated types of $f(\phi)$--singularities, besides the above described power-like  and exponential one, that might be consistent with the regularity of the curvature $R$ (\ref{curvature1}), yet in all examples considered in the main text no other singularities than these two are met.
\par
Relying on these resuts we can summarize the geometric criteria for the classification of isometries in two-manifolds with a metric of  type (\ref{metric-Killing}) which are of the Hadamard type
\begin{description}
  \item[a)] \textbf{elliptic}, if the function $f(\phi)$ possesses a first order zero, i.e. $f(\phi)|_{\phi\,\in\, \mathbb{U}_{\phi_0}}\,=\,a_2\,(\phi\,-\,\phi_0)$;
  \item[b)] \textbf{hyperbolic}, if the function $f(\phi)$ possesses two different leading exponential singularities at $\phi^{(\pm)}_0 \,=\,\pm\,\infty$, i.e. $f(\phi)|_{\phi\,\in\, \mathbb{U}_{\phi^{(\pm)}_0}}\,=\,a^{(\pm)}_1\, e^{\pm \,a^{(\pm)}_2\,\phi}$ and $a^{(\pm)}_2\,>\,0$;
  \item[c)] \textbf{strictly parabolic}, if the function $f(\phi)$ possesses a single leading exponential singularity at either $\phi^{(+)}_0 \,=\,+\,\infty$ or $\phi^{(-)}_0 \,=\,-\,\infty$, i.e. $f(\phi)|_{\phi\,\in\, \mathbb{U}_{\phi^{(+)}_0}}\,=\,a^{(+)}_1\, e^{+ \,a^{(+)}_2\,\phi}$ or $f(\phi)|_{\phi\,\in\, \mathbb{U}_{\phi^{(-)}_0}}\,=\,a^{(-)}_1\, e^{- \,a^{(-)}_2\,\phi}$ and $a^{(\pm)}_2\,>\,0$.
\end{description}
The above characterization yields exactly the same result as the criteria based on the asymptotic behavior of $J(C)$ that have been utilized in the main text and this happens also for such models that do  not lead to exactly Hadamard manifolds, the curvature attaining somewhere also positive values. As an exemplification of the use of the above concepts we briefly reconsider from this point of view the flat models and the constant curvature models presented in \cite{sergiosashapietroOne}.
\subsubsection{Flat models}
The flat metric
\begin{equation} \label{flat2}
  ds^2 \, = \,  d\phi^2 \, + \, (a_2\,\phi \,+\, a_1)^2 \, dB^2
  \end{equation}
in case $a_2\,\neq \, 0$ possesses a coordinate singularity at
\begin{equation} \label{sing1}
  \phi\,=\, -\,\frac{a_1}{a_2}
  \end{equation}
  corresponding to a first order zero $f(\phi)$ at finite $\phi$. According to the above classification this implies that the isometry $B\, \to \, B+ \delta$ is elliptic.
  \par
In the case $a_2\,=\,0$ the metric (\ref{flat2}) becomes
\begin{equation} \label{flat3}
  ds^2 \, = \,  d\phi^2 \, + \, a_1^2 \, dB^2
  \end{equation}
and does not possess a coordinate singularity at all. This implies that the  isometry $B\, \to \, B+ \delta$  is strictly parabolic.
\subsubsection{Constant negative curvature models}
{}~{}~{}~{}~{\bf case A)}
\begin{equation}\label{giroscopioA}
    ds^2 \, = \,d\phi^2 \, + \, \sinh^2\left(\nu \,\phi\right) \, dB^2
\end{equation}
This metric possesses a coordinate singularity at $\phi\,=\,0$. In the neighborhood of $\phi\,=\,0$ at leading order it behaves as follows
\begin{equation} \label{giroscopioA1}
  ds^2 \,\approx \,  d\phi^2 \, + \, \nu^2\,\phi^2 \, dB^2
  \end{equation}
which modulo an inessential rescaling of the coordinate $B$ and a shifting the coordinate $\phi$ reproduces the metric (\ref{flat2}). Hence  its isometry (\ref{isom1}) is  elliptic in this case.
\par
{\bf case B)}
\begin{equation}\label{giroscopioB}
    ds^2 \, = \,d\phi^2 \, + \, \cosh^2\left(\nu \,\phi\right) \, dB^2
\end{equation}
This metric does not possess a coordinate singularity in the finite range of $\phi$, but it has two exponential singularities of the type (\ref{singularityExp1}) and (\ref{singularityExp2}). Hence the isometry (\ref{isom1}) is hyperbolic in this case.
\par
{\bf case C)}
\begin{equation}\label{giroscopioC}
    ds^2 \, = \,d\phi^2 \, + \, e^{2\,\nu \,\phi} \, dB^2
\end{equation}
This metric does not possess a coordinate singularity in the finite range of $\phi$, but it possesses a single exponential singularity either of the type  (\ref{singularityExp1}) or of the type (\ref{singularityExp2}). Hence the isometry (\ref{isom1}) is strictly parabolic in this case.

\end{document}